# Design and Implementation of a Distributed Middleware for Parallel Execution of Legacy Enterprise Applications

Que Thu Dung Nguyen

A Thesis

in

The Department

of

Computer Science and Software Engineering

Presented in Partial Fulfillment of the Requirements

for the Degree of Master of Computer Science at

Concordia University

Montreal, Quebec, Canada

June 2008

**CONCORDIA UNIVERSITY**

**School of Graduate Studies**

This is to certify that the thesis prepared

By: Que Thu Dung Nguyen

Entitled: Design and Implementation of a Distributed Middleware for Parallel Execution of Legacy Enterprise Applications

and submitted in partial fulfillment of the requirements for the degree of

**Master of Computer Science**

complies with the regulations of the University and meets the accepted standards with respect to originality and quality.

Signed by the final examining committee:

________________________ Chair

________________________ Examiner

________________________ Examiner

________________________ Co-Supervisor

________________________ Co-Supervisor

Approved by ____________________________________________
Chair of Department or Graduate Program Director

___________ 20___ ____________________________________________
Dean of Faculty

# ABSTRACT

**Design and Implementation of a Distributed Middleware for Parallel Execution of Legacy Enterprise Applications**

Que Thu Dung Nguyen

A typical enterprise uses a local area network of computers to perform its business. During the off-working hours, the computational capacities of these networked computers are underused or unused. In order to utilize this computational capacity an application has to be recoded to exploit concurrency inherent in a computation which is clearly not possible for legacy applications without any source code. This thesis presents the design an implementation of a distributed middleware which can automatically execute a legacy application on multiple networked computers by parallelizing it. This middleware runs multiple copies of the binary executable code in parallel on different hosts in the network. It wraps up the binary executable code of the legacy application in order to capture the kernel level data access system calls and perform them distributively over multiple computers in a safe and conflict free manner. The middleware also incorporates a dynamic scheduling technique to execute the target application in minimum time by scavenging the available CPU cycles of the hosts in the network. This dynamic scheduling also supports the CPU availability of the hosts to change over time and properly reschedule the replicas performing the computation to minimize the execution time. A prototype implementation of this middleware has been developed as a proof of concept of the design. This implementation has been evaluated with a few typical case studies and the test results confirm that the middleware works as expected.

# Acknowledgements

I am grateful to many people that helped and supported me in order to finish this thesis.

First of all, I would like to express my thankfulness to my supervisor, Dr. Jayakumar, for his invaluable guidance and help throughout my graduate study and research. I am always grateful to his consideration and peaceful characteristic, which helped me overcome difficulties during my life of study.

I would like to send my special thank to Dr Radhakrishnan for his financial support during my research time. His tacit help encouraged and gave me motivation in order to finish my thesis.

In addition, I would like to thank Concordia University in Montreal, especially the faculty of the Department of Computer Science for opening a door and all the advice during my study time.

Thanks all my best friends who stimulated me over the two years of research; especially thanks to Joseph for the discussions about Linux issues.

Finally, this thesis would never accomplish without the never-ending support and love of my grandma, my aunt, my parents, my brother and sister. Thanks Viet, my beloved, for always being there for me.

# Table of contents

# List of Figures

# CHAPTER 1: INTRODUCTION

## 1.1 Context

Typical enterprises use many computers in a local area network (LAN) to conduct their businesses. Most of these computers are used heavily during the normal working hours and are underused or even unused during other times. The goal of this thesis is to develop a distributed middleware, running on each of the computers within the LAN, which can automatically run a legacy application on multiple computers in parallel. Such a middleware, in addition to using the otherwise wasted CPU cycles of these networked computers, can exploit the potential concurrency in the application to improve performance. However, since most of these legacy applications are available only as binary executable code (not in source code form), it is a real challenge to correctly run the binary code in parallel on multiple computers without knowing the potential concurrency, and more importantly conflicts, that can only be identified from the source code.

The idea behind this middleware is very simple. The computers running the middleware periodically broadcast to each other their CPU availability so that all of them know how much slack (unused) CPU power is available in each computer that can be used to run the target application. When the target application is started in one of the computers, the middleware running on that computer (called the server) becomes the controller of that run of the application. The server first starts a process running the binary code of the target application in its host computer. The server also appropriately starts processes on other networked hosts (called clients) running the same executable code, thereby running multiple processes on different computers running the same binary code. Obviously, during the execution, these parallel processes will access the application data in parallel and the middleware should guarantee that all such parallel accesses are safe and non-conflicting.

Thus, the middleware monitors/observes the parallel processes of the target application and tries to capture/intercept all such data accesses. Since it is not possible to do this for access to data stored in main memory, the middleware tries to capture data access system calls that go through the operating system kernels of the involved hosts. That is, the middleware observes all system calls accessing data stored in storage devices. The server and clients coordinate such storage data accesses to guarantee that they are safe and conflict free. Even though this is not possible in general, there are certain classes of applications (for example, applications such as banking system, web crawler, Genomic search, etc.) which perform the same computation on different parts of the data sequentially. That is, the input data to these applications can be partitioned into disjoint parts and the computation can be performed on these parts in parallel. For such applications, the server partitions the data and sends different parts to different clients so that the same computation is performed at the different computers in parallel and the results from these computers are collected by the server, properly pieced together to form the output of the application and stored in the storage device.

The above description of the middleware assumed that only the type and the binary executable code of the target application is available. However, in some cases (for example, in compiling the daily software build by a developer using a makefile), additional information about how the application performs its computation may be available. Such information may also be provided by the user (for example, in a sorting application, the input may be partitioned and each of the partition may be sorted independently and the user may specify how the sorted partitions should be merged to construct the overall sorted data). In such cases, the middleware identifies the independent tasks of the target application by analyzing the additional information available and performs these independent tasks in parallel, executing one of the tasks in each of the participating computer.

Furthermore, the availability of computers within the LAN could be dynamic. Additional host(s) may become available while the middleware is executing the target application and a host currently executing one of the parallel processes may have to perform other computation unrelated to the target application and may not be available to execute the target application process running on it. Thus, the middleware has to manage the parallel processes of the target application in a more dynamic fashion ensuring that processes are properly suspended (when the host computer is not available), activated (when the host computer becomes available), and migrated (when another host becomes available that can run a process currently suspended in a different host), so on. Thus, the required middleware has to perform two major activities: (1) create processes to run the target application code on remote hosts on the LAN and (2) dynamically manage the parallel processes of the target application so that they don't affect the availability and responsiveness of the hosts they are running on.

The thesis assumes that only the binary code of the application is available and no knowledge about the source code is provided. Those assumptions make our approach a general solution but they are also the difficulties to overcome. We try to relax our assumption as much as possible to deal with general cases. However, the more information about the application is provided, the better the middleware is able to support the parallel execution of the application. Those cases will be described with scenarios and following approaches in the latter part of this chapter.

The difficulties of the problem make this thesis stand out from all the researches in the past. The key to successful parallelization is in determining an appropriate structure for the shared object (including the data storage). But such a structure was always fairly obvious given a high-level understanding of the basic source code or knowledge of the application semantics to build patterns to extract concurrency. Without the source code, we can not define parallel computation patterns in the same manner as others have done using locking mechanisms. Without knowledge about the application, we can not define the data access patterns as other semantic researches have done. To overcome this, we combine both an internal (to the application execution—capturing data access system calls) approach and an external (to the application execution—identifying the parallel tasks from additional information) approach and clarify all possible scenarios that might happen in a general application.

However there are applications which should not be applied on our middleware since there is no improvement in performance. Those are applications where the earlier computation updates the data and the latter computation depends on the updated data, such an application is a totally sequential process, that is, data accesses by the computation is not independent of each other. This is not the type of application we're interested in.

## 1.2 Related work

A few difficulties exist in our problem because the source code of the target application is not available and the knowledge about the target application is very limited or almost nothing (except for the application type). Our goal is to auto-replicate the binary executable code of the target application for concurrent execution to achieve a better utilization of the computing resources (hence, better performance) without changing the semantics of the application in spite of these difficulties.

The key to successful parallelization is determining an appropriate structure for the shared data object. Such a structure is fairly obvious from a high-level understanding of the source code or a high-level knowledge of the application semantics. A couple of existing approaches are exploiting this technique. These approaches typically improve the source code to enhance the concurrency performance of the target application. For example, in database management, many techniques use locking mechanisms or data distribution methods by observing transaction status and information [4, 6, 16, 17, 18, 22, 23]. Based on the semantics-based concurrency control, some methods suggest methodologies to extract concurrency from objects [7, 13] or to exploit concurrency at task level [8, 14]. The methods of [13] and [14] gave us a better idea of how to come up with a general approach for our middleware. However, those methods define patterns of application behavior, build up the methodology to extract independent tasks and improve concurrency only at compilation time, when a lot of information about the application is available through its source code.

Alternatively, another way to improve parallel processing is to exploit CPU scavenging, in which an external method (external to the target application) is used to utilize the CPU's availability to provide the concurrency and parallel processing. Grid computing applications [20] and SETI@HOME [19] are the two most famous examples for this method. However, there are a few differences between Grid applications and our approach. A Grid application is already designed to have each task independently performed on any available host. Each task will be downloaded from a server to a client host and the server controls everything from task distribution to data management. The client host only needs to run the given task and return the result back to the server. In our case, although the application type/characteristic is known, we only have the binary executable code of the whole application and the tasks are hidden within the application code. Thus, the middleware has to overcome the difficulties to realize the parallel processes semantically, run them concurrently on client hosts and request data record on server to feed these processes. Besides, a grid environment assumes to have infinite resources, whereas we have an additional requirement to provide the best utilization when the resource is limited in a LAN.

SETI@HOME is a world-wide project in which any free machine in the world available over the internet can perform the processing, but in a simple way that data can be partitioned using a pattern known to the server. Based on the pattern, the server finds a chunk of data and distributes it to any host volunteering for the data processing. In our case, data storage can be of any type, such as file or simple database and the pattern of data access is not known ahead of time. Our solution has to figure out this pattern by observing/learning the behavior of the application during the runtime. We are providing a solution to find the pattern to partition the data at runtime for a more general application.

Besides the above approaches that analyze an application to detect parallel processes, there are a few existing tools for distributively executing an application. Several of them support replication mechanisms. But there are hardly any special facilities to support replication meticulously since such issues need to be entirely dealt within the application. We give a brief comparison here to show that a general auto-replication approach to improve utilization is not considered carefully in most of these tools. Thus, we compare those tools on some criteria that our approach concentrates on to provide better concurrency for the application. Those criteria are auto-replication, shared/dependent data and operation order.

*Chorus* is a tool to provide concurrency at resource level [30]. This tool can be used to assist an application in replicating a service for high availability. This tool is similar to our middleware in many aspects such as managing each host independently and providing global services for group communication. However, Chorus does not provide the auto-replicating feature at runtime; although a new host can join the group, take the replicated resource and start the replicated service, but it does not specify how to do or manage this task. Chorus does not mention about the resource characteristic, such as if they can be partitioned or have dependency within the resource; but treats the resource as a whole. Chorus also does not mention the operation order; instead, uses a multicast protocol to synchronize operation actions among replicas.

*Isis* is a collection of procedural tools that can be linked directly into an application, providing it the capacity to create, join and multicast to process groups, replicating data and synchronizing the actions of group members accessing the data, performing operations for load-balancing and fault tolerance [30]. Each tool can be integrated orthogonally into the application. Replicating data is one of the techniques used by Isis to decrease the accessing time for intensively read data. Same as Chorus, Isis looks at the data as a whole. It concentrates on a virtual synchrony mechanism to maintain the atomic data rather than looking at the dependency among replicas with the partitioned data. Like Chorus, a new replicated service can be created but Isis does not mention how to manage it automatically, as well as the operation order.

*Harp* presents a lazy replication mechanism and exploits the semantics of distributed services to improve the availability [30]. Based on the semantics of the application, users can specify the operation order. Those replicas are statically created at the beginning. Users know their locations ahead of time. Harp does not support multi-operation transactions, but it provides methods to define each atomic operation so that the processing can be sped up. There is no dependency among operations or the shared data storage discussed in Harp.

*Paxos* is a tool to build a fault-tolerant database using the Paxos consensus algorithm [30]. The significant aspect, when Paxos operates on a replicated database, is to apply repeatedly the consensus algorithm on a sequence of input values on each replica. If the same sequence of operations is applied to the local database on each replica, eventually all replicas will end up with the same database content. Since this tool concentrates on building a fault-tolerant database, it does not look closely on how to speed up the application processing time by the replicating method. Like other tools designed for fault-tolerance or high availability, Paxos uses static replicas and try to make those replicas consistent or backup if one of them fails. Furthermore, Paxos does not concentrate on operation order or shared data.

Neither the current approaches nor the existing tools mention a general approach to auto-replicate an application in an appropriate way in the absence of source code or knowledge about the application.

## 1.3 Conceptual approach

As mentioned, the key to successful parallelization is determining an appropriate structure for the shared object (including the data storage) from a high-level understanding of the source code or a high-level knowledge of the application semantics to figure out a strategy to identify and exploit concurrency. In our case, since only the binary executable code of the target application is available, such approaches are not possible. Thus, we enforce some assumptions about the type of target applications and narrow down the scope of the problem from a general case to scenarios that can be solved or solved better, as described in the following.

We consider cases in which the data storage of the application can be split into disjoint partitions and each replica of the target application can operate on different data partition without any conflict. Moreover, we use the *transaction model* of data access, since the computation of each replica is isolated from the other replicas. It means that one replica performing the computation on one data partition/element is segregated from the other replicas, even if they operate on the same or different data element.

***Strategy 1**: No knowledge about the application is available; but data storage can be partitioned.*

The whole application will be replicated to idle hosts in the network. Since the data storage can be partitioned into independent chunks, we assume there is no dependency among these data partition. Each chunk of data will be given to one of the replicas. The middleware observes the data accessing operations of a replica for synchronization and directs the write back of results from each replica to the server to collect and store the result of the computation. However, since we do not know the characteristic of the application, the merging results from different partitions at the server may face some difficulties. For some application such as updating records of a database or a web crawler, the merging mechanism just needs to put all the processed records together. But for other applications such as sort, search, etc., some other actions need to be performed to merge the results. This issue can be overcome if more information about the application (as in Strategy 3) is known which helps the middleware in correctly merging the results.

***Strategy 2:** No knowledge about the application is available and data storage can not be partitioned.*

When the data storage can not be partitioned, we assume there are shared global variables among the data elements or the computation (by each replica) has to be performed on the whole data storage. In this case, the data storage should not be partitioned but kept on the server side. The middleware observes each replica at low level operations. When a replica accesses a shared variable, the middleware directs the access to the server. Similarly, every data request of each replica has to be made to the server in order to synchronize it. In this case, only applications with high levels of computation can achieve significant improvement since each replica utilizes the local idle capacity for processing its computation and only waits for data from the server. However, there are some applications, such as sort or search application, in which the computation has to be performed on the whole data storage. Even if the data storage can be partitioned (as in Strategy 1), the results from all replicas have to be merged at the end. If the information about how to merge the result is not available, the middleware still maintains the appropriate operations of the application but it does not fulfill the semantics of the original application. Thus, this case needs Strategy 3.

Strategy 1 and Strategy 2 are considered the *internal analysis* approach and Strategy 3 below is considered as the *external analysis* approach.

***Strategy 3****: Some knowledge about the application in provided.*

The knowledge about the application usually is the *structure of the application*. This information can be obtained either from the user/programmer (for example, how to statically merge the results from the replica) or other control information about how to run the application (for example, makefile indicating how to compile a software build). From this external information, the middleware constructs a priority dynamic tree of tasks, allocates the independent tasks to appropriate hosts in the network depending on their priority. The middleware continues to observe each task to determine whether the internal replication method can be applied on the tasks accessing the data storage. Thus, the middleware at the server host applies the external analysis approach while some of the client hosts apply the internal analysis approach. This combined approach gives a fine grain control and improves the performance in many possible ways. The result is much better than applying only one approach, as in Strategy 1 or Strategy 2.

In this thesis, we design the middleware for Strategy 1 and Strategy 3. Strategy 2 is mentioned in Chapter 6 as a possible future work. We do the implementation and testing for Strategy 1 on a couple of applications and the results show significant improvement in utilization and performance. The results are presented in Chapter 5.

The contributions of this thesis range from conceptual ideas to implementation, concentrating on the following points:

1. Develop two approaches—internal analysis and external analysis—to deal with target applications and exploit as much as possible the ability to enhance concurrency to improve application performance.
2. Create a dynamic scheduling approach for all hosts in the LAN to dynamically exploit the idle capacity of hosts in the LAN.
3. Develop a middleware which can wrap the binary executable code of the legacy target application and build up the server side with extended distributed features, combining two techniques: a wrapper and remote procedure call (RPC). The wrapper is to provide an interface to extend the limited features of the current application and RPC is used by the server to support concurrency, synchronization, and scalability.
4. Partition the data storage in an indirect way and maintain the synchronization during concurrent processing among replicas.

The rest of this thesis is organized as follows. Chapter 2 presents the solution approach in detail and justifies the design choices. Chapters 3 and 4 describe the middleware design and implementation details. Chapter 5 shows the test results and performance improvements. Finally, Chapter 6 concludes the thesis mentioning possible future work.

# CHAPTER 2: SOLUTION APPROACH

The last part of Chapter 1 (Section 1.3) presented the outline of the approach used in this thesis. This chapter gives more details on the approach, such as the assumptions, the characteristic of the target application and the conceptual view of the approach. Also, discussions on and justifications for the design choices are given to strengthen our points.

## 2.1 Technique

As described in Section 1.1., the required middleware has to perform two major activities:

(1) Create processes to run the target application code on remote hosts on the LAN and observe the data storage accesses of these processes by observing kernel system calls in order to guarantee safe and conflict-free accesses.

(2) Dynamically manage the parallel processes of the target application running on different hosts in the LAN so that they don't affect the availability and responsiveness of the hosts they are running on while improving their utilization.

The middleware is designed using the client-server model. Remote Method Invocation (RMI) is used to invoke functions implemented on server side. Communication between server and clients uses TCP/IP protocol. Furthermore, the middleware is designed for platform independence. The middleware can run on a LAN either in Linux or Windows environment.

## 2.2 Conceptual view

As described in Section 1.1, when the target application is started on a host, the middleware running on that host creates the server for that application which will control the application during this run.  This server first analyzes the characteristics of the target application to select the strategy to follow to run the application on multiple client hosts in the LAN, running the available binary executable code in each of the clients. The server also monitors the replica to observe and guarantee safe and conflict-free data storage accesses.

### 2.2.1 Server side components

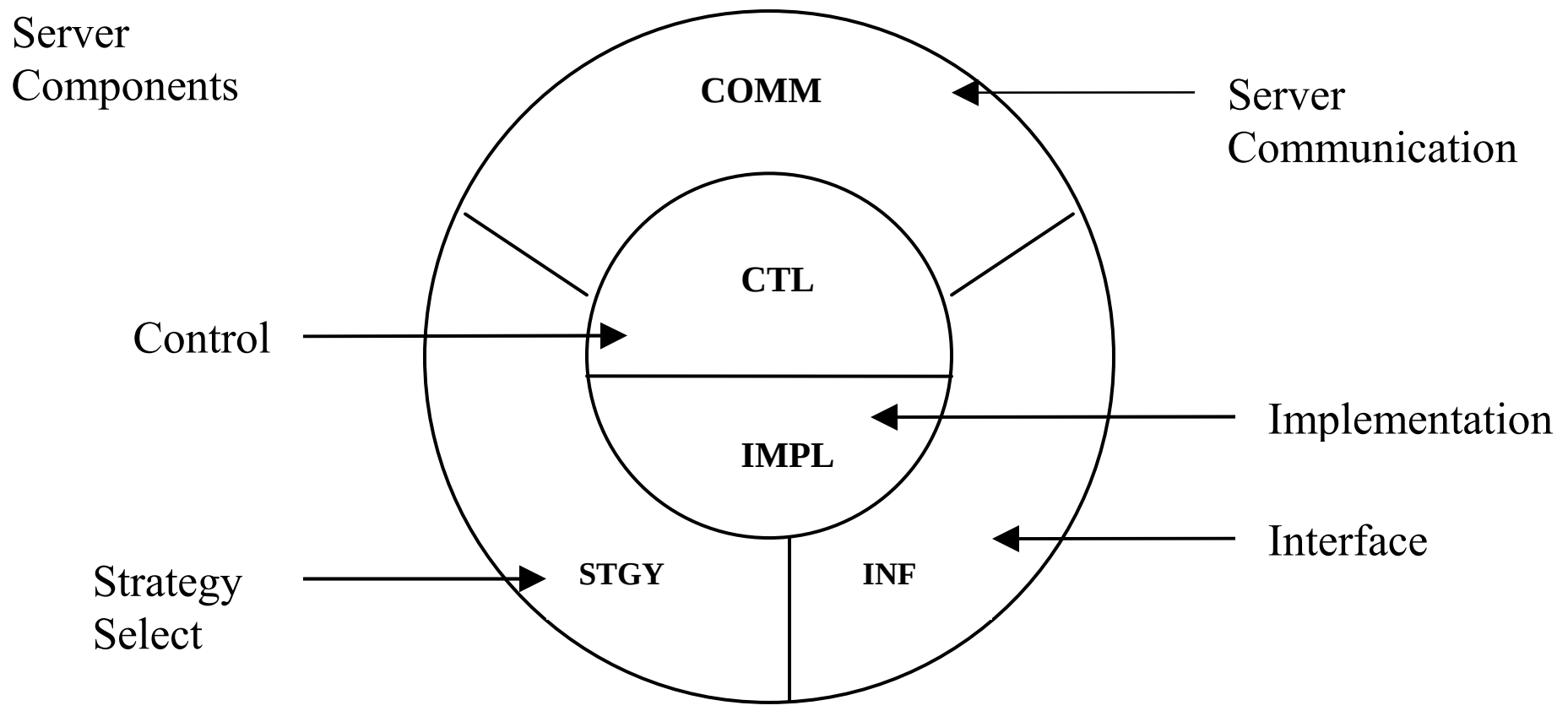

**Figure 1: Conceptual view of the Server model**

**Strategy Select component**

This component analyzes the application characteristics to select the strategy to run the target application on multiple hosts in the LAN. Once the strategy is decided, it will invoke other components to monitor and control the run of the application. This component will control the deployment of the application until the application completes its execution and terminates.

**Server Communication component**

Server Communication receives messages from client hosts to the server, such as registering the client host to join the middleware and current host's status, parses the client message and processes these messages on the server. Server Communication component also sends server commands to clients to request changing client host's behavior to improve performance of the middleware, such as launching a new replica, activating or suspending a process.

### Server Interface component

Interface component stands between implementation for data processing on the server and the data request from clients. Any client data request will be directed to the Interface component which will return the next available data record to the requesting client.

### Server Implementation component

The storage device/database is isolated and only accessed by the server host. Server provides an interface to clients to 'ReadData' from and 'WriteData' to the database. Server implements the 'ReadData' function by reading the database and returning the next sequential record. Server implements the 'WriteData' function by writing requests from the clients sequentially to the database. Server Implementation component is in charge for data management and data synchronization for the storage device/database.

### Control component

Based on the updated information from clients and the existing information, after a specific period of time, the Control component will re-schedule the computation and issue appropriate commands. These commands will be sent to clients through the Communication component.

### System component

System component observes system calls on host where the replica of the application is running.

## 2.2.2 Client side components

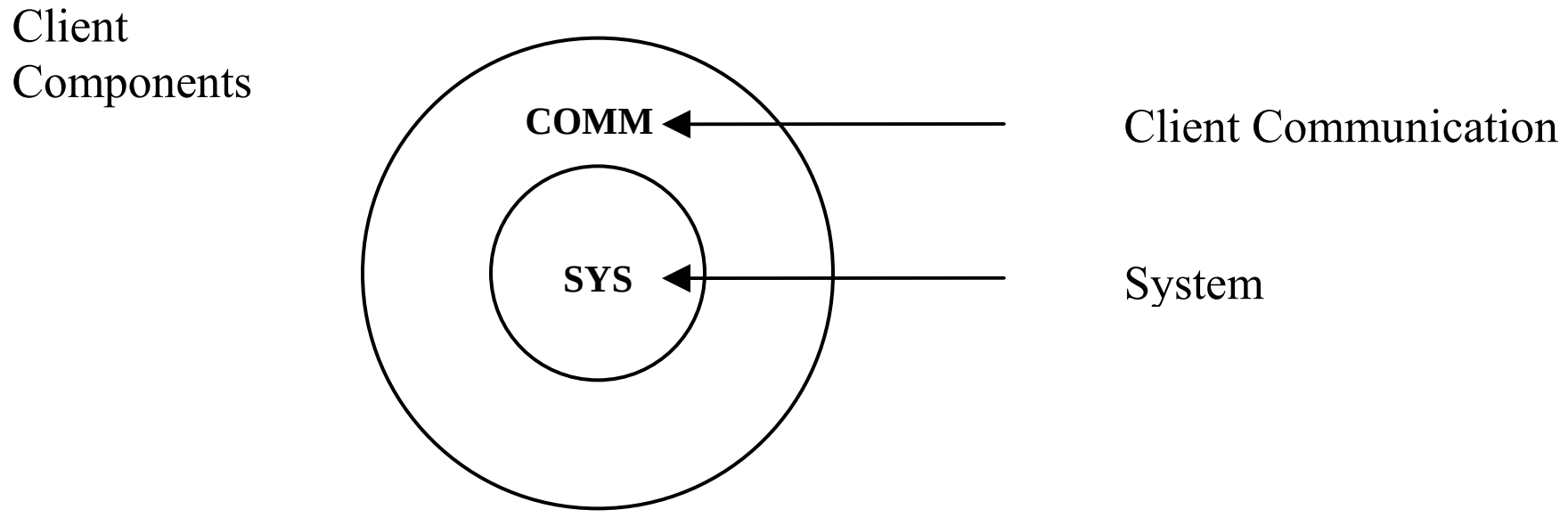

**Figure 2: Conceptual view of the Client model**

### Client Communication component

Client Communication component sends the CPU status to the server after a specific period of time and performs the command received from the server, such as download the replica, launch new replica, activate or suspend process.

### System component

System component observes all system calls on the client host. If a read system call was captured, it holds the current process and directs the read operation to the server by calling the ‘ReadData’ function implemented in the server, receives the returned record and resumes the waiting client process.  If a write system call was captured, then it directs the write operation to the server by calling the ‘WriteData’ implemented in the server.

## 2.2.3 Interactions between components

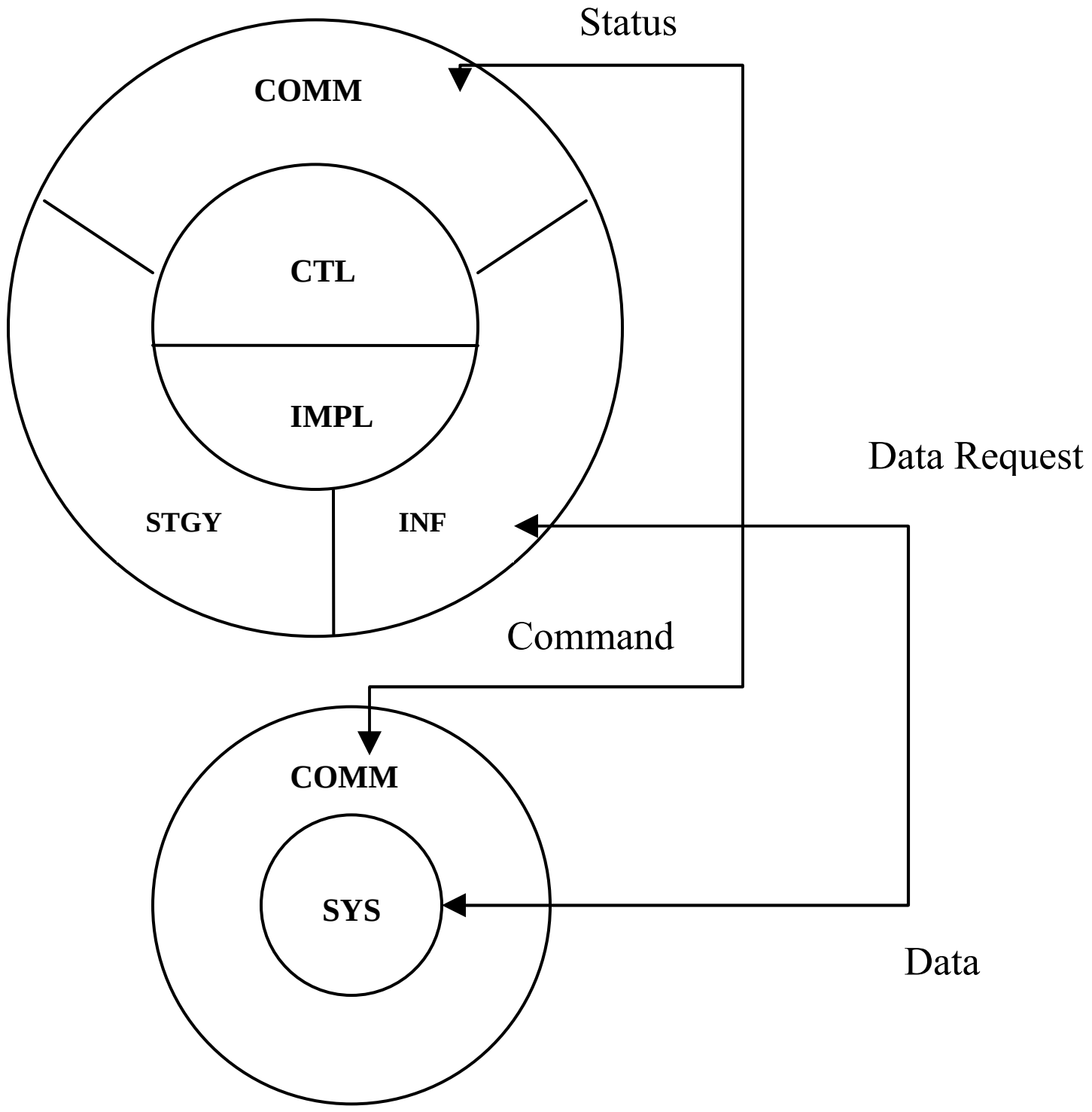

**Figure 3: Conceptual view of the middleware**

### Strategy Select component – Control component – Communication component

Depending on the selected strategy, the Strategy Select component invokes Control Component to check the hosts' status and to launch the appropriate data partition to deploy.

### Server Communication component – Client Communication component

The Server Communication component always listens to one specific port for messages. It receives only the *status message* from the other client hosts, sent by Client Communication component. It processes these messages after a specific period of time. From the *map* of current status of hosts it decides to launch the replica on the client hosts. It sends out *two* types of replica messages: if the replica has never been launched on a client host, then the server sends the *LAUNCH* message to that client host which will download the executable binary code of the target application from the server and will launch it on the client host. Otherwise, the server sends the *INVOKE_REPLICA* message to that client host which will invoke the replica to start. Besides these two messages, Server Communication also sends three other messages, namely *ACTIVATE*, *SUSPEND* and *MIGRATE*, to control the LAN performance. All those messages are issued by Control component.

The Client Communication component sends only the status message to the server. It receives either the *LAUNCH* message from the server and then downloads and launches a replica or the *INVOKE_REPLICA* message from the server and invokes the replica. When it receives the *ACTIVATE*, *SUSPEND* or *MIGRATE* messages from the server, it performs the indicated operation on the replica running at the client host.

**Control component – Server Communication component**

After receiving a signal from the Server Communication component informing that the LAN status has changed, such as number of available hosts or hosts' CPU availability, the server checks the capacity of each host, performs a rescheduling and then issues a command to *LAUNCH*, *INVOKE_REPLICA*, *SUSPEND*, *ACTIVATE* or *MIGRATE* to the clients. This decision is also made based on the scalability of the Server Implementation component on the server side so that the maximum number of running replicas and requests to Server Implementation component don't decrease the performance of the whole system. After Control component issues its decision, Server Communication component is invoked to send this message to client hosts.

**System component – Server Interface component**

This component is installed in all machines in the LAN. It acts as a middleware layer of the machine. The mission of this component is to

1. Capture all system calls occurring in the machine (for our purpose, we capture fork, write, read system calls);
2. Maintain a map of the current status of all the observed processes (maintaining the running and suspended queue);
3. Block the process whenever a system call is captured, perform the necessary tasks for this system call and then resume the process; and

4. System Component directs the captured system call along with its parameters to the Server Interface to implement these system calls on server data storage and to get the result back.

**Server Interface component - Server Implementation component**

Any function provided by the Server Interface component is implemented by the Server Implementation component. Mainly, the real unique data storage is located at and accessed by the server. The Server Interface provides data access functions with the parameters passed from the System component. The Server Implementation component will use all these parameters to form the system calls exactly as they were captured by the System component. Those system calls will be applied on the data storage on the server.

# 2.3 Justification of design choices

In the following, the design choices are justified by focusing on three aspects of the design: the general model, the dynamic scheduling and the partition issue. As mentioned in Section 1.3, the data storage of the application can be split into disjoint partitions. Each replica of the target application can operate on one data partition without any conflict. In addition, the middleware uses the *transaction model* of data processing. It means that one transaction performing its computation on one data element/partition is segregated from the other transactions, even if they perform on the same or different data element.

## 2.3.1 General approach

**Why two approaches and the combination of two?**

The external analysis approach provides an overall look on the application which helps to carry out any independent tasks of the application at other idle hosts in the LAN thereby distributing the processing. This is quite an old technique. However, from each subtask we continue the analysis to determine if the internal analysis can be applied. The internal analysis goes deeply into application operations (instead of tasks) to further enhance the concurrency. When the combination of these two analysis techniques is successfully applied on a task, there will be significant improvement in utilization/performance as will be shown in Chapter 3.

**How is the application wrapped to provide new features?**

First we give some background information about the wrapper technique and then we justify why we choose this technique in our design. In order to distributively execute a legacy application on a LAN, many new features (to communicate and control) need to be added to the application. There are two possible ways to do so, either using a toolkit or a wrapper. A toolkit provides powerful, prepackaged features, but a legacy application has to be recoded to use this toolkit. It can be hard to glue an object from a toolkit into an arbitrary code. This method is not suitable in our case because we do not have access to the source code of the application. A wrapper is a better approach to follow. A wrapper is a technology to overcome the limitations of an application by intercepting events at some interface between the unmodifiable technology and the external environment [31], replacing the original behavior of that interface with the extended behavior which confers a desired property on the wrapped component, thereby extending the interface itself with new functionality [30].

One of typical wrapping techniques is to wrap with interposition agents and buddy processes. This technique involves replacing the connections from an existing process to the outside world with an interface to a buddy process that has a much more sophisticated view of the external environment [30].

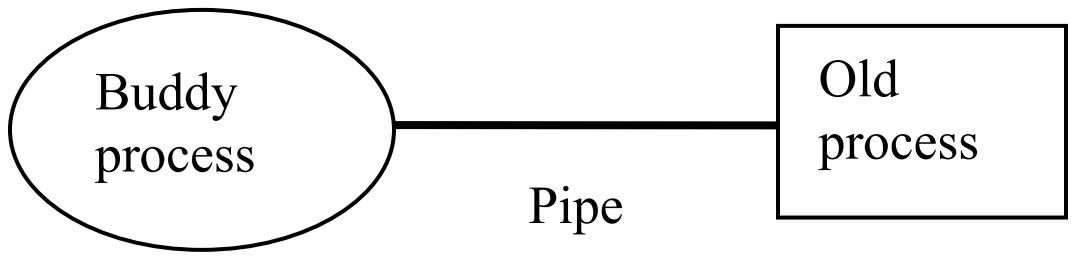

**Figure 4: Wrapping with Interposition Agents and Buddy Processes**

Inspired from this technique, we replaced the buddy process by the server interface to access the server services and the pipe by a RPC to direct requests from the wrapper to the server interface. On the other side of the server, we build the implementation for the server data storage. Thus, our wrapper works both as a monitor that captures system calls accessing the data storage and as a connector to the outside world to transfer this information to the server side. The transferred information helps to control the data access pattern. The connection replaces the requested access with a RPC to the server, gets the data from the server and returns it to the original request. Thus, the request gets data from the server instead of its local data storage. In our approach, the wrapper is not only a connection to the external world to exploit the sophisticated distributed technique of RPC, but also a transporter of information to build the server data storage and for data accessing at run time.

**Why replicate the binary application code instead of data storage?**

The justification for replicating the application code instead of the data storage is given now. There are a couple of ways to enhance the concurrency, such as improving the locking mechanisms, replicate the task and execute them in parallel. We choose to replicate the binary application code to make the design with less coupling. Each replica is independent and provides self-management to realize internal tasks and directed to the server data storage for data management. By making each replica independent the design has low coupling. On the other hand, the data storage can be replicated and delivered to client hosts. But since the status of available hosts in the LAN is dynamic, client hosts can join or withdraw from the middleware any time. If the data storage is replicated at the client hosts, it increases the cost for maintenance. Thus, we apply the model of desktop grid applications, in which client hosts trust the central system. The main data storage is kept on the server side and all services, such as maintenance, security, partitioning, are processed on the server side.

**Why keep the data storage at the server instead of partition and distribute to hosts?**

There are a couple of ways to partition data storage, such as vertical partitioning (as in relational databases [15]) or horizontal partitioning (data storage to sub data storage). Note that the data storage in our case is not a relational database, but data storage to store consecutive data elements needed to be processed. It is better to think of it as an archive file or an intermediate file to store data records. Thus, horizontal partitioning, in which the data storage is partitioned into sub data storage and distributed to any available hosts, is a better choice.

The major benefit of this approach is to reduce the bottleneck on the server side when there are a lot of requests to the server waiting for the returned data. It is also good to reduce the communication traffic in the LAN. Moreover, waiting time for a local request for data is also less than waiting for data returned from the server. On the other hand, the approach we are using is to exploit the CPU availability. When a host is free and willing to join the middleware, it will register with the server. Since available hosts can join and leave any time, the model is not stable if we partition the data storage into subsets and distribute them to any client hosts. When the number of joining and leaving hosts increases, the data storage is partitioned into more pieces. The cost of synchronizing small subset of data storages is higher than the queuing time of data requests to server, since the cost of inter-processes communication is higher than the cost of accessing data storage.

We choose to centralize the data storage on the server so that it is easy to maintain all the mechanisms, such as synchronization, scalability, and partitioning. The client hosts can freely join and leave any time. As long as hosts are available, they can contribute their availability to the middleware. This model is similar to the grid computing approach.

### 2.3.2 Scheduling approach

**Why divide into two phases?**

In this scheduling part, we give information on relevant scheduling mechanism and explain why and how we divide our scheduling into two phases. Our goal for scheduling is to achieve minimum execution time for the application using CPU scavenging and idle hosts in a LAN. Since there are some similarities between our model and the grid computing model, in the following we compare the two and fine tune the scheduling approach for our case.

## Comparison with grid computing

A grid application typically consists of *T* independent, identical tasks. The grid model comprises a *master* and *clients*, in which the master keeps the application and coordinates all computation. A client asks for a task, downloads the task from the master and submits the result back to the master. Similarly, our applications also repeatedly perform these actions: request for data to process, do some computation and produce the result. We call this combination of actions a *task*. Thus, a task in our case can be compared to a task in a grid application, because the tasks are independent and identical to each other.

However, since our middleware does not know the source code of the application, it can not let clients download the task to the client machine instead of downloading the whole application. The server in the middleware is the host keeping the data storage and managing it to provide data record in synchronized and concurrent fashion. Client launches the whole application as a replica and utilizes its local availability to do the computation. But instead of accessing its local data storage (stored along with the replica), the middleware directs data requests of each replica to the data storage on the server for data synchronization. So instead of a client asking for a task from master as in a grid application, the client asks for a data record from the server in our middleware.

Since all requests are directed to the data storage on the server, and each task needs a data record on the server, we can consider each data record on the server as a remaining task needed to be processed. The number of remaining data records on the server, the number of available hosts in the LAN and the CPU availability of each host will be the parameters for scheduling. During the deployment process, depending on the changes of these parameters, the middleware has to do the re-scheduling.

The difference between the two models is the resources. While a grid environment has infinite resources, our model has limited resources in the LAN. Moreover, our middleware has suspended processes, does replicas of the whole application and supports multithreaded applications. Thus, we have to re-calculate our application's execution time, complexity and modify the existing schema for better scheduling.

**Categorization of the cases for scheduling**

Since our middleware executes the target application in a way similar to a desktop grid application, we can apply the scheduling algorithm of desktop grid for our middleware. As in [1], there are two main cases: when the number of tasks is greater than or equal to the number of available hosts, a greedy scheduling algorithm can be used; and when the number of tasks is less than the number of available hosts, the execution time prediction with exclusion algorithm should be applied.

However, due to the differences stated in the comparison above, this thesis is building an environment which provides the best utilization/performance with limited resources, unlike the Grid environment where the resource is unlimited. In our model, there are active and suspended replicas. Thus the two main cases are divided into smaller cases to match with the real situation.

**During Phase 1 scheduling, why there is replication but not process migration?**

In the first phase, the number of tasks is significantly greater than the number of available hosts. In this state of scheduling, greedy scheduling is applied to complete as many tasks as possible. A new replica will be launched as soon as a new host joins the middleware or only after all suspended replicas in a host are re-activated. A suspended replica will be kept in the same host and only be resumed when the CPU availability of that host increases. Since the tasks should be completed as quickly as possible and the cost of launching a replica is less than the cost of migrating a process to another host, there is no migration for process in this phase of scheduling. Details and state machine of this phase will be given in Section 3.2.

**During Phase 2 scheduling, why there is migration but not replication?**

When using the first-come-first-served (FCFS) scheduling, the performance will increase significantly until a steady state is reached. In the state where the number of tasks is less than the number of available hosts and about 90% of tasks are completed, the performance increases very linearly and then reaches a plateau. The remaining 10% of tasks finish in a time which is double the time used for the 90% of the tasks at the beginning [1]. The reasons are due to task failures at the end of the execution and waiting for slowest hosts to complete their tasks.

The desktop grid scheduling uses an execution time prediction with exclusion strategy. The scheduler predicts the application execution time and justifies the timeout that a task should be completed. If there are tasks remaining at the timeout, then the scheduler picks up machines with the highest availability priority and replicates these tasks. When replicating the task (that is, running the task again from the beginning and discarding the one with timeout), since the time to re-start those tasks on other idle hosts will make the execution time shorter than waiting for them to finish on the current low availability hosts. The algorithm concludes with 12% waste of resource, this schema brings application completion to within a factor of 1.7 of the optimal for all application sizes in their experiments.

Besides applying the execution time prediction with exclusion algorithm, our design modifies the existing schema by suspending the timeout process and statically migrating it to other idle hosts in the LAN. Our design works better since the middleware does not need to store the checkpoint for a replica and restart the replica from the beginning, but just wait for the migration time to other host. Moreover, this will improve the performance since the host a replica is migrating to has higher CPU availability. In general, the cost of resuming a suspended process is less than the cost of launching a replica, which is less than the cost of statically migrating a process. Details will be given in Chapter 3.

The relationships among the number of remaining tasks, the number of available hosts and the number of current replicas divide the scheduling into two smaller cases. Besides the fact that the activated replicas always get higher priority to finish the remaining tasks.

Case 1: When the number of remaining tasks is less than both the number of available hosts and the number of current replicas, suspended replicas are migrated to another idle host and re-activated to finish their tasks. The reason is that all active replicas get all the chances to finish the remaining tasks; and the rest of the replicas including the suspended ones just need to finish their tasks. Moving the suspended replicas to idle hosts is a way to give these replicas a chance to finish their tasks independently and release other working hosts with higher availability.

Case 2: On the other hand, when the number of remaining tasks is less than the number of available hosts but greater than the number of current replicas, a new replica is launched to the idle hosts to hasten the processing time.

If the type of the data storage is dynamic, meaning the data storage is extended with new records which makes the scheduling status change, this scheduling algorithm is easy to change from one case to another as mentioned above. This is a dynamic scheduling.

### 2.3.3 Partition approach

**How to preserve the semantics of the application?**

Now we explain and justify our choice to partition the data storage while supporting the characteristics of the application, such as semantics, multi-task operation, concurrency and scalability. The middleware uses the component *Ptrace* to observe all requests made from the application to its local data storage. Ptrace works as a wrapper which directs the observed requests to the Server Interface which is the interface of the data storage on the server side. These directed requests are then applied to the data storage on the server side by the Server Implementation. Generally, the middleware learns what has happened with the local data storage and applies these actions to the server data storage.

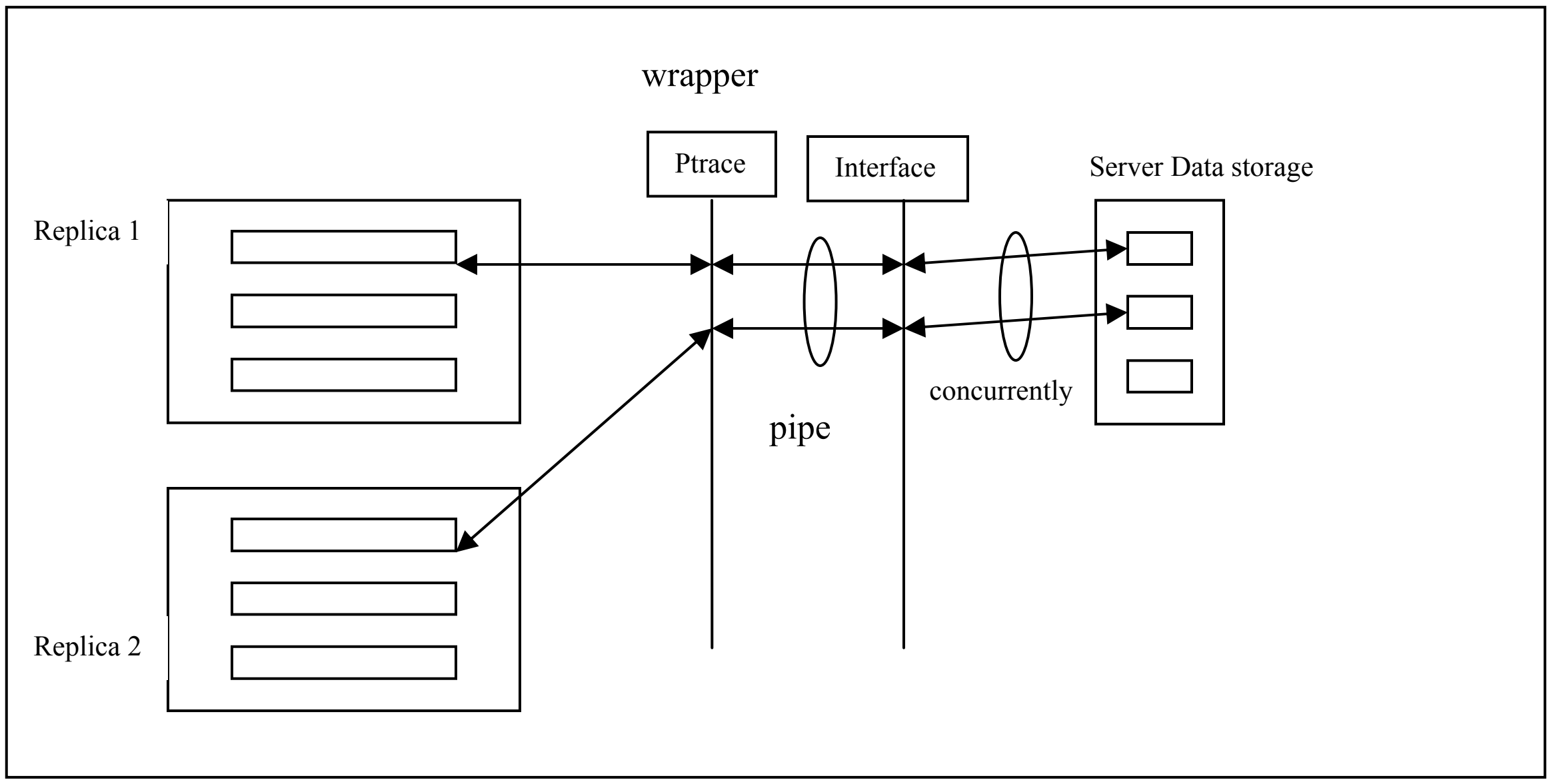

**Figure 5: The middleware reserve the semantic for the application**

The reason to make all concurrent replicas running as one application is to feed requests for data by next data records from the server as if each of the requests obtains data records from its local data storage. Since we only have the binary code of the application, the only way to do this approach is to observe system calls of the running application at the low level. When we do the replication, the Server Implementation will receive a lot of requests from all the running replicas directed by Ptrace. There will be requests that the Server Implementation has to integrate and perform only once, such as open or close calls. The first *open* call from one replica received by the Server Implementation will open the server data storage to start the data processing on server. If another replica sends the next open call, Server Implementation will filter this call and data storage will not be affected. Similarly, the last *close* call received by Server Implementation will close the server data storage. On the other hand, each read (and write) request will be directed and the data record returned from the server will be sent to the wrapper and overwritten on the buffer of the system call. The task gets the new data record from the server for the computation as though from the local data storage. Similarly, *write* system calls are also directed to server with their parameters to be written to the data storage. Since everything is done at the low level transparently, integrated in a cooperating way, nothing is changed in the logic of the application. Thus, the logic and data access are still preserved on the application side.

**How is the design appropriate for both single-threaded and multi-threaded applications?**

The middleware runs the replica concurrently. Thus, if the middleware replicates a single threaded application, the requests sent to the Server Implementation will be concurrent among the replicas and sequential among tasks of the same replica. On the other hand, if the middleware replicates a multithreaded application, the requests sent to the Server Implementation will all be concurrent, both among requests in the same replica and request among different replicas. For either case, whenever a request is made to the local data storage, it is captured by the middleware and directed to the server. By the design of the middleware running concurrent replicas, there is no difference between two requests from two replicas of either a single-threaded or a multi-threaded application, since they are both concurrent requests from two concurrent replicas. The Server Implementation manages these concurrent requests by returning data records to continue with the computation. This design is appropriate for both single threaded or multi-thread applications.

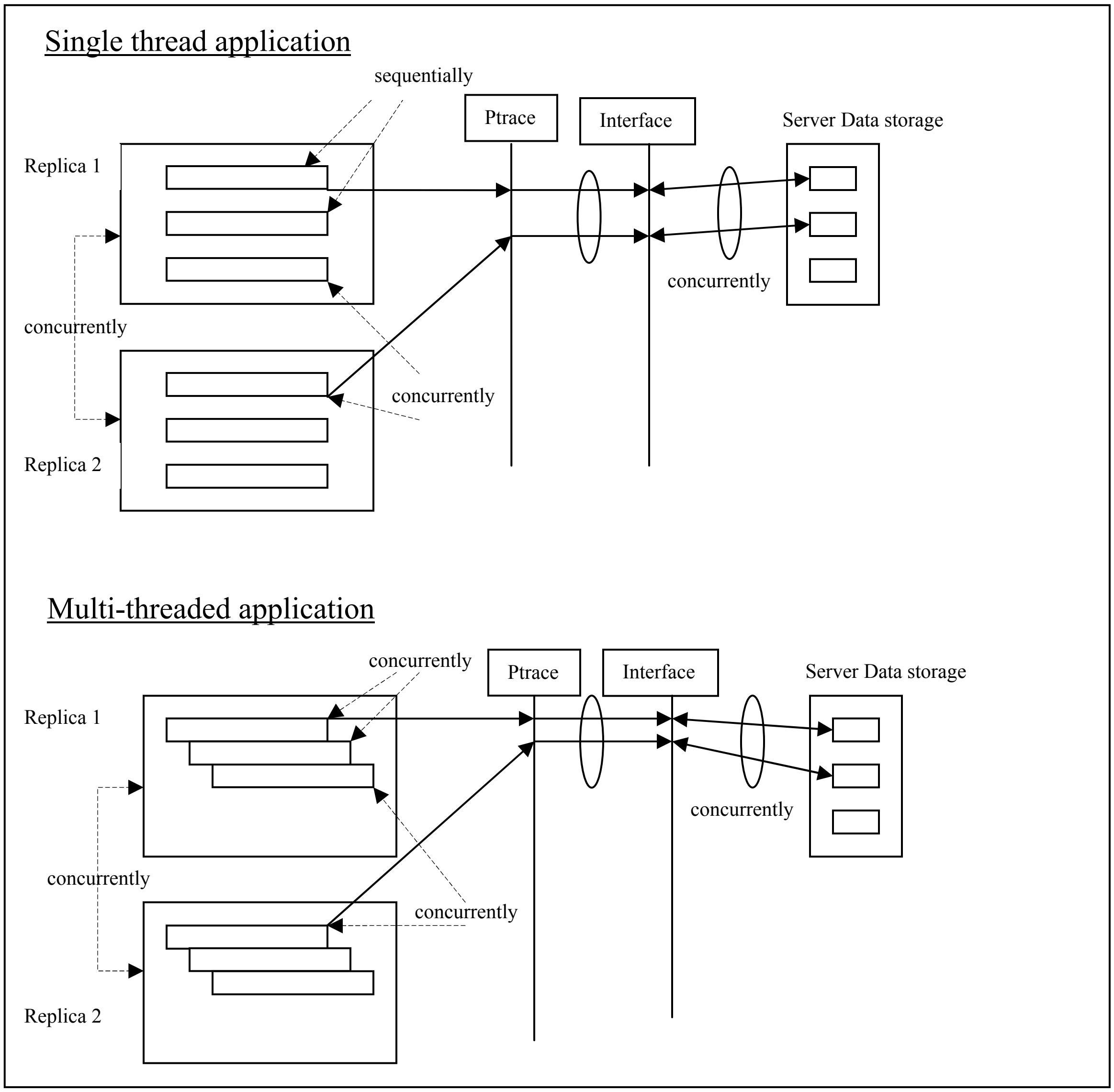

**Figure 6: The middleware satisfies both single-threaded and multi-threaded application**

## How does the design support concurrency?

We are trying to provide a model with less coupling and high cohesion as much as possible. Each replica runs concurrently and independently of the other. Each replica has a Ptrace component to observe activities of the application, direct requests to the server and process returned data records from the server. The Server Implementation queues all data requests and provides synchronized mechanisms when it gives the next unprocessed data record to queued requests. The two main calls invoking data storage are the *read* and the *write* system calls. As the Server Implementation has to read the whole next unprocessed data record from the server, a synchronization mechanism is definitely provided for read/read request. But synchronization mechanism is not always necessary for read/write requests depending on whether the data storage can be extended or where the result of the computation should be written to. Thus, a framework is defined for the type of data storage and operations, and it will be detailed in Sections 3.3.1 and 3.3.2.

**How to balance between concurrency and scalability?**

As mentioned above, although the data access requests are sent concurrently from all replicas, the Server Implementation really processes requests sequentially. Because the data storage is a file with sequential data elements, to get the next data element the Server Implementation has to finish the current data access so that the file position can be advanced to the end of that data element. This makes the Server Implementation process requests sequentially. To improve concurrency, the middleware has to return the data to the requests as soon as possible. Hence, the middleware is modified with an Index Buffer, which works as a data buffer for the Server Implementation. The Index Buffer is an array of elements, each of which is a data record fetched ahead of time from the data storage to memory. By moving forward one element in the Index Buffer array will be much faster than waiting for the synchronization of reading one data record from the data storage. This mechanism not only improves the performance of the middleware but also provides the scalability for the middleware.

**Can the middleware handle applications that do not satisfy the application characteristic?**

Our middleware is not appropriate for the following type of applications.

- Applications with dynamic but sequential data record of data storage:
  In this type of application an earlier computation updates the data and the latter computation depends on the updated data. This type of application is totally sequential, in which each computation is not independent of the other. This is not the type of application we are interested in.

- Data storage cannot be truncated:

  The middleware can still handle applications in which the data storage cannot be truncated or cannot index each data element. By reading information directed from the wrapper to the server, the Server Implementation can form the request, return the result to the wrapper and sent it to the original request. This request is only processed by the time it comes to the server, but the server cannot predict and process the request in advance as in the case where the size of each data record is fixed and known. In this case, Index Buffer can also be applied, but the Index Buffer counter advances for every request. Compared to applications where the size of data is known, the Server Implementation takes a bit less time to return data record than for this kind of applications. The model is useful for this kind of application if the computation phase of this application is much longer than the request time for data record.

# CHAPTER 3: DESIGN

Chapter 2 provided justifications to our design choices. The two highlights of the middleware design are dynamic scheduling and partitioning the data storage. In this chapter, the thesis gives full details of the design concentrating on the middleware features, the scheduling and the partitioning algorithm.

## 3.1 Features provided by the middleware

### Auto replication

When the middleware observes that a machine in the LAN is idle, it sends a message to the idle host to download the executable file of the application from the server. If the executable file has already been downloaded, the middleware requests the idle host to run the executable file, to observe its behaviors and to report the system calls accessing data storage back to the middleware. Once the executable file is downloaded to a host, it is kept in the host's repository for later use when the host needs to launch another replica. The method of downloading once and activating later by passing messages in the middleware between the server and the client hosts helps to reduce the message traffic in the LAN.

### Provide concurrency

When the middleware launches a new replica, the newly created replica runs in parallel with the existing replicas on different hosts, thereby supporting concurrent execution. This feature can be clearly seen when the target application is single threaded. We have noted in our case studies (reported in Chapter 4) that when the middleware creates one replica of an application in a host, it reduces the processing time almost into half of the time compared to the case where there is only one running application to process the entire data storage. The processing time reduced due to the fact that each replica processes half of the data. In the case where the application is a multi-threaded one, the processing time reduces directly proportional to the number of threads in the application and the number of launched replicas in the LAN. The problem in exploiting concurrency this way is to properly partition the data storage and to provide concurrent data management.

### Data partition

When the middleware creates replica of the target application, it has to ensure these replicas can run and collaborate with each other so that different parts of the data storage can be processed by different replicas. This fact leads to partitioning the data storage and maintaining the synchronization among data requests sent from different replicas and within one replica.

### Scalability

Although each host operates independently of the others, the single server processing data requests from all the replicas becomes the bottleneck. If the server is slow in processing and returning data, all requests have to wait and be queued for data synchronization. So even though the middleware supports concurrent execution of the target application, a single server may hinder the effort to improve performance. Hence, the server data storage needs to provide not only a synchronization mechanism but also the scalability feature.

### Utilize idle CPU capacity

Client hosts always keep track of their current CPU availability. This information is sent as a heartbeat message to the server. Based on this information, the server will do appropriate scheduling and may decide to launch a new replica on an idle host or suspend a running replica on an overloaded host to suit most the current status of the hosts. By applying this method in the LAN, the CPU capacities of all the hosts in the LAN are utilized to improve the utilization and performance.

**Dynamic Scheduling when number of hosts change**

This feature goes hand in hand with the utilization of the CPU capacity. The server is notified when the status of hosts in the LAN changes. The server maintains a replica pool which lists all the current hosts and replicas along with the status of each host. The middleware provides dynamic scheduling which achieves the best utilization/ performance considering the number of available hosts, the hosts' status, the number of tasks to be processed, the number of replicas, and suspended and active replicas.

**Environment flexibility**

The conceptual design is considered for an independent environment. The Communication component, the Observer component, and the RPC component all can be implemented on either Linux or Windows running either in a LAN or a WAN. To achieve the best performance, the middleware should be deployed in a LAN.

**Type of application**

As already mentioned, the target application can be deployed as concurrent replicas if the server data storage of the application can be split into disjoint partitions and each replica of the target application can operate on each data partition without any conflict. The application will get a higher improvement in performance if the time to execute the computation is much greater than the time to make a read and a write requests to the server data storage. On the other hand, the application might not work correctly if the data storage cannot be split into disjoint partitions. Each element in the data storage is dependent on other data element(s).

## 3.2 Partitioning

In our design, there are two approaches to support data partitioning. If there is no knowledge about the application, internal analysis is applied to detect the partitions. On the other hand, if some information of the application is known in advance, external analysis is used to widely enhance the concurrency of the application. During the external analysis, internal analysis approach is combined to further improve the concurrency performance of the application.

### 3.2.1 External Analysis

Based on the additional information of the application, a task map for the whole deployment process is built. The task map defines the priority of tasks and the relationship among the tasks to be executed during the deployment time. The task map is good for allocating tasks to host in order to achieve the best performance. The information needed for the task map is how the computation of the application is organized. This information can be obtained from the application designer or from automated application handling details as specified by scripts or makefiles. Since there is not so much difference among topologies [5, 34, 35], the task map uses tree topologies to easily express the relationship among tasks. After the tree map is built, the external analysis process decides which node can be replicated and which host should the node be assigned for the best performance.

**Building the tree map**

The following rules describe the building of the tree map based on the dependencies among tasks as in a makefile.

Rule 1:

*The main function is the root node.*

Rule 2:

*In a statement, the target is an ancestor and the dependency is a descendant.*

Rule 3:

*The construction of the tree map finishes when all the dependencies of any target are marked.*

**Identifying the replicas**

Rule 4:

*If a partitionable data storage is given to a node in the tree map, this node can be replicated and each replica can process one data partition.*

Rule 5:

*Any node applying Rule 4 can be combined with the internal analysis approach to enhance the concurrency performance.*

Since tasks satisfying Rule 4 have patterns of data access as described in internal analysis, if internal analysis approach is combined with the current approach, the performance is much improved. This is the improvement for the performance, which makes the analysis more fine grained when the internal analysis is combined with the external analysis.

Rule 6:

*If a node needs to operate on any data element in the data storage just once, the output of all the replicas just need to be put together.*

*If a node needs to operate on all data elements in the data storage, the output of all the replicas need to be processed again with another iteration of the replicas.*

**Relationship among nodes in the map**

In the makefile, since each dependency of a target is independent of the other, those dependencies can be processed simultaneously.

Rule 7:

*In each tree, descendants of the same ancestor can be processed concurrently. An ancestor and its descendants have to process sequentially.*

Rule 8:

*Leaves need to be processed first.*

*An ancestor is only processed when all its descendants finish their computation.*

Once this tree map is defined, it is passed to the scheduler to allocate defined jobs to hosts to achieve the best performance.

We now illustrate the construction of the tree map using an example. Figure 7 shows the makefile of a Sort application. The user provides the information that the data storage can be split for independent sorting and the result can be merged from the all the sorted sub results.

```
all: main

main:  run1 run2
       @echo "Deployment is done."

exec: sch.c
       @gcc sch.c -o sch

run1: exec
       ./sch largefile

run2: exec
       ./sch file1 file2 file3
```

**Figure 7: Makefile of Sort Application**

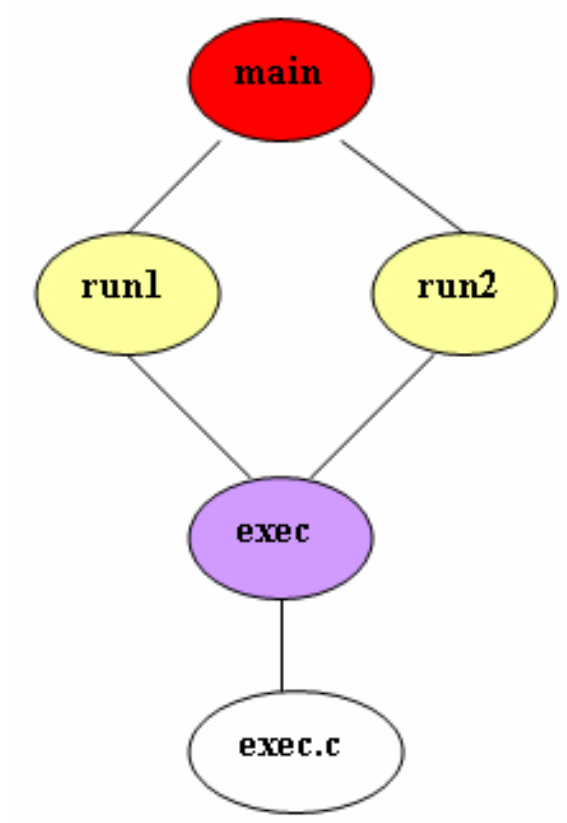

**Figure 8: Tree map of the makefile example**

The tree map is built based on the first three rules. Each node represents a task to be executed. Node exec.c has the first priority to run its task. The two child nodes run1 and run2 have the same third priority. Those two tasks can run concurrently. The root node main has the lowest priority. The job of the node main can run only after those two jobs of node run1 and run2 finish. Once the tree map was built and the priorities of node are defined, the tree map is passed over to the Strategy component for the middleware to assign tasks to hosts to run in the defined orders.

The middleware starts to observe the running target application for each node. The middleware notices that at node exec, the task accesses the data storage. Then the Strategy component applies the internal analysis approach to this node. The host where this node runs becomes the controller/server of the running binary executable program. Then the server replicates this task to other clients in the LAN and starts to observe the running task. Based on the additional information provided by the user, eventually the server collects the result from all the replicas, merges them and puts back the result to the server.

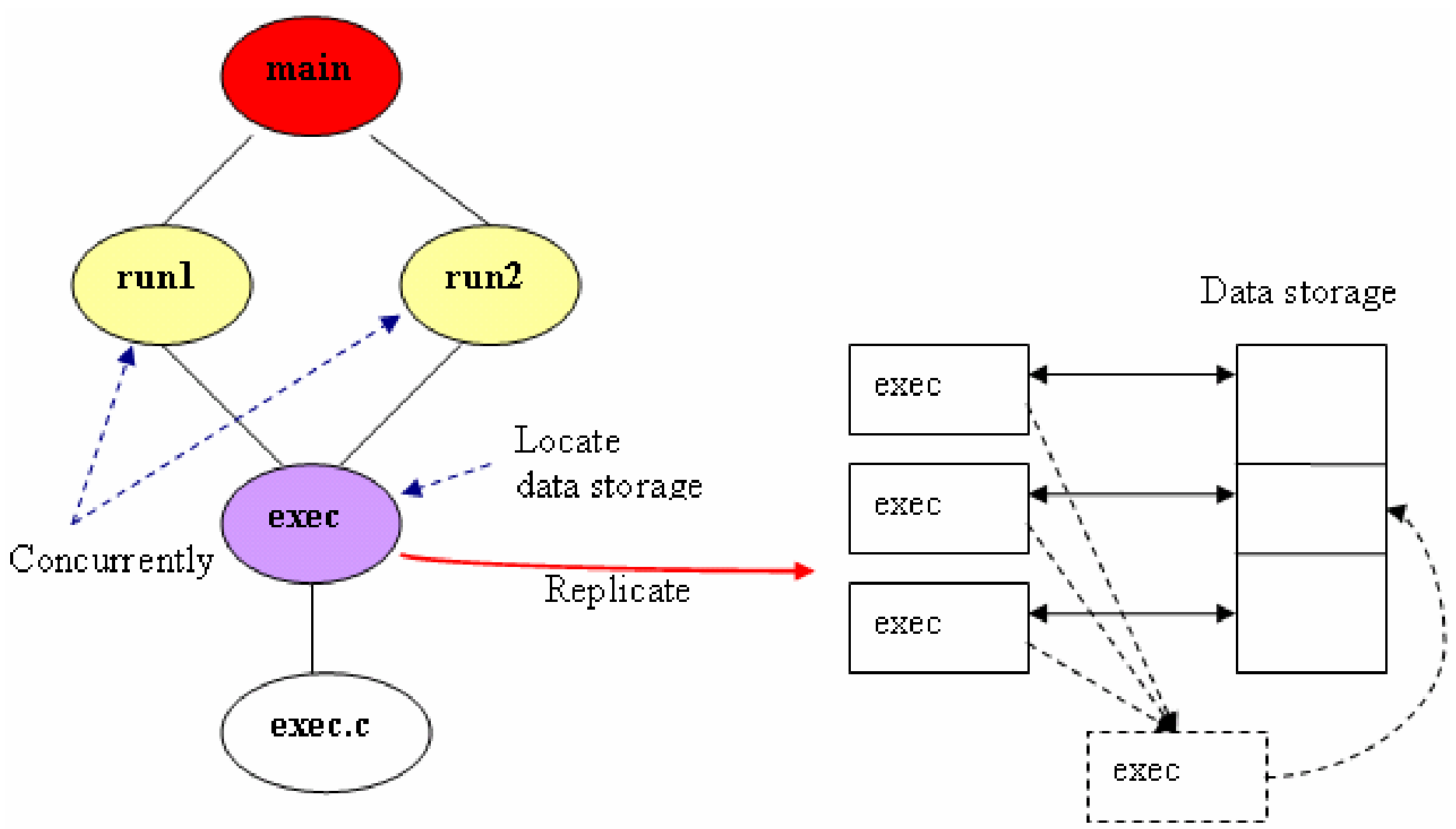

**Figure 9: External analysis for makefile example**

## 3.2.2 Internal Analysis

This method provides a fine-grain approach to detect and exploit concurrency while running the target application. This section shows the issues, patterns to be defined, partition and then Index Buffer mechanism for this internal analysis approach.

### 3.2.2.1 Emerged issues

Now we mention some requirements that the design has to satisfy.

1. Synchronized data access

During the experimental implementation, sometimes the result returned by a task printed out unreadable characters. It means that there was an error during the read data elements process. If a read request comes to the ServerImpl while it was busy processing an earlier read, the file position was confused and got errors between the two consecutive requests. The wrong data record value was returned as the unreadable result to the latter request. Hence, the data storage has to be managed in a concurrent way and the data elements should be correctly accessible by multiple concurrent requests.

2. Concurrency

The target application run by the middleware can be one of the following two types.

*Single-threaded application*

It is obvious that this type of application processes each task sequentially until all data in the data storage are processed. Observed at the system call level, the middleware will see request data system calls (read/write) generated sequentially. Corresponding to each task, the data storage will receive data request, read from the disk and return to the request sequentially as shown in Figure 10.

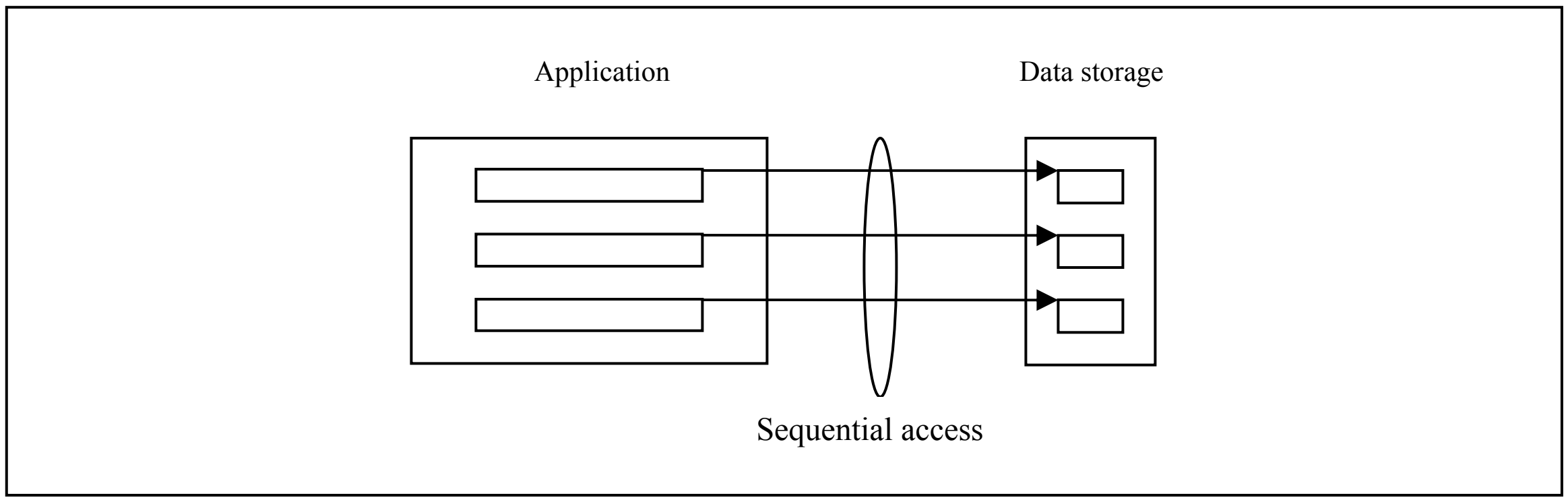

**Figure 10: Overview for single-threaded application**

*Multi-threaded application*

In this kind of application, each task is represented as a thread. Although all the tasks process concurrently, they may still face a race condition to acquire the lock to access the data storage. Thus, it comes to the point that there are multiple requests for the data storage, but the application synchronizes and lets only one of them pass by one at a time as shown in Figure 11. Certainly the middleware can not see all the logic performance of the application, but it can see one system call sending to the data storage once at a time in a corresponding way.

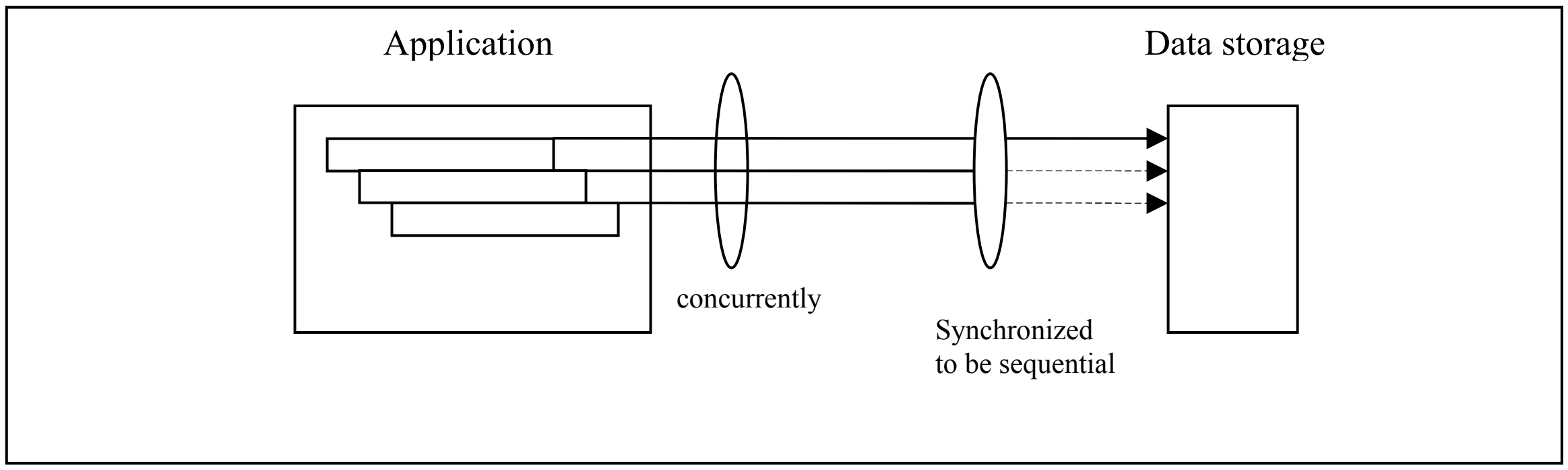

**Figure 11: Overview for multi-threaded application**

Over all, irrespective of whether the application is single threaded or multi threaded and whatever the logic performance of the application might be, by catching the system calls on the application side and performing these system calls on the server data storage, the middleware can only see the system calls to the data storage sequentially. In addition, the fact is that accessing a physical disk can be performed one at a time; meaning that the server implementation processes the data requests sequentially.

3. Scalability

The server implementation resumes any I/O system call captured by the middleware and implements them on the server data storage. As mentioned above, although data access requests are sent concurrently from all the replicas, the server implementation really processes them sequentially. In addition, the data storage is a file with sequential data elements. Thus, to get the next data element, the server implementation has to finish the current data access so that the file position can be advanced to the end of that data element. This makes the server implementation process the requests sequentially. To support concurrency, when there are multiple requests sent to the data storage in a concurrent manner, the server implementation has to respond to these requests by providing data access points readily. Thus, besides the synchronized mechanism for concurrency, the middleware must provide the scalability for multiple requests. This characteristic not only support concurrency feature but also enhance the middleware's performance.

### 3.3.2 Defining patterns for data partitioning

When data requests are directed from the wrapper to the server, the middleware analyzes this information and builds up the server's operations at runtime. In this part, we define the patterns that affect the data partitioning at the server. We need to consider the following.

**Data storage**

*Data storage type*

Data storage is the pool of data shared by all the replicas of the application. There are two possible cases: data storage is a physical storage, i.e. a file or a folder of files stored in hard drive; or data storage is a shared variable which is fetched into memory during the deployment process.

*Data storage state*

Data storage can be static or dynamic. The application only needs to perform on a static data. When all data in this file is processed the computation is done. Some applications have to refresh data. The later computation will update the data storage. The deployment only finishes when there is no more update and all data in the data storage is processed.

*Data in data storage can be truncated /cannot be truncated*

Data can be truncated: Each record is independent, i.e: a line in a file/consecutive data chunk. Every time, data can be read in as a single record or a chunk of data consisting of a multiple of the record's size.

**Accessed Object**

*Accessed Object type*:

Each element needed for every computation of the application is defined as an accessed object.

*Accessed Object size:*

Each accessed element is of a fixed size or of variable size.

**Operation on file**

*Operation type*

Data to be read is read only / read and write.

Data to be read and written to the same file / different file

*Operation access type*

Every action to perform the access to the data storage can be observed with each system call.

When the application performs the access to data storage, it reads into buffer and fetches into memory. When each computation requests data to process, it reads a specific chunk of data from the memory and returns to the requesting computation. Thus, only the first read access to the data storage can be observed as a read() system call. But when the computation requests data, it seeks for the next chunk of data in memory instead of reading from the data storage. In this case, there is no read() system call to perform. We can only catch this action through the mmap() system call. When all data in the buffer in memory are consumed, then the application performs another access to data storage, then read() system call can be seen again.

### 3.3.3 Realized case for partitioning

Based on the above categories, we come up with different combinations (among data storage, accessed object and performing operations) that can and cannot handle the partitioning task.

Data can be truncated

*Each record is fixed size*

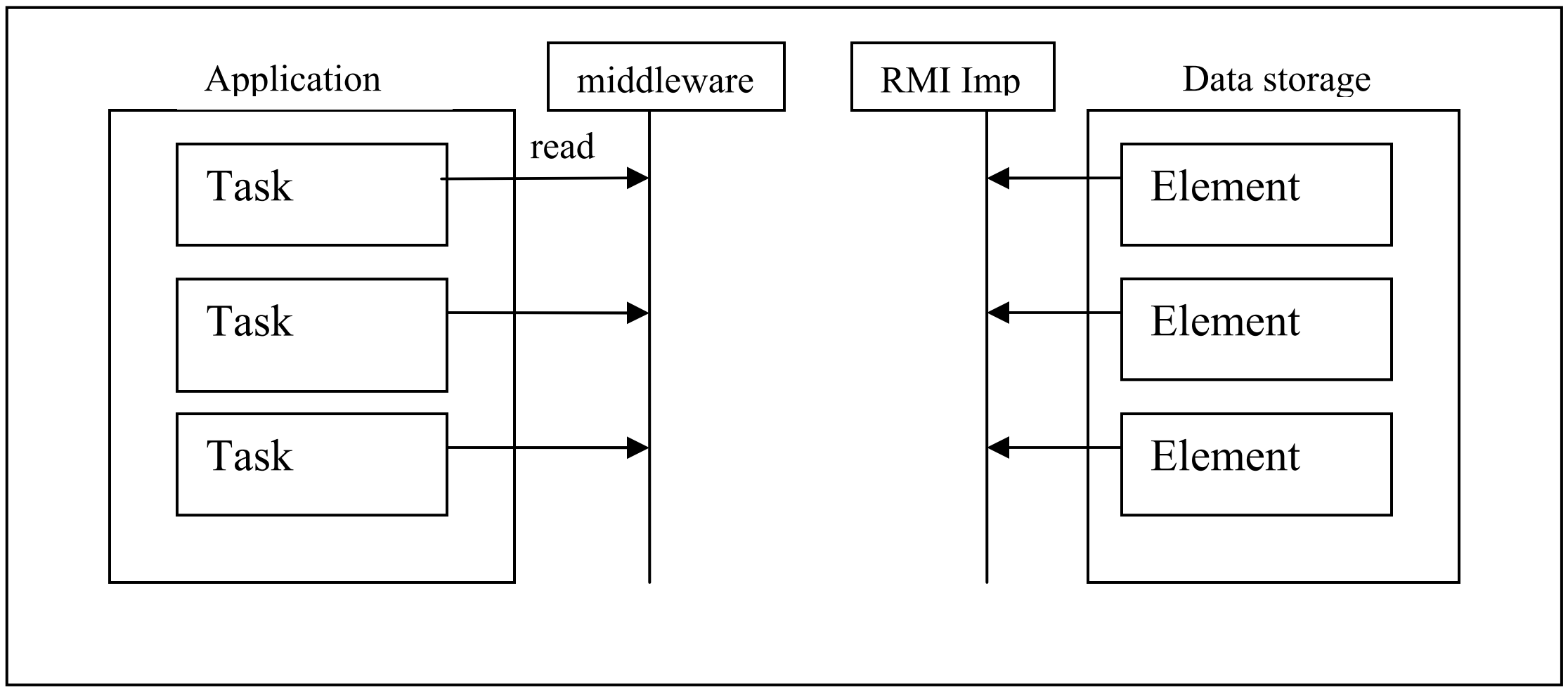

**Figure 12: Data storage can be truncated with fixed-size records**

*Each record is not a fixed size*

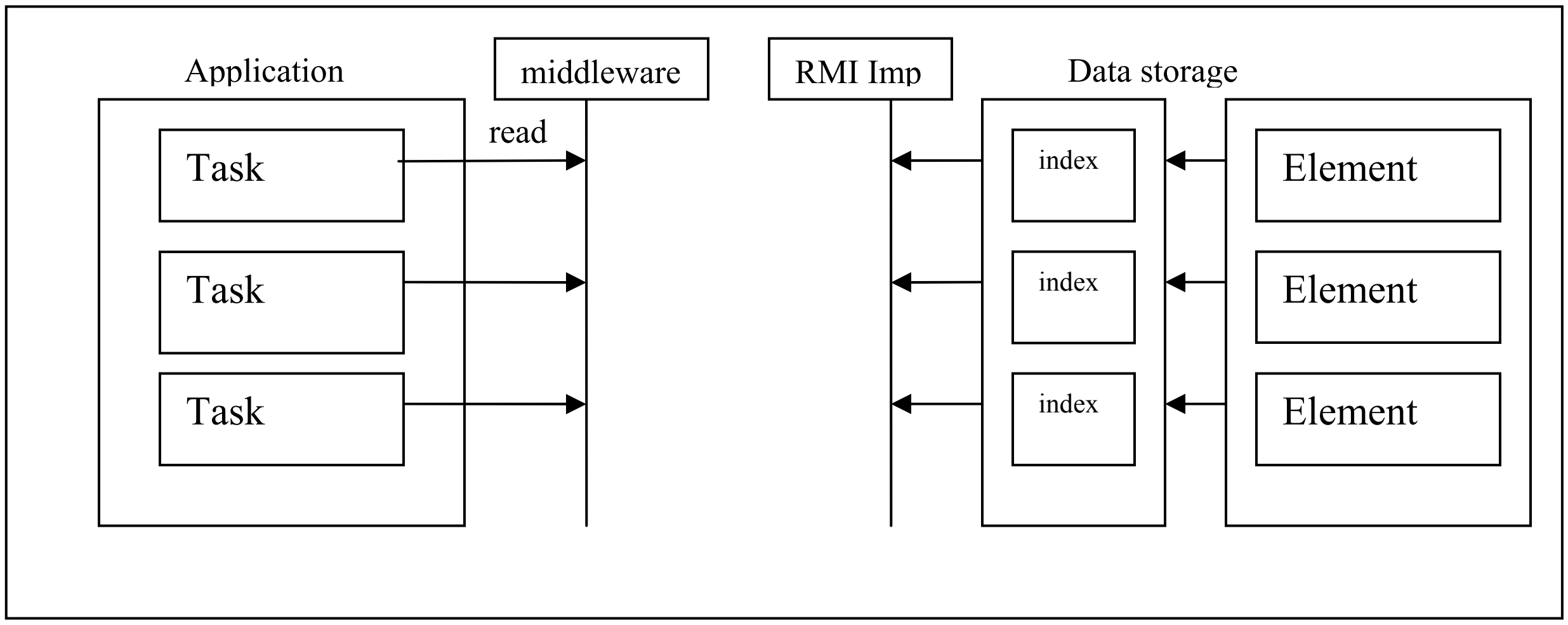

**Figure 13: Data storage can be truncated with fixed-size records**

The index interface helps to truncate data into records. This is good if there are some accesses that cannot be performed as one read system call.

*Dynamic data*

In this case, the data storage is to be updated. As mentioned at the beginning, all actions performed on the observed file will be applied on the file at the server. Thus, the file on the server is updated along with the local file for each replica. Moreover, the index will be updated along with data written/appended to the file. In this case, both the cases with fixed or variable size records are still tolerated.

*Dynamic but sequential data*

In the type of application where an earlier computation updates the data and a latter computation depends on the updated data, the application is totally sequential process in which each computation is not independent on the other. This is not the type of application we're interested in.

*Data storage cannot be truncated.*

This can only happen if the computation reads the whole file and the result of the computation updates the data file. Thus, the next task reads in the updated data and performs repeated computation. Otherwise, if no data is updated and only one is read in, the application does not have repeated tasks and we are not interested in this type of application.

### 3.3.4 Partitioning mechanism

Since the application source code is not available, the only way the middleware can observe the data access is to observe the system calls and capture them to perform appropriate actions. The Ptrace interface is a wrapper of the application code. Ptrace is used to observe and capture some specific system calls of the running application. If a *read* or *write* system call is made on the data storage of the application, then Ptrace extracts parameters of this system call and makes a requests with all those parameters to the ServerImpl interface. This is how the system calls of the application are wrapped and directed. These actions will be applied on the server data storage so that the server data storage is presented as unique data storage for each replica.

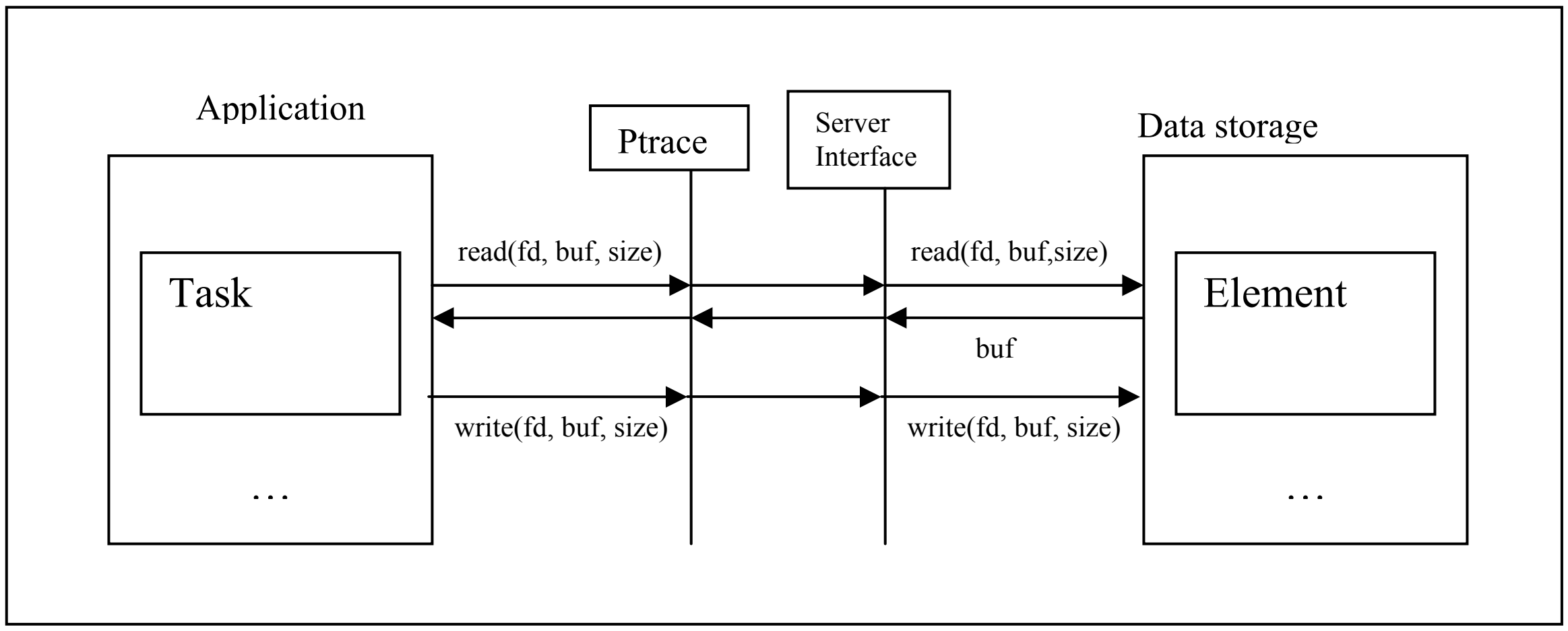

**Figure 14: Design of data partition operations**

*read(int fd, char* buf, int size)*: overwrites the return result.

When this system call is captured, and the file descriptor is the fake data storage we are observing, then we let this system call perform its action, meaning, going to the fake data storage, reading the data element and writing into the buffer *buf*; but in the same time, Ptrace gets all the parameter, directs the information to the ServerImpl and gets the returned data element from ServerImpl and overwrites this result onto the buffer *buf*. Thus, before this system call returns its control to the application, the appropriate data element fetched from the server has already been given to the system call so that it can be ready for the computation in the task.

*wite(int fd, char* buf, int size)*: pass to ServerImp

Writing case is easier. In the same way as read, when the write system call is captured, Ptrace gets all the parameters and directs the call to the ServerImpl. Those requests will be put on the queue to write sequentially into the data storage on the server side to protect the mutual exclusive context.

*mmap()*

There are some languages, such as Java and Perl, which for every read system call read in more than one records instead of a single record as in C/C++. The returned data chunk is kept in the local memory and shared by other processes of that replica. The middleware can observe this case due to the size of the first data request and the appearance of the *mmap* system call. Since each replica works independent of the other and the shared memory is accessed by processes in a replica only, there is no shared memory among replica. The design helps to overcome the difficult issue and reduces coupling among replicas.

The server implementation is the interface on the server data storage side. The server implementation receives requests with system call name with all parameters from the Ptrace wrapper. Then it forms a system call with those parameters and sends it to the server data storage. This is to perform exactly what the system call does with the data storage going along with the application; but now we direct all the requests on the data storage on the server. Thus, the main responsibility of the ServerImpl is always to listen from Ptrace and to learn the system calls performed on the application side on its fake data storage and make it in the same way on the data storage on the server side. For this reason, when we do the replication, the ServerImpl will receive a lot of requests from all running replicas directed by Ptrace and returns the new data element. The reason to make the ServerImpl always return new data element is because the ServerImpl always keeps the open connection to the data storage and remember the current file position.

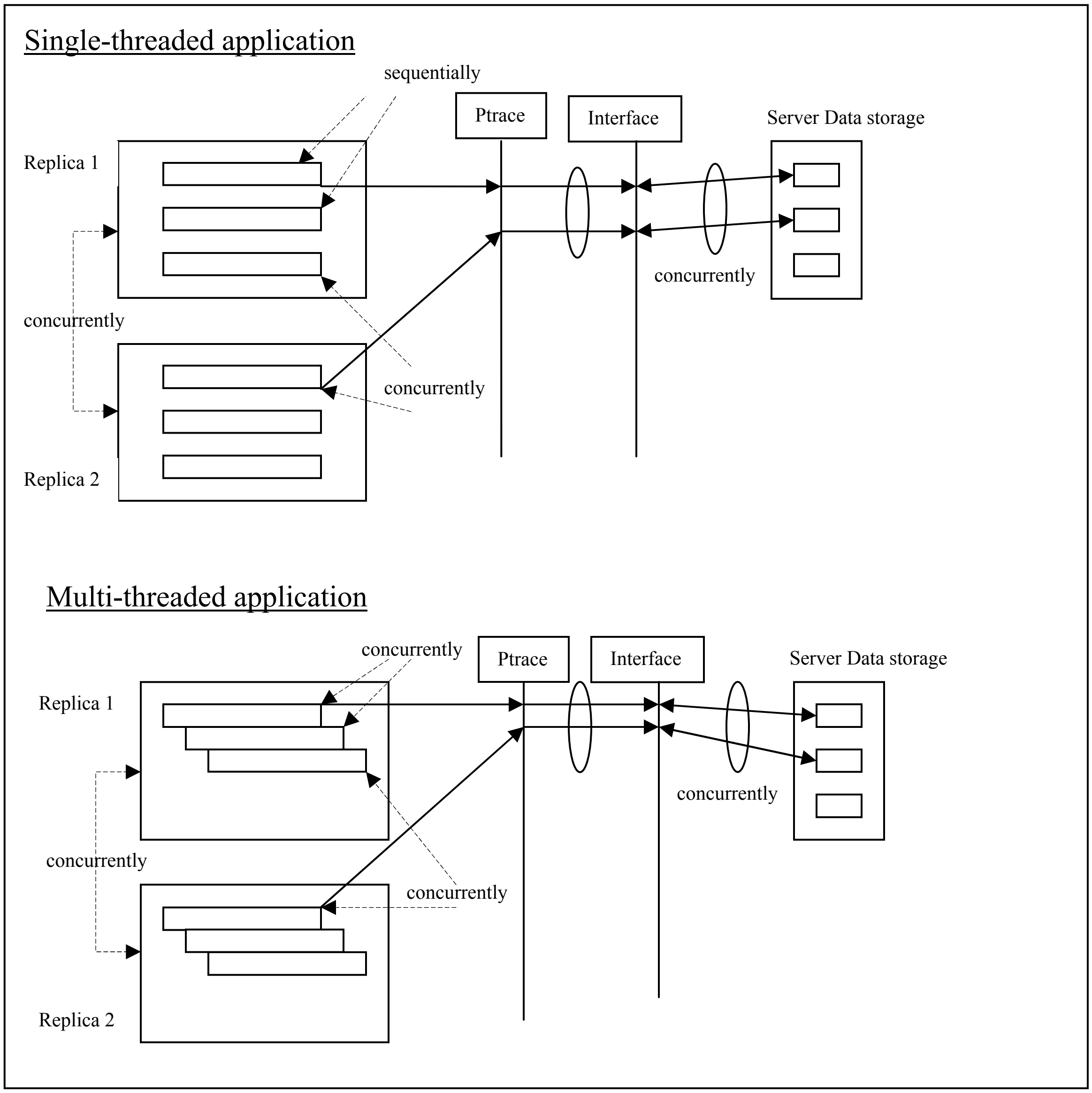

**Figure 15: Synchronization for concurrent requests**

This strategy preserves the context of a single-threaded or a multi-threaded application because when data is requested, the middleware directs the request and overwrites the result into the buffer; giving the application a feeling of reading from its real data storage. Beside this wrapper, nothing is changed in the logic of the application. Thus, the logic and data access are still preserved on the application side. Since the ServerImpl always listens, the data element is always ready to be given to requests and there is no conflict in returning the request data element concurrently.

A banking application that updates all account information at the end of the day is presented below as an example to describe how the middleware performs the internal analysis. In this simple application, each replica of the application processes all the updates for an account number sequentially. The replica reads transactions made during the day on the account from the log file, does the computation for the balance and writes back the balance to that account on the data storage. The data storage is a collection of elements. Each element is an account along with its information, such as account number, balance, etc. Each element account does not have any relationship with other accounts.

The middleware starts up. The target application is launched on the server. The middleware starts to observe system calls of the target application. The application makes a request to the data storage for reading the account. The wrapper wraps and tunnels this request to the server data storage. The server splits the data storage into every account, takes the first account and tunnels this data element to the wrapper of the target application. The wrapper returns the requested data to the target application. The target application does the computation and makes a write request to the data storage. Again the wrapper wraps this request and tunnels it to the server data storage so that the server can write to this account when there are no other requests holding a lock on this account.

A host becomes available and the server notices this availability. The server starts a new process on that client host and runs a replica of the target application on the client host, concurrently with the target application on the server. The replica begins to process and makes a request to the data storage for another account. The wrapper of the client wraps and tunnels the request to the server data storage. In the same time, the target application makes a request for this second account. The two requests might come to the server at the same time, while they are running concurrently. Since the server has already split the data storage and keeps track of the currently processed account, the server returns the second data element on the server through RPC. If the wrapper of the replica gets the result before the target application; it returns the request to the replica. The replica then gets the second account and the target gets the third account. The replica and the target application will write back to the server with the corresponding accounts.

### 3.3.5 Index Buffer

The Index Buffer structure is really dependent on the data storage type and data elements in the data storage and the size of each data element. Based on the data element characteristics, we can build the Index Buffer if the following conditions are satisfied.

- Each data element is of the same size.
- Otherwise, each data element is independent; meaning the data access point of the next data element is not dependent on the current data access point. In this case, a formula can be derived for any data element.
- Files in a folder.
- Data elements are with different sizes but follow a predictable access pattern.

A large chunk of data from the data storage is read and fetched into the memory ahead of time and stored in a place called Index Buffer. When data requests come to the server implementation, they can be directed to the Index Buffer and get the result. Since the Index Buffer is stored in memory, multiple concurrent requests can be performed at the same time, without waiting until one request finishes and then process the next one. So the middleware now is scalable and provides real concurrency feature.

The Index Buffer is an array to store data fetched from the data storage. Each element of the array is a data element given to each request. The server implementation maintains two variables, a *lock* and a *counter*. The lock is to synchronize the request. The counter is used to advance the address of the next data element in the Index Buffer. At first, the counter keeps the beginning address of the Index Buffer. When a request comes, it contains the process ID and the size of the data requested. If a request is at the top of the queue, the server implementation gives the lock and the information *dataElement(counter, size)* to the request. When a request obtains this information, counter is advanced to the next data element's address. In this case, *counterAddress = currentCounterAddress + size*. The ServerImp can move to the next request and does not have to wait for the previous request to finish reading the data element. With the lock, the Index Buffer can respond to multiple data requests at the same time.

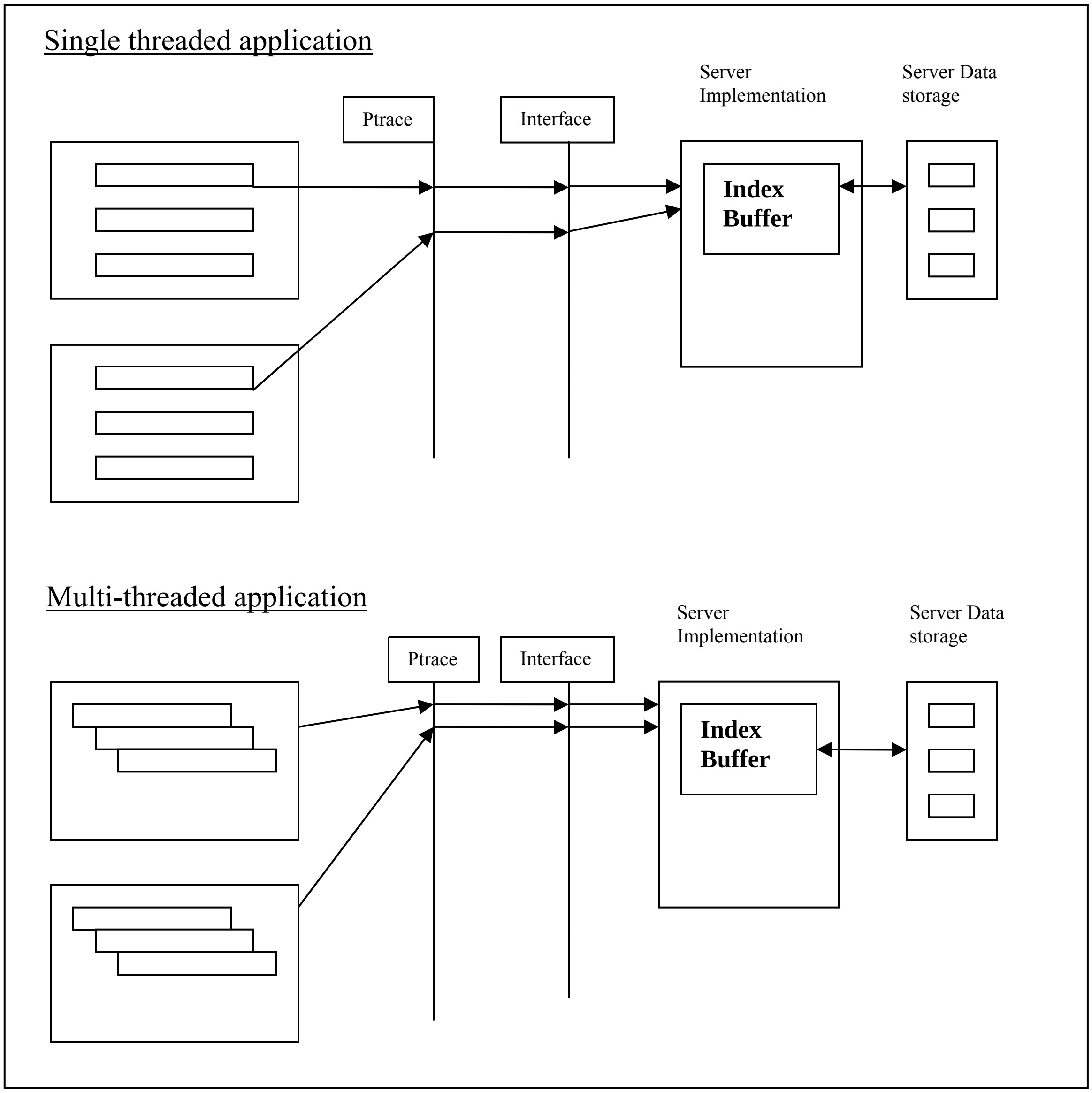

**Figure 16: Index Buffer**

With this design, even when one request has not finished with the data accessing, the next request can still come and access the next data element. This helps data element to be requested and returned concurrently instead of waiting for the current request to finish. Moreover, when the data accessing point is ready, it also makes the middleware scalable because multiple requests can be sent and results can be received from the server implementation in a period of time.

## 3.3 Scheduling

In this section, we will discuss how our middleware dynamically schedules the concurrent replicas of the legacy application to improve the utilization/performance in the LAN. Our goal is to achieve the minimum execution time for the application by using CPU scavenging of the idle hosts in a LAN.

Our middleware is deployed on a LAN. Taking advantage of the characteristics of a LAN, we assume two things. First, we assume that we know the characteristics of all the machines on the LAN. The characteristics of the machines on the LAN are understood by conducting a measurement over the LAN to evaluate their CPU availability [1, 10, 11, 27]. As described in [1], CPU availability is a percentage value that quantifies the fraction of the CPU that can be exploited by a desktop grid application. This characteristic helps the middleware perform the best scheduling for all the replicas running under the middleware. Second, we ignore the communication cost between hosts in the LAN, since the message latency in the LAN can be assumed to be low.

Due to the similarities between our middleware and the desktop grid model described in Section 2.6, we apply the desktop grid scheduling with some modifications.

**Categorization of the cases for scheduling**

The scheduling is done differently for the two analysis approaches. In the external analysis approach, based on the task map, tasks are allocated with their priorities, as described in Section 3.3.1. Since this scheduling is in a local network, we simplify the delay due to the distance among hosts. Also, we assume the status of hosts change during the runtime; thus we exclude the release time of hosts in the algorithm.

In internal analysis approach, since tasks have no order and have the characteristic of a desktop grid application, we can apply the scheduling algorithm of desktop grid into our middleware. As in [1], when the number of tasks is much greater than or equal to the number of available hosts, a greedy algorithm is used. When the number of tasks is less than the number of available hosts, the execution time prediction with exclusion algorithm should be applied. However, as our middleware has suspended processes, replicates the entire target application and supports multithreaded applications; we have to re-calculate our application's execution time and complexity, and modify the existing schema for a precise scheduling. Besides the scheduling overhead, our schedule also concerns about the cost for resuming/migrating suspended replicas or launching new replica.

Since the state of a suspended replica is currently in memory of the host, sending a signal to wake up the suspended replica costs Rp = O(1). The cost of launching a new replica on a host is Lp = O(δ) + O(γ), where O(γ) is the time to prepare a process for running, such as allocate memory, stacks, registers; and O(δ) is the time for launching the replica from the server to the client host. If we assume that there is a copy of the replica on each host, this cost can be reduced to Lp = O(γ). Similarly, the cost of a statically migrating a replica to a host is Mp= O(α) + O(β), where O(α) is the cost of taking the current state of a suspended process and O(β) is the cost of resuming that state of the host where the process is migrated to.

To simplify the calculation of the complexity, we assume that the time for preparing a process for running, the time for taking the current state of a suspended process and the time for resuming this state on another host is a constant $h$, since they almost do the same thing as going through the low level state of a process. Thus, the cost of resuming a suspended process is Rp = O(1), the cost of launching a replica is Lp = O($h$), and the cost of statically migrating a process is Mp = O(2$h$). Since $O(1) < O(h) < O(2h)$, the cost of resuming a suspended process is the least for scheduling. The cost of launching a replica is less than the cost of statically migrating a replica.

### 3.3.1 Case: Allocate ordered jobs to hosts

In this case, tasks have different order. Thus, when the number of hosts changes, it does not affect this case since the scheduling has to locate as much tasks as possible following their priorities. Thus, in this case the following rules are used for the scheduling.

Rule 7:

*In each tree, descendants of the same ancestor can be processed concurrently.*

*An ancestor and its descendants have to process sequentially.*

Rule 8:

*Leaves need to be processed first.*

*An ancestor is only processed when all its descendants finish their computations.*

As in [33], hosts are divided into groups with the same availability and the scheduler searches the allocated host from maximum to minimum availability, in the hope that a high priority task can finish as soon as possible and give the host back to the next priority task. If there are tasks with the same priority, allocate the one assigned with data storage or the one with big size as shown in the following.

```
for hosts with availability from max to min
        if (number of tasks with same priority > 1)
        if task assigned with data storage
                allocate task with internal analysis approach scheduling
        else if task with bigger size
                allocate task
        else
                allocate any task with same priority
    else
            allocate the most priority unallocated task
  end
```

There are some nodes that can be automatically replicated for concurrent processing which needs to apply the internal analysis approach for scheduling, which are listed below.

### 3.3.2 Case: The number of tasks is greater than or equal to the number of available hosts.

This case happens when the application is just started by the middleware. If the number of hosts in the LAN is small, say less than fifty hosts, this case can occur until half of the tasks are processed. Otherwise, if the number of hosts in the LAN is greater, say about a hundred or more machines, this case might happen until late in the deployment process. In this case, since the number of tasks is so huge, we apply the greedy scheduling. This algorithm helps the task to be processed as fast as it can be. We need to consider the following situations.

*New host joins*

If a new host just joins the middleware, a new replica will be launched on this new host.

*CPU availability increases*

When the middleware gets notice that the CPU availability of a host has increased, it will resume suspended replicas on this host as long as the host performance is still within its availability. In the case where all suspended processes are already re-activated and the CPU is still available, the middleware will command the host to launch new replicas, but maintaining the performance under the CPU availability. The work of resuming suspended replicas gets more priority than launching a new replica. There are two reasons for this decision: first, whenever a replica is created, it has to finish anyway; second, the cost to resume a suspended process on the same host is less than the cost to launch a new replica on a different host.

*CPU availability decreases*

During the deployment process, if the middleware notices that CPU availability on some hosts is below the threshold, it will suspend some replicas in the host until the performance of a host just gets above the threshold so that the deployment is guaranteed to be processed effectively.

In summary, in this state of scheduling, a suspended process will be kept in the same host and only be resumed when the CPU availability of the host is improved. A new replica will be launched on a new joining host or only when all suspended processes of a host are re-activated. There is no migration for process in this state of scheduling.

```
if number of tasks> number of hosts
        if new_host_comes
                launch new replica
        else if CPU_host_increase
                resume suspended processes
        else if CPU_host_decrease
                suspend processes until host achieve balance state
        end
end
```

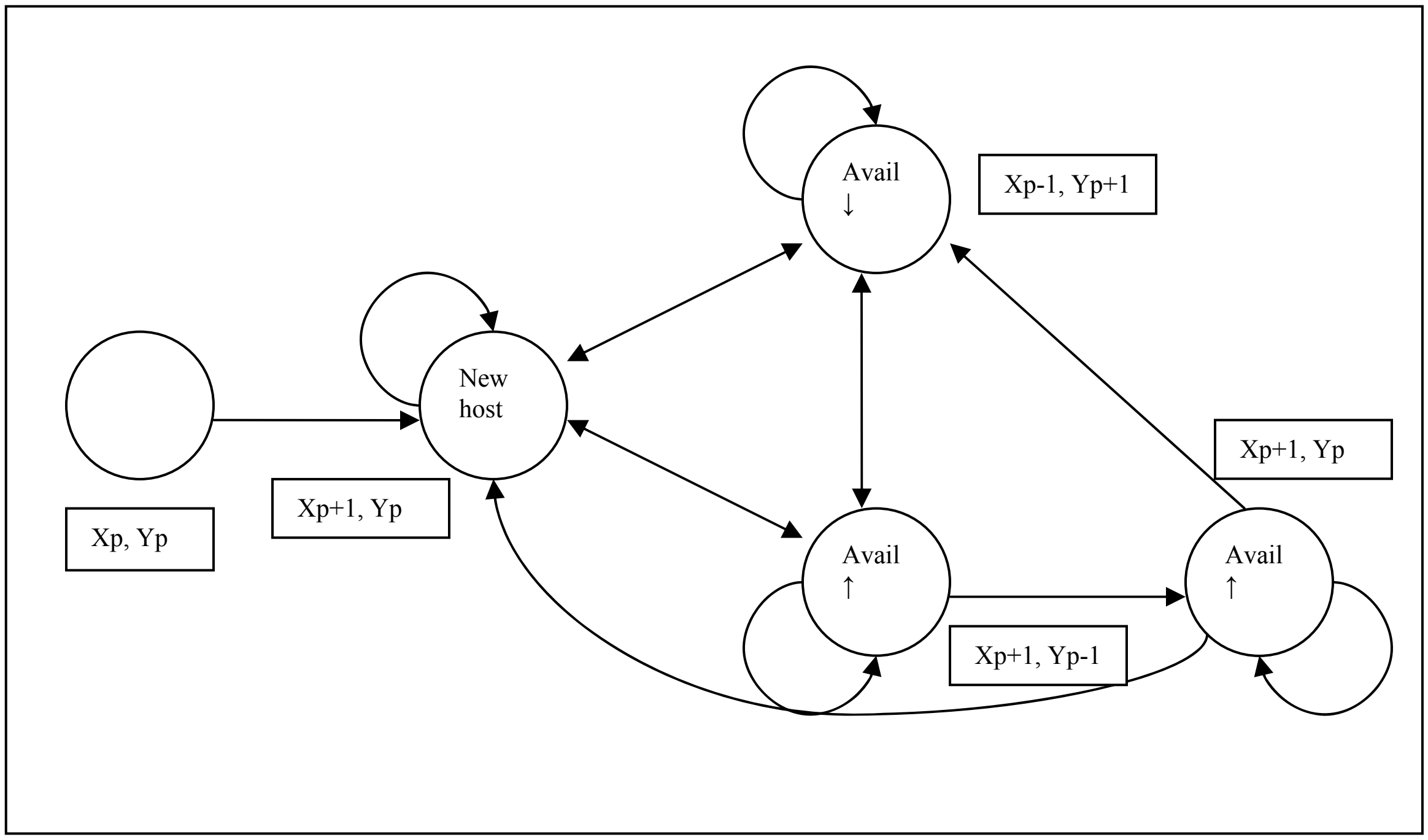

**Figure 17: Machine state for dynamic scheduling when number of tasks is greater than number of available hosts.**

Xp: Number of active processes. Yp: Number of suspended processes.

(Xp+1, Yp): launch one replica

(Xp+1, Yp-1): activate one suspended process

(Xp-1, Yp+1): suspended one running process

### 3.3.3 Case: The number of tasks is less than the number of available hosts

In Phase 2, when the number of tasks is less than the number of available hosts, the scheduler applies the execution time prediction with exclusion algorithm, as mentioned in Section 2.6. Moreover, instead of restarting the timeout task on another host, our design modifies the existing schema by suspending the timeout process, statically migrating it to another idle host in the LAN. Our schema works better since the middleware does not need to store the checkpoint for processing replica and restart the replica from the beginning, but just waits for the migration time to the other host. Secondly, the host to which the replica is migrated to will increase its CPU availability to improve the performance.

During the deployment process, there are some replicas suspended due to the unavailability of CPUs. Although the middleware re-activates those suspended processes in a priority manner whenever the CPU availability of a host increases, still there might be several remaining suspended replicas on the host due to the low CPU availability of host in the first phase. Note that in the first phase, whenever a host is available, the middleware tries to give tasks to those hosts maintaining its CPU availability. Since the availability is relatively high when a host just joins the middleware, it tries to launch as many replicas as possible. And along the deployment process, the availability decreases due to the large number of tasks, some or many replicas are suspended. Some of them might never get re-activated, but these processes have to be finished anyway. The question is when these replicas get re-activated.

All suspended replicas will be re-activated in this phase, when the number of remaining tasks is less than the number of available hosts. The reason is, in this phase the LAN has more idle hosts with relatively high power. These hosts will be the place to migrate suspended processes. Also, the static process migration will be done concurrently with the idle hosts while other hosts are still processing tasks. Processing time for migration will overlap with application deployment time. This helps to utilize idle resource but does not affect the execution time of the application.

We already considered the number of remaining tasks and the number of available hosts. But we have not yet mentioned about the relationship between the number of the processes/replicas, the number of remaining tasks and the number of available hosts. The number of replicas includes those active replicas which are running and those suspended replicas. As explained above, in the first phase the middleware tried to launch as many replicas as possible to consume data records to process. It can likely cause the number of replicas to be greater than the number of data records. On the other hand, there could be a lot of hosts in the LAN and the CPU availability is quite low; then less number of replicas will be launched. So when it gets to the point where the number of remaining tasks is just less than the number of available hosts, in the prior case, the middleware will have the number replicas greater than the number of remaining tasks; and in the latter case, the number of replicas is less than the number of tasks. Since the active process will consume data record first, but suspended processes also have to complete the rest of their tasks. Understanding the relationship and the role of each of them helps us to have a better scheduling algorithm.

#### *3.3.3.1 The number of replicas is greater than the number of remaining tasks*

There are two cases:

*Case 1: active processes >> suspended processes*

It means the number of active processes is much greater than the suspended processes. It also means the number of active process is greater than the remaining tasks. The middleware simulates the data storage on the server to the data storage of the application, whenever there is no data record to process, the data storage on the server will signal the empty data when any data request comes. Any data request gets response of empty data from the data storage server, it will get notice to the replica that it belongs to, that all data records are processed and the application should be finished. In this case, this replica will do the rest of the work of the application and terminate. Since the active processes is greater than the number of remaining tasks, some of the replicas will get data record to process. Those replicas will continue until the task finishes and terminates the replicas. Those replicas which didn’t get any data will wrap up with the rest of the work of the application and terminate as described above. Concurrently with this process, suspended processes will be migrated in a static manner to other idle hosts in the LAN. After being re-activated, the process will be treated as those replicas which did not get any data record, wrap up and terminate. Thus, the execution time of the application totally depends on the processing time of those processes which got the data records to process.

*Case 2: active processes < suspended processes*

In this case, the remaining tasks will be consumed by active processes and some suspended processes. In the priority manner, active processes consume available data records on the data storage server. Concurrently, suspended processes will be migrated and re-activated on the idle hosts. Hosts with the least CPU availability will have suspended processes to be moved. Any re-activated process will consume data record and process until there is no available data record. The reactivated processes help the middleware to have more tasks to consume data record. Hosts which have increased CPU availability will re-activate any suspended process remaining in hosts to contribute to the deployment process.

```
if number of tasks< number of hosts
        if number of replica > number of remaining task
                if CPU_availability_increase
                        active suspended process
                elseif CPU_availability_decrease
                        suspend processes until CPU availability balance
                end
                if process passes its predict time
                        suspend that process
                end
                if host_is_idle
                        static migrate suspended processes to idle hosts
                end
        end
```

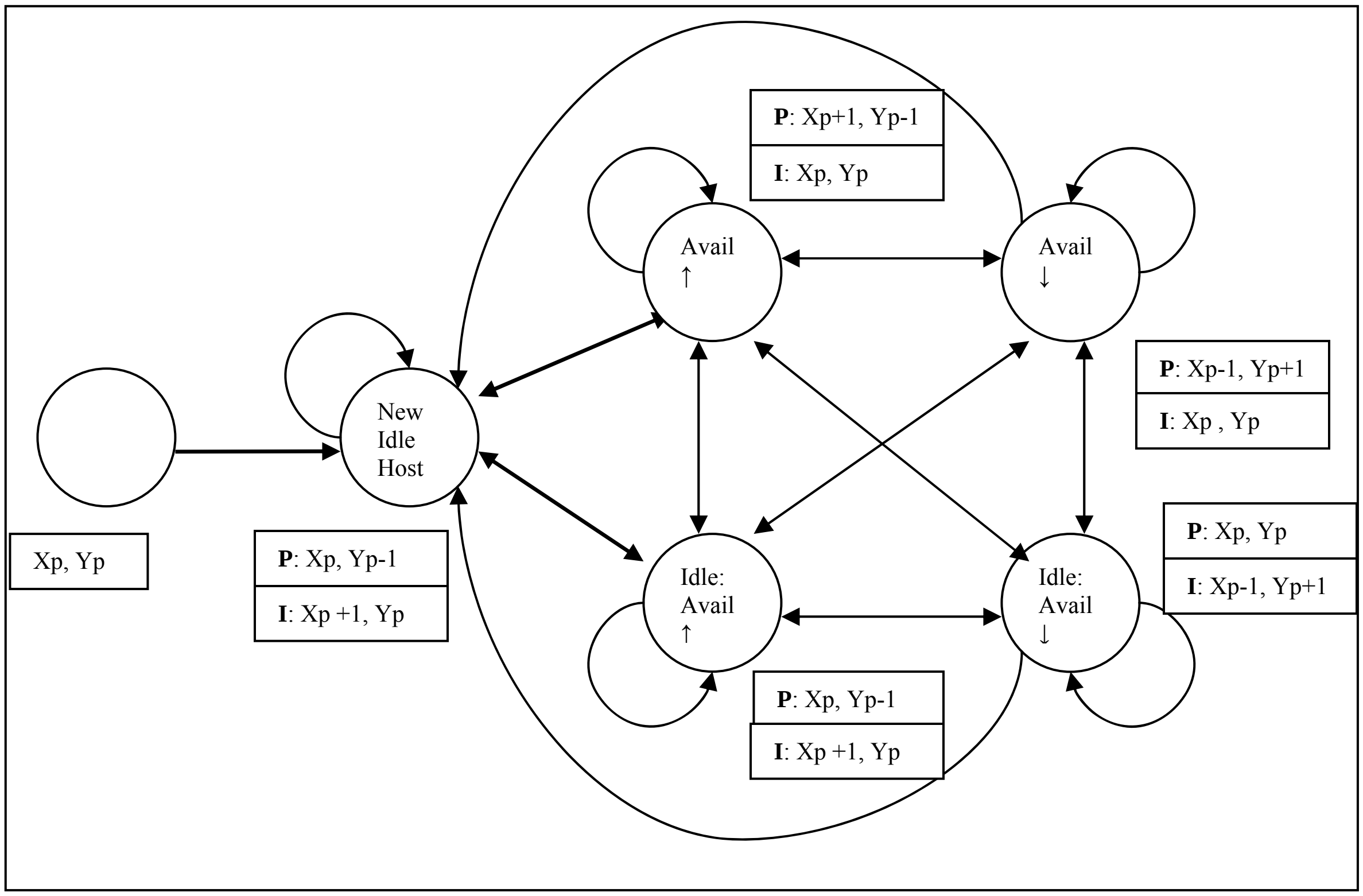

**Figure 18: State machine of the dynamic scheduling in Phase 2a – the number of replicas is greater than the number of available hosts.**

P (Xp, Yp-1) move suspended process on processing host to
I (Xp+1, Yp) migrate and active it on idle host

P (Xp, Yp)
I (Xp-1, Yp+1) suspend an running process on idle host

P (Xp+1, Yp-1) active an suspended process on processing host
I (Xp, Yp)

P (Xp-1, Yp+1) suspended a running process on processing host
I (Xp, Yp)

In either case, when the number of tasks is less than the number of available hosts, suspended process will be migrated immediately in a static manner to idle hosts. In this phase, the LAN will have more idle hosts with relatively high availability, migrating the suspended processes to those idle hosts will not affect the resources of the hosts. Also, the migration will be done concurrently to the idle hosts while other hosts are still processing the application. This helps to overlap the execution time of the whole application. In this case, even if the CPU availability increases, the host will be reserved to activate any suspended process to wrap up the deployment process. No more replication is allowed, since there are more replicas than the current data recorded needing to be processed.

#### *3.3.3.2. The number of replicas is less than the number of remaining tasks*

Same as the above case, but with a little modification, any active process will consume the data record first. Concurrently, in any idle hosts, new replica will be launched to hasten the deployment process. Obviously those new replicas will get data records, since the current number of replica is less than the number remaining tasks. This will continue until there is no data record to process. Also whenever the CPU availability in any processing host increases, the hosts will be reserved to re-activate any suspended process to contribute to the deployment process or wrap up and terminate if no data record needs to be processed. In this case, launching new replica on new idle hosts gets more priority than static migrating suspended process since the cost of launching is less than the cost of migrating.

```
if number of tasks < number of hosts
        if number of replica < number of remaining tasks
                if host_is_idle
                        launch new replica
                elseif CPU_availability_increase
                        active suspended process
                elseif CPU_availability_decrease
                        suspend processes until CPU availability balance
                end
                if process passes its predict time
                        suspend that process
                end
        end
```

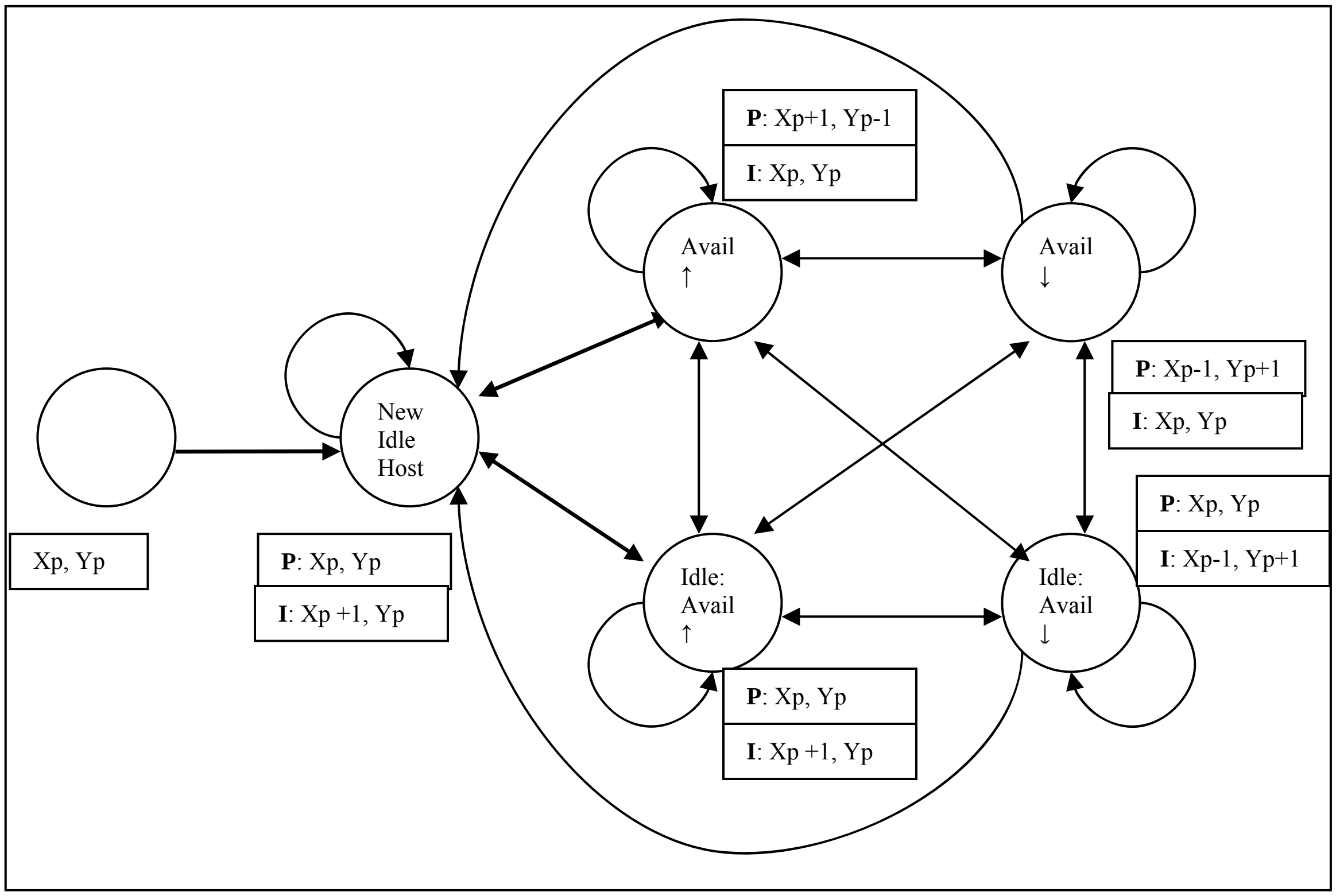

**Figure 19: State machine of the dynamic scheduling in Phase 2a – the number of replicas is less than the number of available hosts.**

P (Xp, Yp)

I (Xp+1, Yp) launch one new replica on idle host

P (Xp+1, Yp) activate one suspended process on processing host

I (Xp, Yp)

P (Xp-1, Yp+1) suspend a running process on processing host

I (Xp, Yp)

P (Xp, Yp)

I (Xp-1, Yp+1) suspend a running process on idle host

**Summary of the dynamic scheduling:**

The dynamic scheduling approach described so far can be summarized as follows.

| | | |
|---|---|---|
| tasks >= hosts | FCFS | |
| | New joining host: | launch new replica |
| | CPU availability increases | • priority1: re-activate suspended processes<br>• priority 2: launch new replica |
| | CPU availability decreases | suspend some processes |
| tasks < hosts | exclude CPU with less availability | |
| | replicas > tasks | • processing hosts: active processes grasp data to process or finish task with no data<br>• idle hosts: migrate and re-active suspended process to finish task |
| | replicas < tasks | • processing hosts:<br>1. active processes grasp data to process<br>2. re-active suspended process if CPU availability increases<br>• idle hosts: launch new replica |

# CHAPTER 4: IMPLEMENTATION

This chapter describes the implementation of the design detailed in Chapter 3, including use cases, class diagrams and scenarios. The implementation was done to run concurrently multiple replicas of the same binary executable code of the application on different hosts in a LAN to share the same computation performed on different disjoint partitions of the data storage. The utilization of idle CPUs in the LAN is exploited as much as possible in this design. Thus, the more CPU cycles on different hosts in the LAN are utilized, the better performance is expected. The total deployment of the application should be decreased significantly.

## 4.1 Use cases

<u>Replicate application</u>: Replica (string filename)

This function makes a replica of the application at runtime. When a host receives a message from the server to launch a new replica, the middleware checks if the host already has the executable file of the application. If not, it downloads from the server and keeps a copy in the repository for latter replication. The middleware invokes the fork system call to generate a child process and passes the replica file name to execute a new replica of the application.

<u>Catch the system call</u>: TraceProcess (pid_t pid)

This function starts a trace system call of a replica. The middleware passes the process id to this function to observe the system calls made by a replica. The process id of a replica is obtained when the middleware generated a new replica. Usually, this function is called right after a new replica is launched. For our purpose, the middleware only processes the following system calls: fork, exec, wait, terminate, trace, read, and write. If the system calls are for file access, such as read or write system calls, the middleware directs them to the Interface object on the server and waits for the result. If the system call is fork, the middleware continues to observe the child process of this replica.

Schedule

This function is used to re-schedule hosts performance in the LAN. When notified by the Communication object for re-scheduling, the function gets the number of currently available hosts, current status of the data storage on the server, current status of hosts in the LAN, and based on the dynamic scheduling algorithm the scheduler makes the decision and sends command to client hosts and any appropriate change in performance.

Suspend Process: SuspendProcess(pid_t pid)

This function is used to suspend a running process in a client host. When notified by the Communication object that a host performance is under its CPU availability, this function will send a message to the client host to suspend a running replica and get back the process id of that replica. This information will be updated in the pool of hosts on the server.

Migrate Process

This function is used to statically migrate a suspended replica on a host to an idle host. In the second phase of the deployment process (when the number of remaining tasks is less than the number of available hosts), and notified by the Communication object of new idle hosts, the scheduler decides which host having the suspended replica to be migrated. The scheduler sends a command to the host to obtain the current status of the suspended replica. Then the scheduler sends a command to the idle host to resume this status on that host. Finally, the scheduler updates the status of the two hosts in the pool of hosts on the server.

Process a client Message: ProcessHeartBeatMessage(String message)

This function allows the server to process messages sent from client hosts. The Communication object receives and parses the messages sent from client hosts. If a new host joins the middleware, this function registers the new host as a member and updates its status in the host pool on the server. If the message announces the current status of a client host, this function updates the current status of the host for later use by the scheduler. If the newly joining host has increased/decreased CPU availability, this function notifies the scheduler to do the rescheduling.

Request Data: RequestData (pid_t pid, string buffer, int size)

This function is used to direct a data request from a replica to the interface for data access on the server. The returned data record is written over the buffer.

Write Data: WriteData (pid_t pid, string buffer, int size)

This function is used to direct a data write from a replica to the interface for data access on the server. The content of the buffer to be written to the data storage on the server is passed to the interface.

Buffer Server Data to Memory: Index Buffer()

This function is used to buffer data from the data storage on the server to memory for scalability and concurrent access. The data fetched into memory will be accessed by functions in the Interface object.

# 4.2 Class Diagram

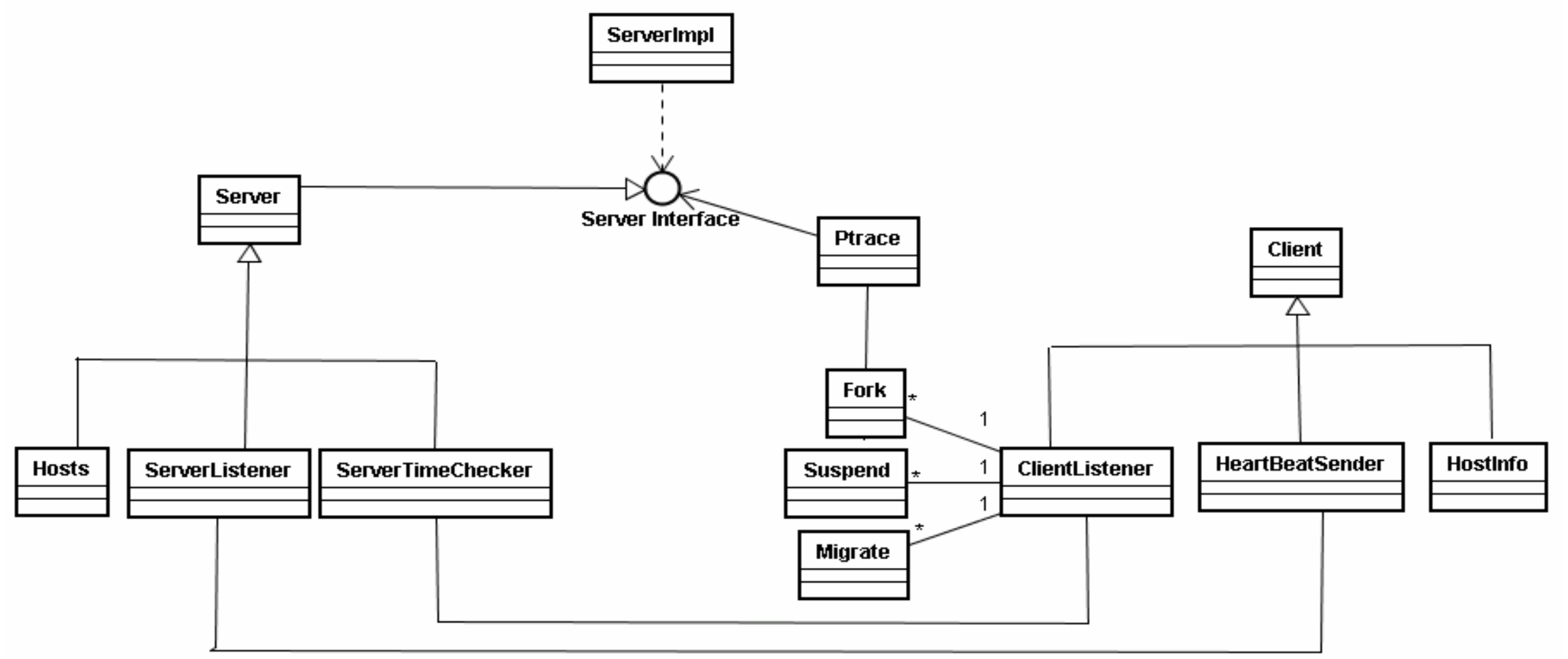

**Figure 20: Class diagram**

**Server**

This class is the main control of the middleware. It's similar to the Client class but is integrated with more functionality for the server side. This class has three inner classes: Host class, TimeChecker class and HeartBeatSender class. Server class initializes and registers the Server.

**Server.Hosts**

Host class is an inner class of the Server class. This class stores information about the current status of each host in the LAN. The information in this class is updated by Server.ProcessHeartBeatMsg() method when client hosts send their current information to the Server. The information of this class is retrieved by Server.TimeCheck() method to get current information of a client host for the server to do the scheduling.

**Server.Listener**

Listener is an inner class of the Server class. ServerListener is a thread living as long as the middleware is running. ServerListener always listens for messages send from hosts in the LAN and parses the messages to the server to continue processing.

**Server.TimeChecker**

TimeChecker is an inner class of the server class. Server.TimeChecker is a thread initialized by the server class. After a specific time, TimerChecker will collect the current information of hosts in the LAN and do the scheduling based on its own algorithm.

**Server.HeartBeatMsgSender**

HeartBeatMsgSender is another inner class of the server class. HeartBeatMsgSender is also a thread initialized by its container class. After a specific time, HeartBeatMsgSender sends the current status of the host that it is running on to the Server to update its status.

**ServerInterface**

ServerInterface provides the interface for all the access features of the data storage on the server. These features will be invoked by hosts in the LAN to access the data storage.

**ServerImpl**

ServerImpl implements all the access features provided by ServerInterface. ServerImpl performs the actual access to the data storage on the server side. ServerImpl is initialized and registered by the server when the server is first started.

**Client**

Client plays the role of each host in the LAN. This class performs the responsibilities of a host in the middleware. Client class receives commands from the server and controls the performance of replica on the host where Client is running on. Client has the following two inner classes.

**Client.HeartMsgBeatMsgSender**

HeartBeatMsgSender is an inner class of the Client class. HeartBeatMsgSender is a thread initialized by the Client. It performs the same responsibilities as the class HeartBeatMsgSender on the server. After a specific time, HeartBeatMsgSender sends the current status of the host it is running on to the server to update the host's status.

**Client.Listener**

Listener is an inner class of the Client class. Listener is a thread initialized by theClient. Client.Listener always listens to commands sent from Server.Listener. It parses the commands and passes them to the Client class to do appropriate actions.

**HostInfo**

HostInfo stores the information of a host, such as its CPU availability, IP address, and current status.

**Ptrace**

Ptrace performs as a wrapper for the running process it is observing. Ptrace traces some specific system calls of the process it is observing. When these system calls are captured, Ptrace directs them to ServerInterface, gets the result and passes it as the result of the original system calls.

**Fork**

Fork performs two tasks. Fork does the replication for the current application by spawning a new process based on the binary code of the application. On the other hand, Fork also invokes Ptrace and passes the process ID of the new process to Ptrace to detect this new process.

**Suspend**

Suspend suspends a specific process.

**Migrate**

Migrate resumes a process on a host, provided the process's status.

# 4.3 Scenarios

**LaunchAReplica**

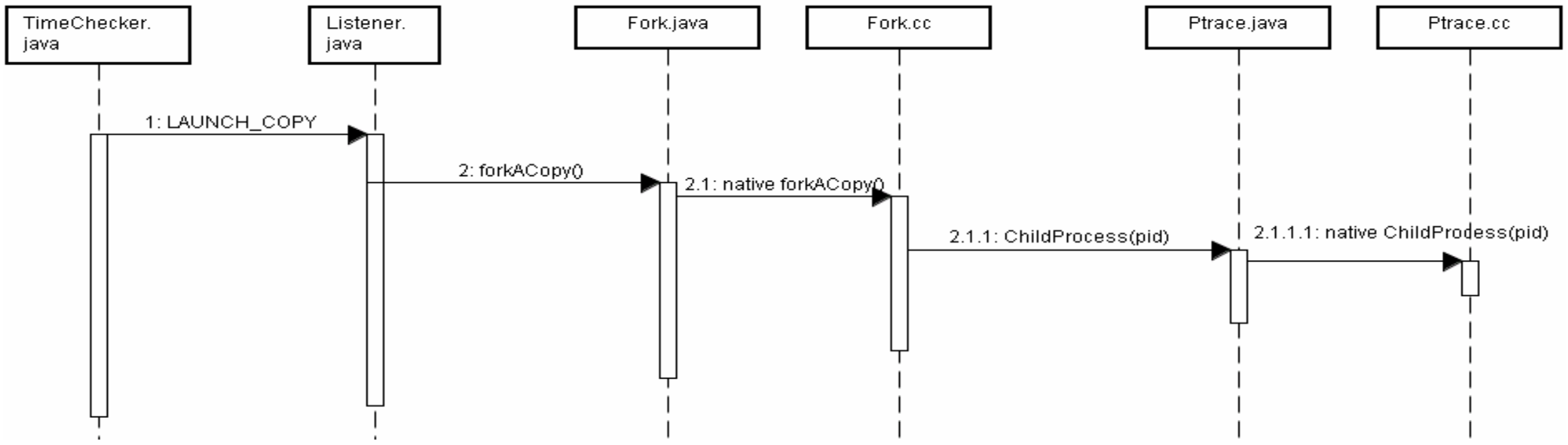

**Figure 21: Launch a Replica Scenario**

This case happens when the Server Core component does the frequent scheduling and decides to launch a new replica. Server.TimeChecker class does its frequent scheduling by calling Server.TimeCheck() and decides to launch a new replica by calling Server.Fork.forkAReplica(). This method forks a new process of the application and passes the process id of the clone process to Ptrace.traceProcess(pid).

**ServerScheduling**

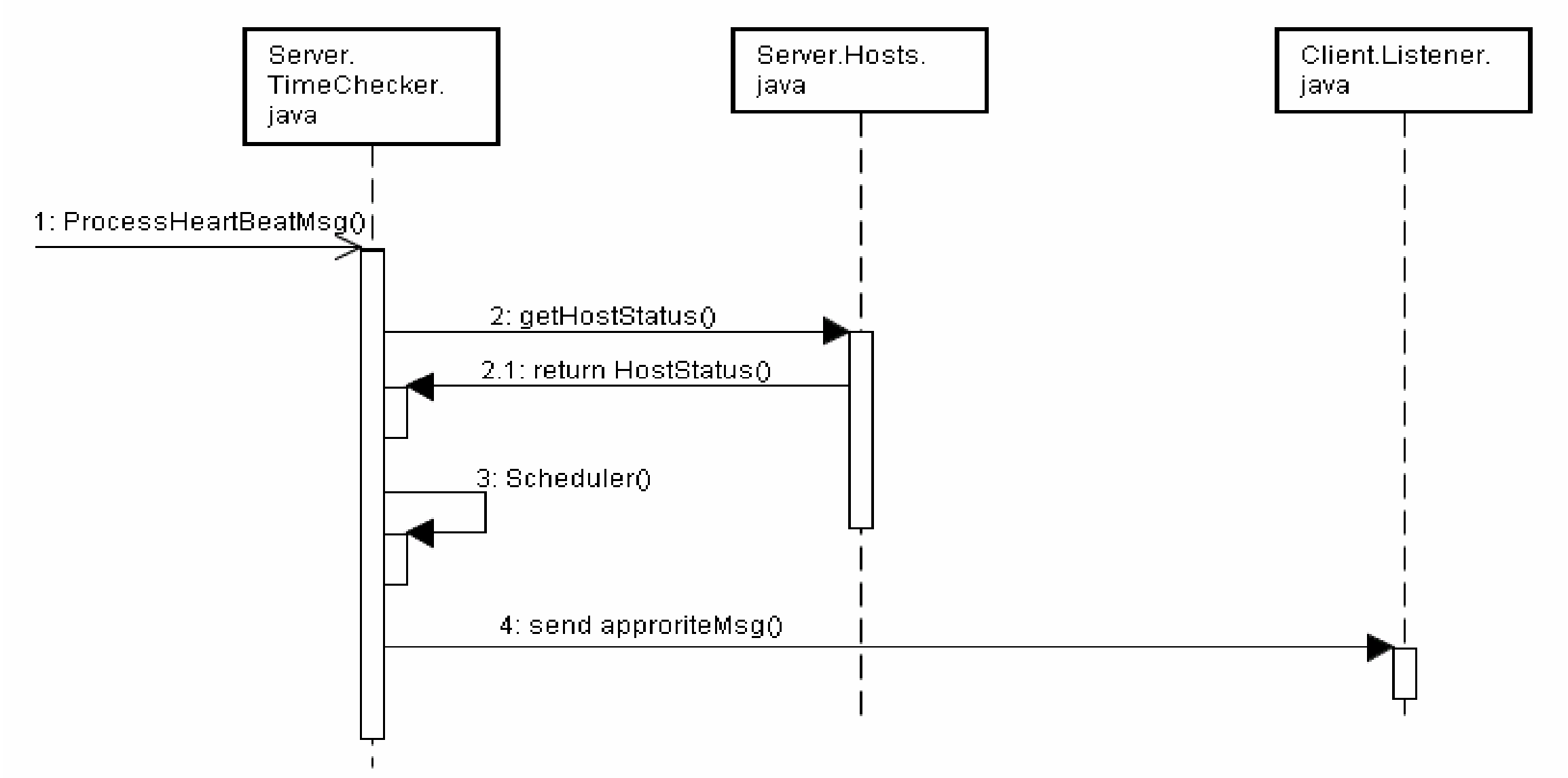

**Figure 22: Server Scheduling Scenario**

Similarly, after a specific time, the Server does its scheduling. The Server grasps the current information of each host kept on the Server by invoking Server.Host class. Based on the information, the Server produces appropriate command and sends this command to the Listener on Client side.

## SuspendProcess

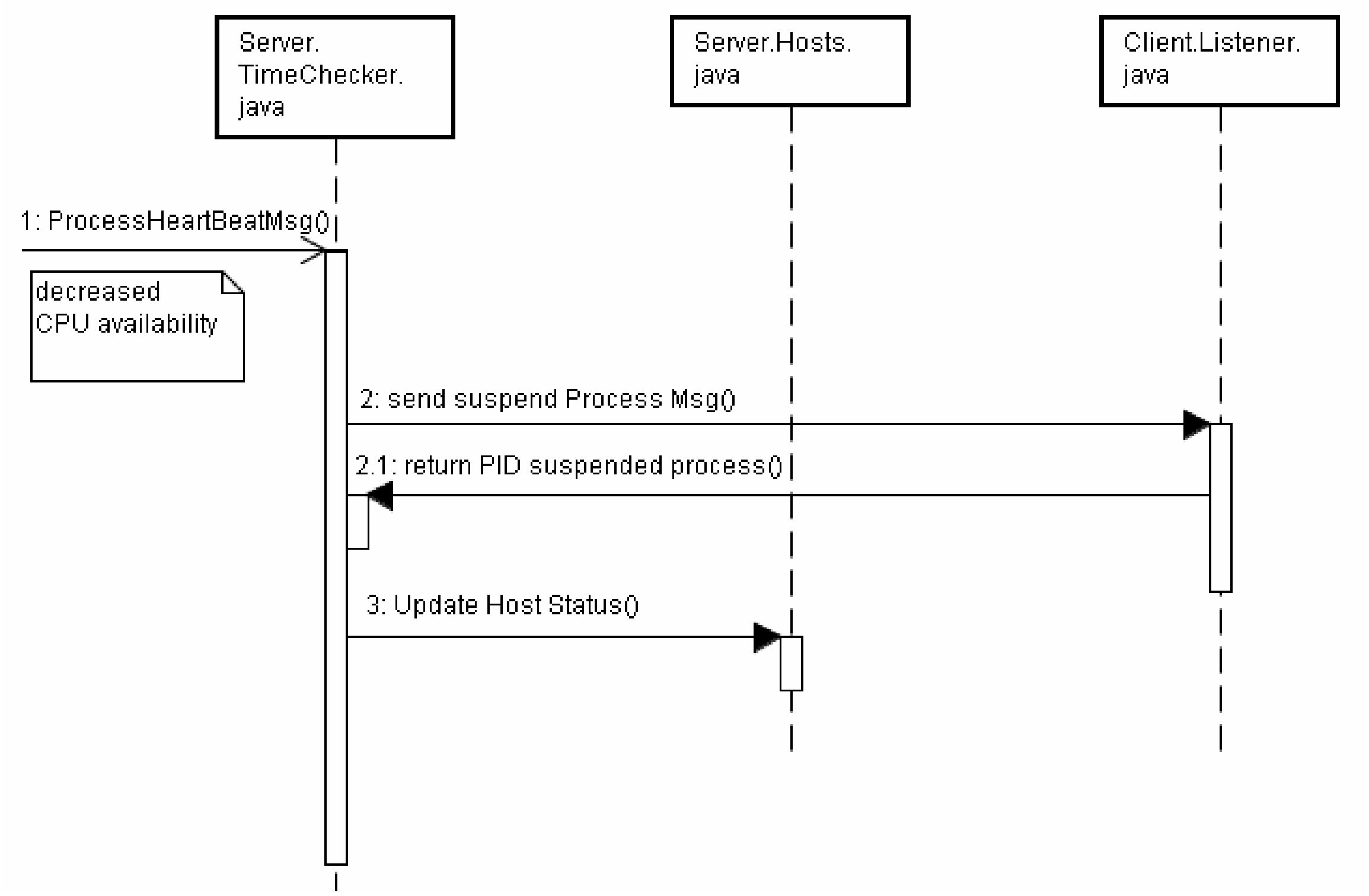

**Figure 23: Supend a Process Scenario**

## MigrateProcess

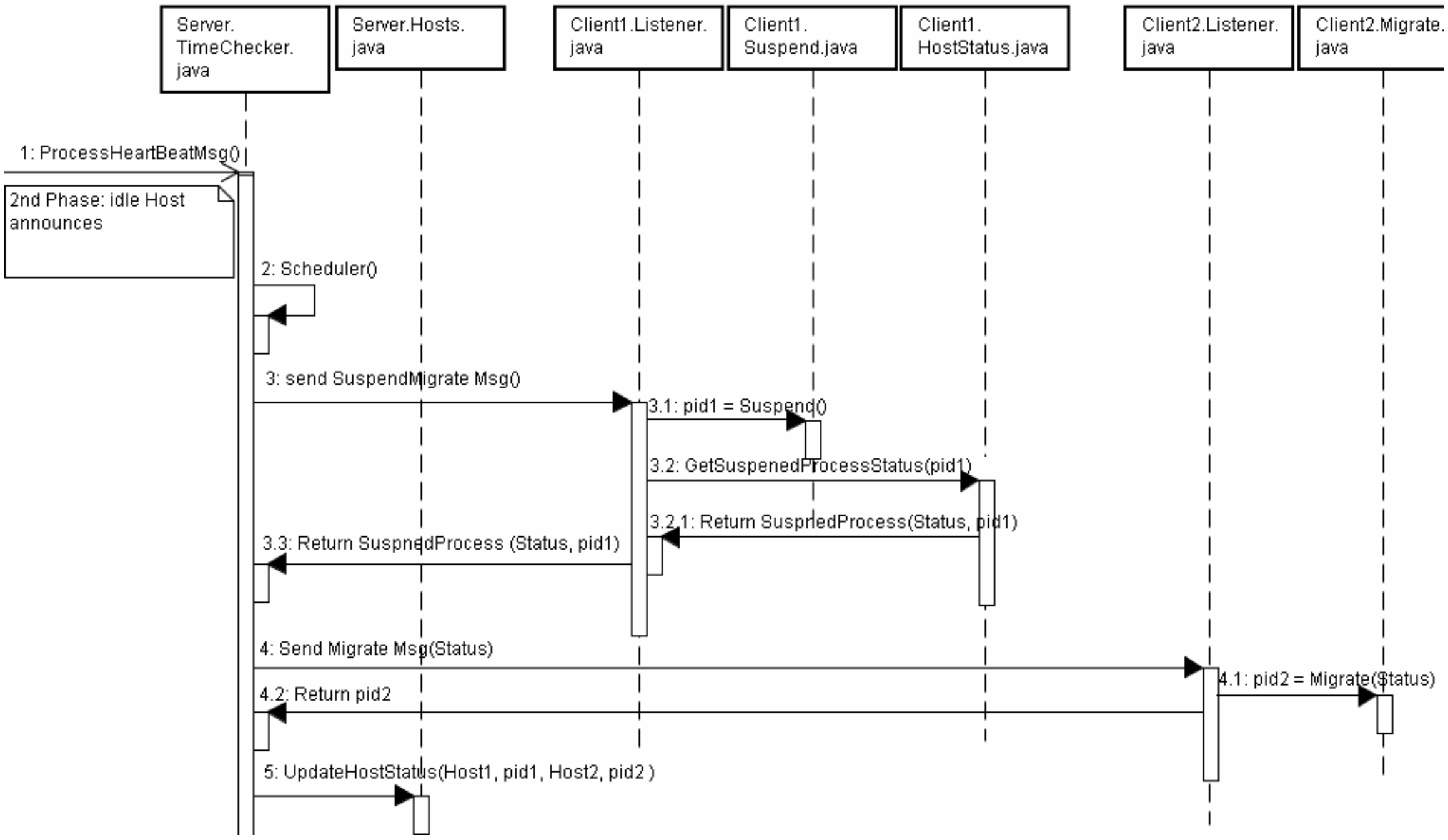

**Figure 24: Migrate Process Scenario**

### HeartBeatMessage

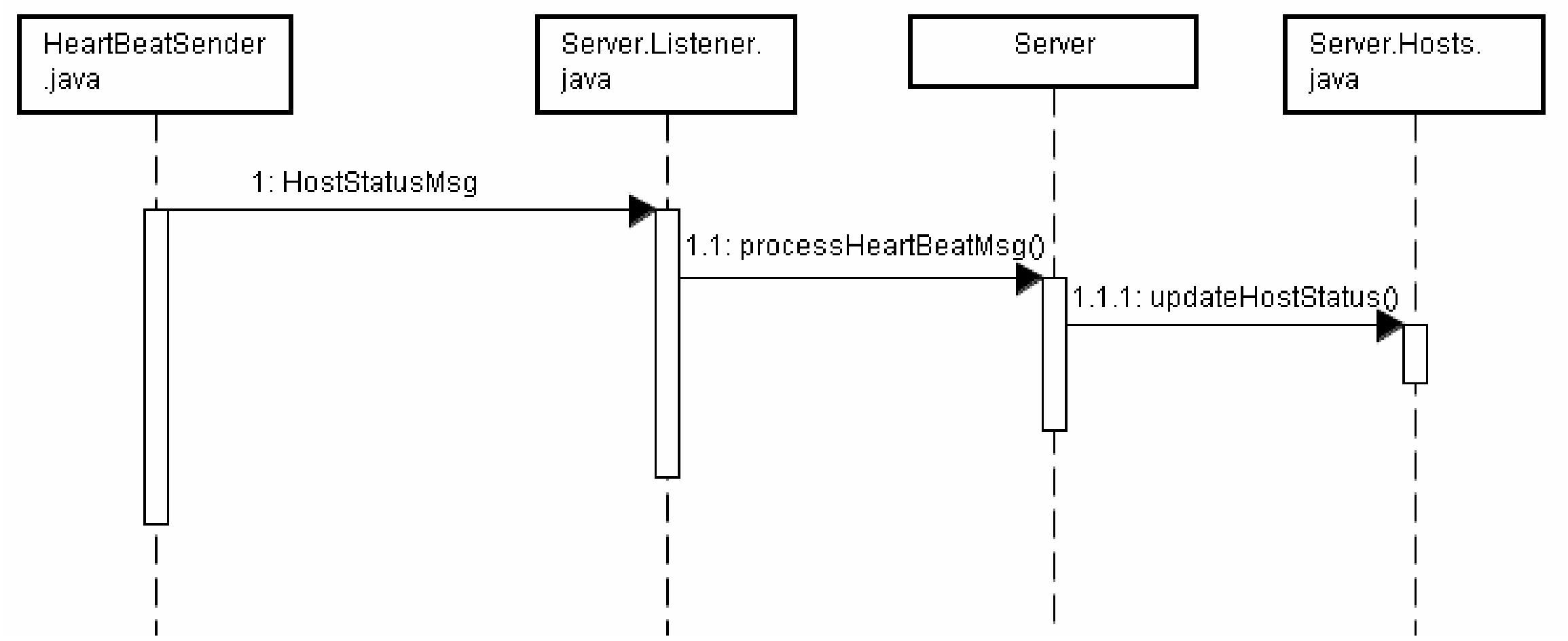

**Figure 25: Heartbeat Message Scenario**

After a specific time, each host sends its current status to the server by invoking HeartBeatSender(status). Server.Listener class receives the message, parses and passes it to Server.ProcessHeartBeatMsg(status) to process hosts' information. This function updates the information of each host kept on the server by calling the Server.Host class.

### RequestDataFromServer

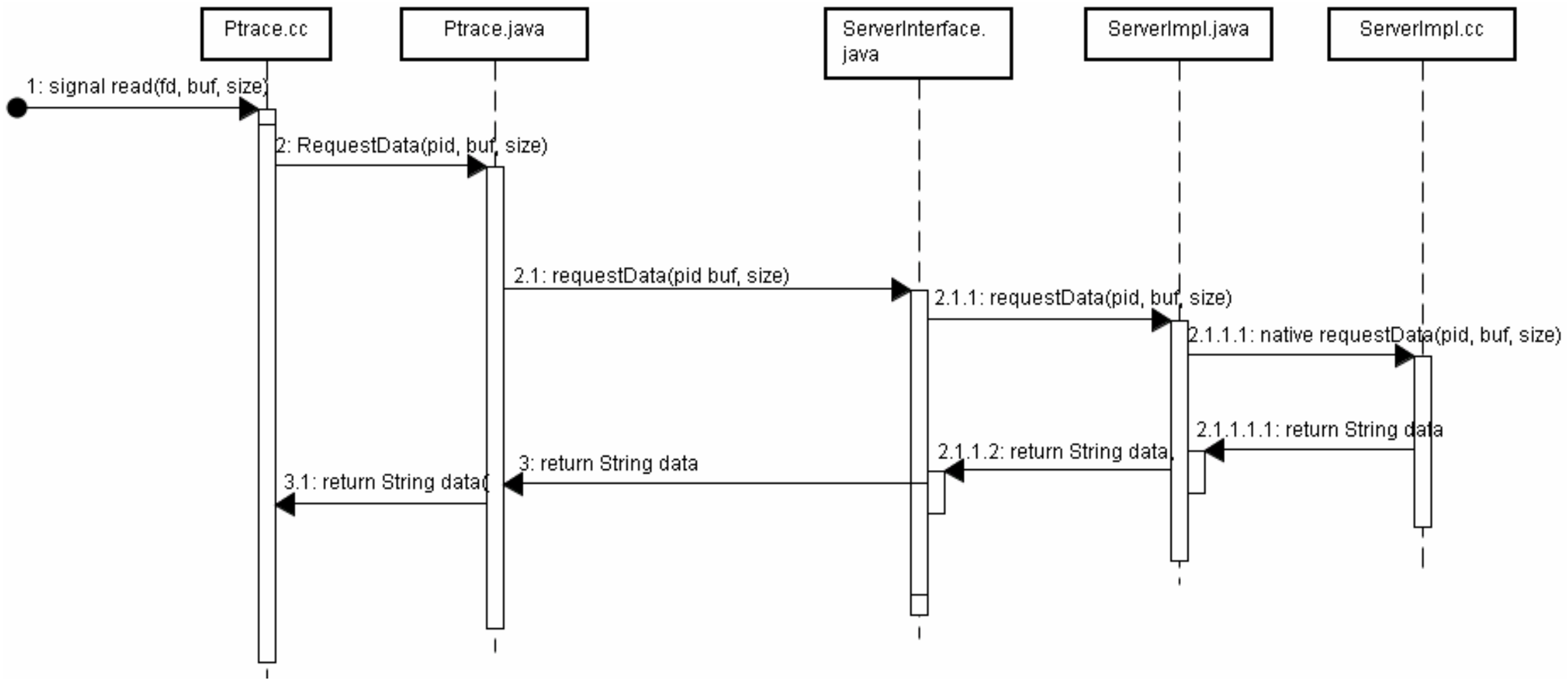

**Figure 26: Request Data from Server Scenario**

When Ptrace detects a read system call, it takes all the necessary parameters and makes a RMI request to ServerInterface. The real handle is executed in ServerImpl. After processing, ServerImpl returns the result to ServerInterface, which returns back to Ptrace. Ptrace will do the appropriate actions to overwrite the read system call and returns the control back to the read system call.

It's the same scenarios for write system call, but the returned result is to let Ptrace know whether the write system call is written successfully on ServerImpl.

**WriteDataToServer**

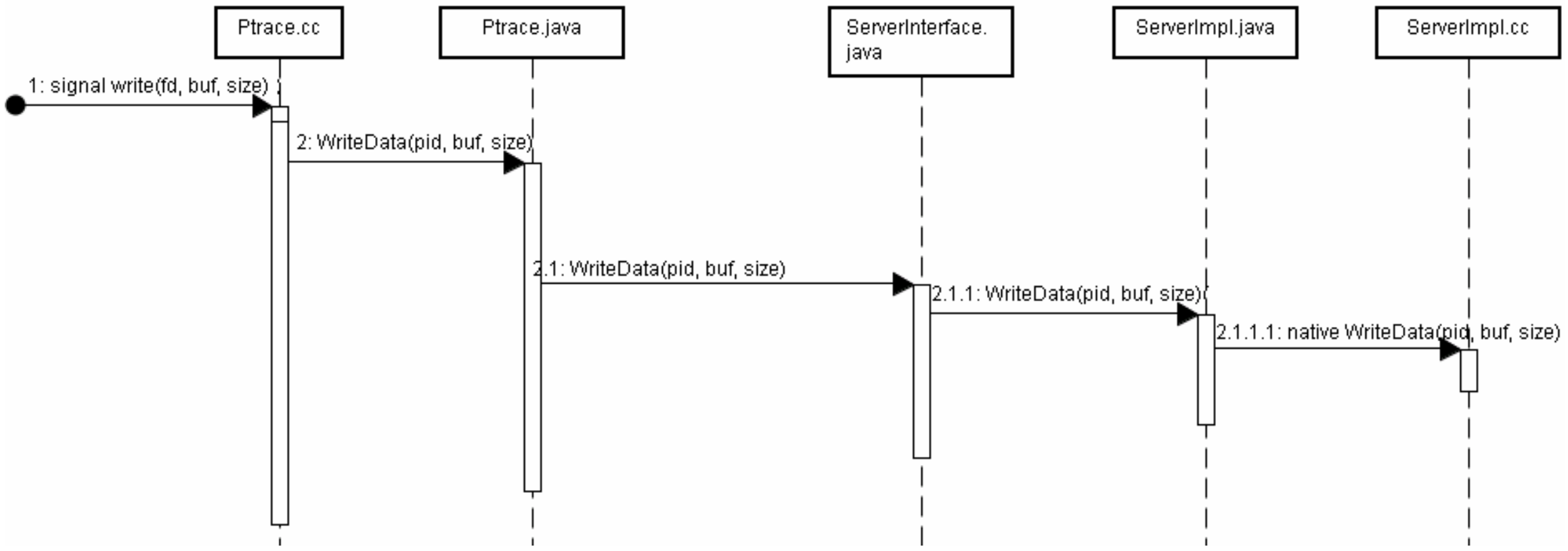

**Figure 27: Write Data to Server Scenario**

# CHAPTER 5: EXPERIMENTS

The previous chapters detailed all aspects of the design, ranging from concrete ideas, design and implementation. This chapter reports some experiments and case studies as the proof of concept for our idea. The techniques used, the difficulties encountered during the experiments, the results, observations and judgments are presented for each experiment.

### Why Java, CNI, Linux?

Observing system calls needs programming at low level. The C programming language can be used for this purpose. But we also want to make our middleware platform independent; using Java is one of choices. Thus, we need an interface between these two languages. The Java Native Interface (JNI) and the C Native Interface (CNI) are possible choices. We use CNI since it is more efficient, more convenient and less complicated than JNI, although CNI is less portable than the standard JNI. Also, Linux is easier for kernel programming, so we chose Linux for the experiment environment. All implementation is in Java. Only when a new replica is launched, a Java native method is invoked, which actually calls a C method, to observe the new process replica.

### Difficulty

Java programs are compiled using the `javac` command and C programs are compiled using the `gcc` command. We can also use the `gcj` compiler from GNU to compile both Java and C programs. Also, `gcj` makes it easy and efficient to mix code written in Java and C++. Another advantage is that the `gcj` compiler can combine all file export from RMI to make the final executable file [9]. Another difficulty, which is significant for our experiment, is that Java supports JDK 1.6 or more whereas GNU supports up to JDK 1.4 only due to the current version of GNU 4.0. Thus, a lot of features of Java 1.6 are not used. Before this issue was realized, the Communication component and the System component were written separately. While the Communication component used a hash table to store and retrieve information for each host or process. When compiling all the components, a lot of errors were generated due to the incompatible GNU compiler. The Communication component had to be rewritten using arrays to be suitable for the compiler environment.

**First error while doing experiment**

In the System component, we observe running processes to catch system calls. When a system call invokes the data storage, the middleware will notice it, take the actual data needed and write this data into the buffer in the read system call after returning from the call. Due to the concurrency it is possible that, if we don't observe carefully using the file descriptor, we might overwrite onto the buffer of the wrong file. This led to crash of the middleware.

**Second error while doing experiment**

To set up a LAN, connect all machines, say a maximum of four machines, into the router. Now we need to name the machines, if they have not been named yet. Also, use the *ifconfig* command to find the IP address assigned by the router for each host in the LAN. Modify file */etc/hosts* with three columns: IP address – domain name – hostname. Then try to ping among hosts in the LAN with their names or IP addresses to make sure they can contact each other. Once the network is set up, we can deploy the Communication component on all the hosts in the LAN to test if the hosts can talk to each other. There is still the case that those hosts can not communicate to each other, even though they still can ping other hosts. For communication purpose, in the Communication component on each host, there is one port registered to listen to messages sent from the other hosts. The first case is where this port is used for other application. Solution one is to change to another port; or the other solution is the Communication component should detach if this port is available before registering it for the communication purpose. The second case is the firewall locks this port for security purpose. We can open this port by the GUI interface, by going into System → Admin → Security → Open another port.

In the following, we describe three case studies conducted by running our middleware on two hosts.

## 5.1 Case study 1: Accounts Update, a simple banking application

### 5.1.1 Overview application

The data storage is a file in which every line is a record storing the information of a client's account. During the day, all transactions on an account are stored in a temporary file. At the end of the day, when there is no transaction made and there are more idle hosts in the LAN, the application will be run to update all the accounts in the data storage incorporating all the transactions made during the day. Thus, the application reads in a line containing the current balance of an account, does some computation based on the transactions that occurred during the day, and writes back the new balance into another file. The application continues to do this until the end of the data storage when there is no more record to process. When the application finishes updating, the file in which the new balances are written becomes the new data storage, and the data storage where previous balances are store becomes the checkpoint file. This application is written in C. This is a single-threaded application, it processes every account sequentially. In this case, the data shared between all tasks is the original data storage.

### 5.1.2 Middleware operations

A copy of the data storage is put on the server and is observed by the middleware. To simplify, a record just stores the current balance of an account. The host in which the application starts is considered as the server. It loads the main core of the application, starts the TimeChecker thread responsible for the scheduling, registers the ServerImpl on the server side and registers a port for the Listener thread to start listening for messages from all the hosts in the LAN,. Besides these duties, the server also performs other tasks for other client hosts in the LAN. It starts the HeartBeatMsgSender thread to send the host status to the server after a fixed period of time. This communication is maintained during the operation of the middleware. Then the scheduler finds that the current server host is idle, it decides to launch the application into the middleware. The data storage is copied into the server. The Index Buffer is fetched with a big chunk of data. Address of the Index Buffer is assigned to the variable counter. The application starts to operate. The data storage is opened for accessing. Any system calls made for the file descriptor to this data storage will be observed. Since we know the size of each data element, we expect each data request to be of the same size. If the first data request to the data storage for the size of each element, then the middleware can observe every data request as a read/write system call. On the other hand, if the first data request comes and asks for the size of data bigger than the size of one data element, then we expect the middleware observe *mmap* and *read* system calls for every data request coming later. In this case, the middleware observed the first read system call asking for data from the first data storage.

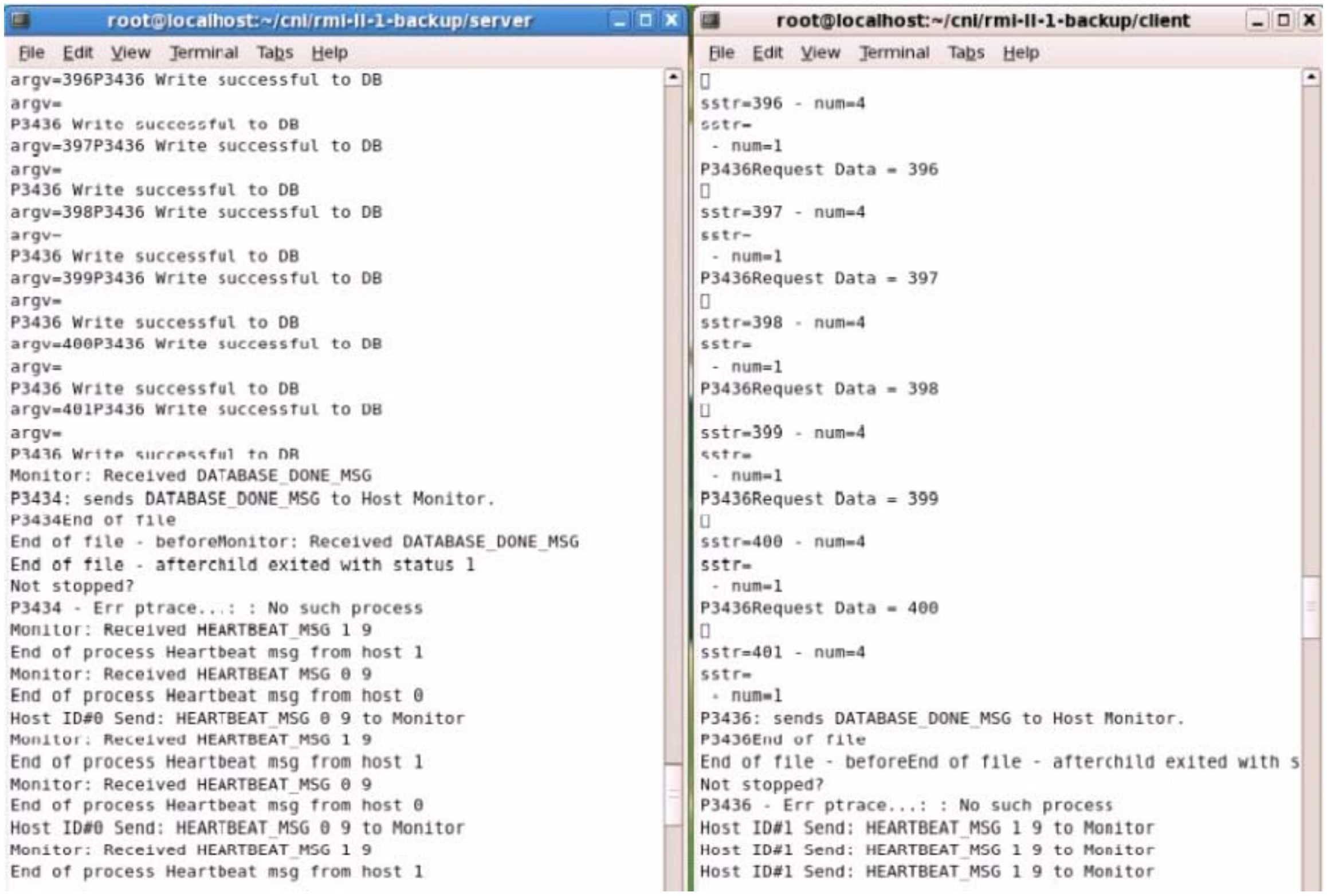

**Figure 28: Two replicas run concurrently: One is on server side. The other is on client side**

The middleware sees that the first read system call asks for the data with the exact size as that of one data element. The middleware will apply the first strategy for later observation. The middleware directs the request to ServerImpl, gets the data from Server Interface, returns to the request and overwrites the buffer of the read system call. Read system call gets the new synchronized data element from the server, feed it for the computation. This routine continues for the next iterations.

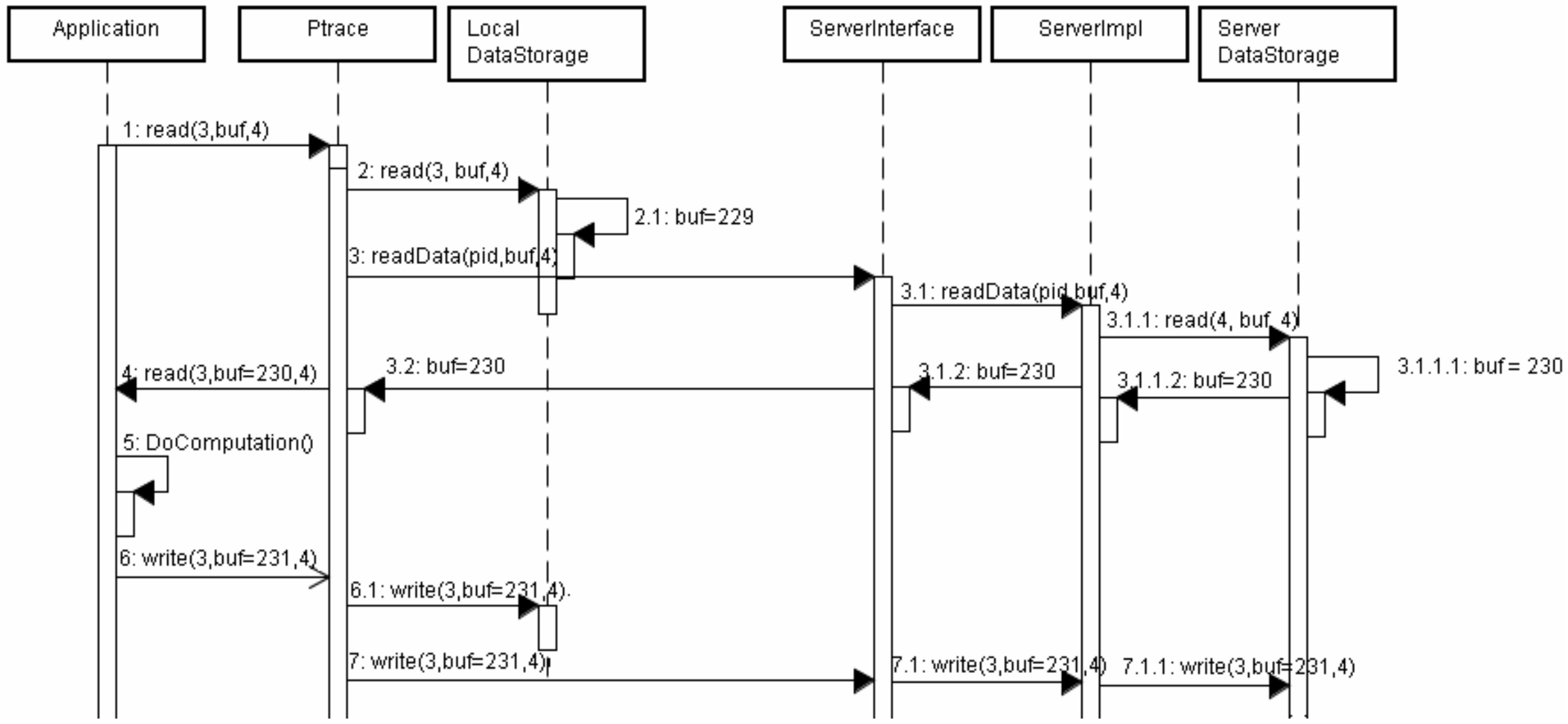

**Figure 29: Processing of data request**

## 5.1.3 Result

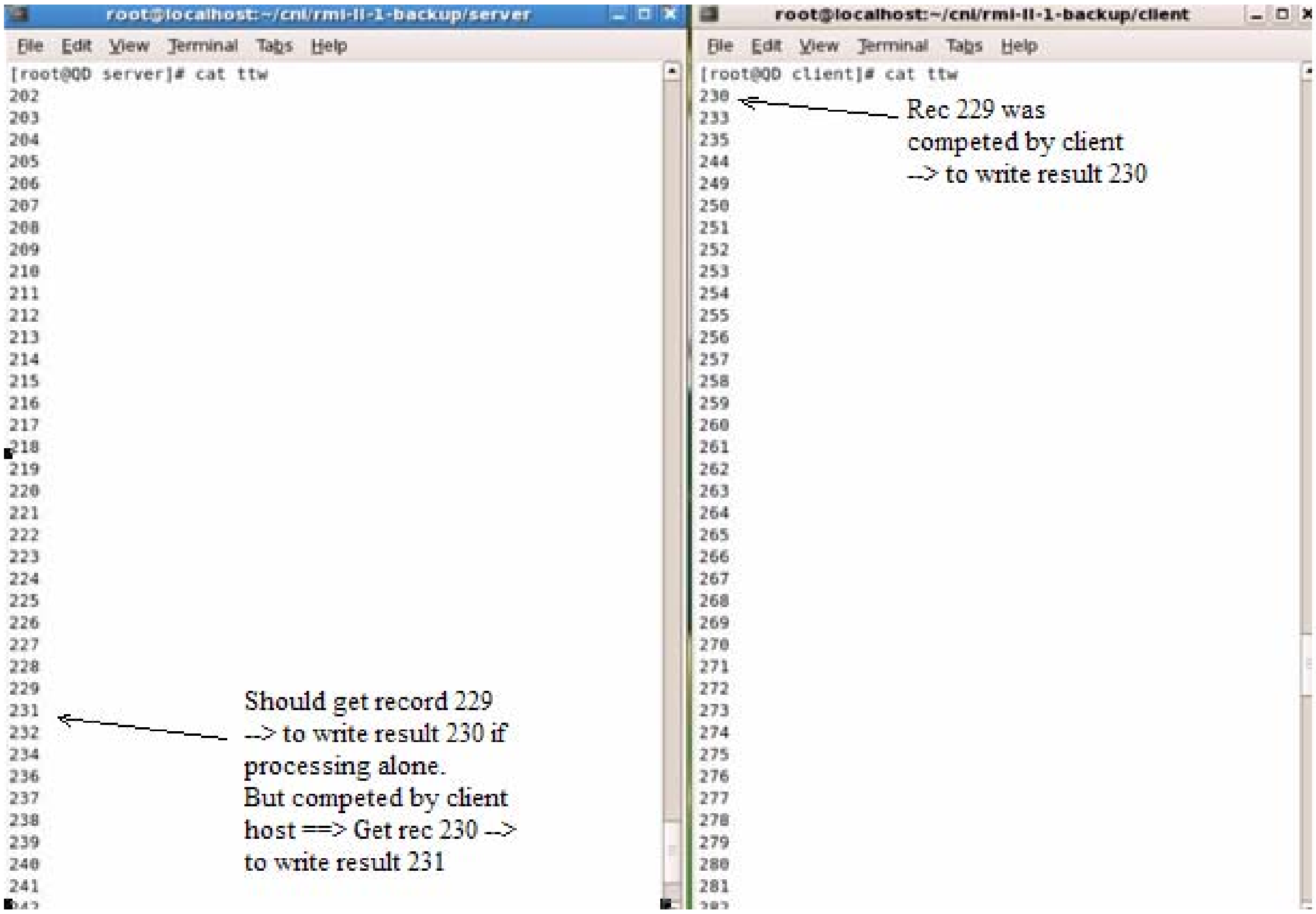

**Figure 30: Result of Experiment 1 for each replica**

Figure 27 shows records consumed on host #0 on the server side and host #1 on the client side. On host#0, at the 29th record, if this host reads from the local data storage, the host should get the 229th data record, which is next to the 228th record processed previously. But since this request was directed to the server data storage and the 229th record was given to the client host#1, host#0 took the 230th record and did the computation producing the 231st record as the result. Then host#0 wrote this result to its local data storage and the request was also directed to write to server data storage. So the results written to local data storage were records processed by that host. But the server data storage still gets the full results for all processing. As the experiment showed that host#0 processed 43 records and host#1 processed 157 records. All results were directed to server data storage for read and write. The server data storage was full updated with 200 records.

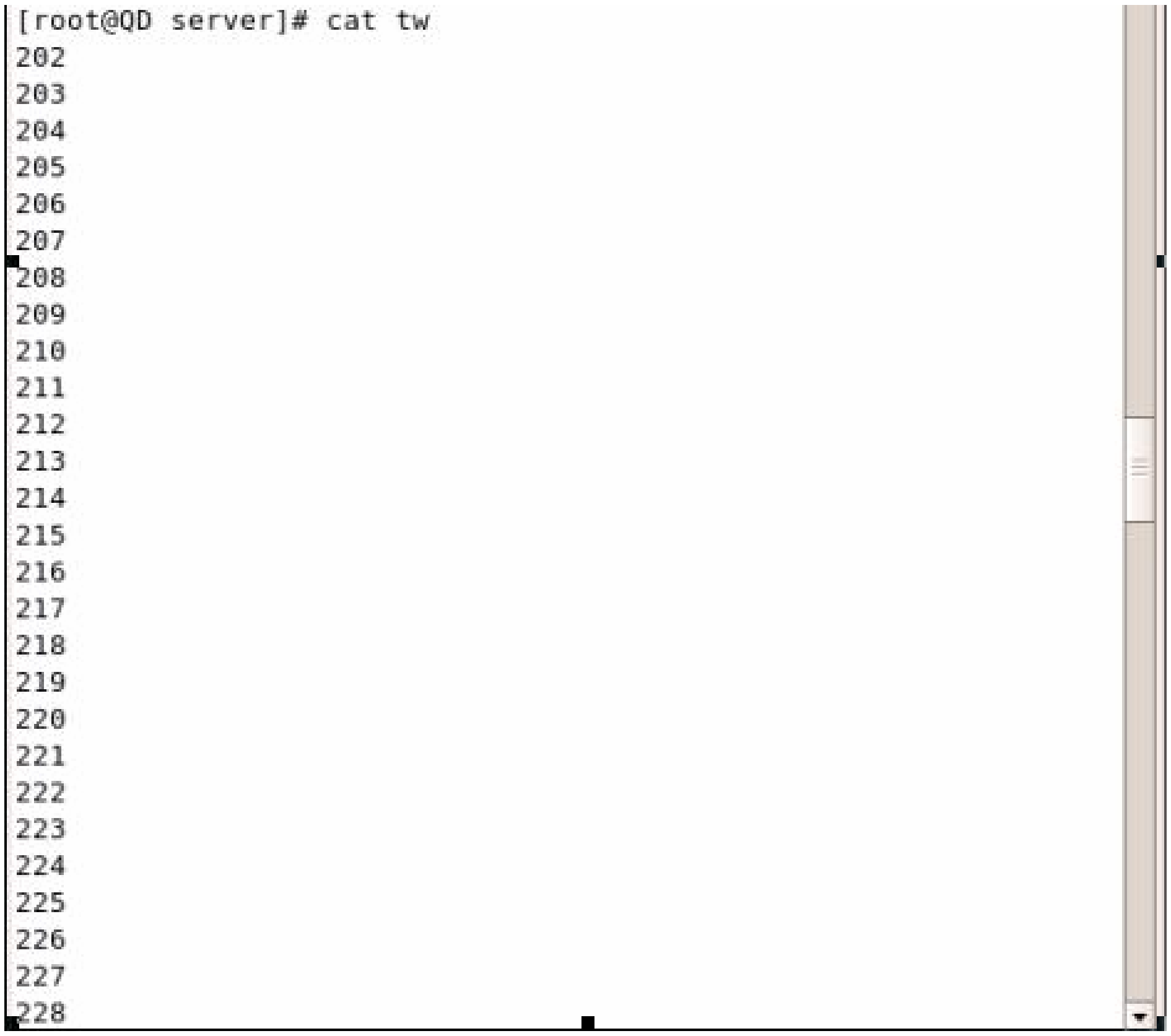

**Figure 31: Case study 1 - The server data storage after all replicas finished.**

It showed all records are fully processed, synchronized and written to the server data storage as if they were processed by one replica.

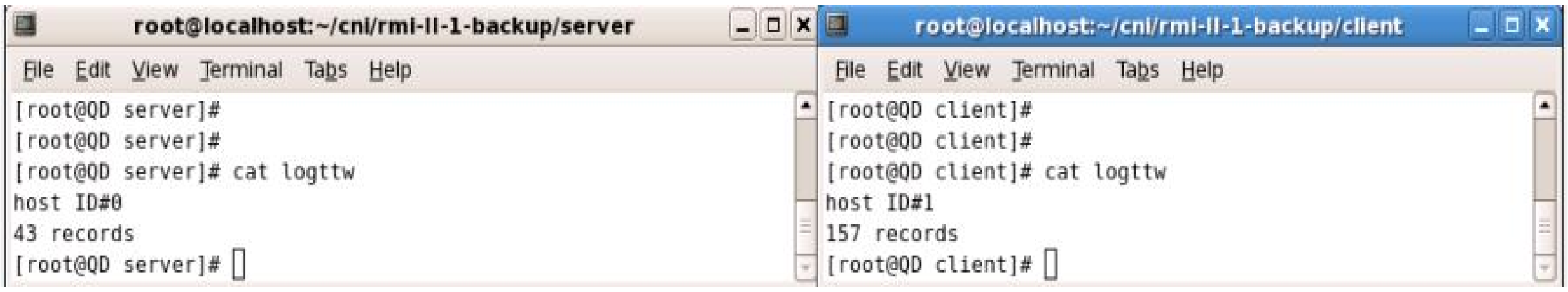

**Figure 32: Case study 1 - Statistic for each replica**

The statistic for the times running this experiment: The total records need to be processed are 200 records. Host #0 on the server side consumed 43 records. Host #1 on the client side consumed 157 records.

Due to the difficulties met in the setup phase, as we did not have the privilege to install the middleware in a controlled fashion on multiple hosts at the same time in the LAN, we manually setup only two machines in our lab. Thus, this case study was done with the replicas running on two hosts only. But there should not be any problem in running this experiment on multiple hosts. If more than two machines are used, the performance should be better, i.e., the total deployment time should decrease when the number of available hosts joining the middleware increases.

### 5.1.4 Statistics

The following diagrams show the comparison for two cases.

Case 1: Host 1 runs server-core processing and one replica. This is exactly as the original application, before we provide the middleware, exploits on one host. The result of this experiment is illustrated with the *light/blue bar*.

Case 2: Host 1 runs server-core processing and one replica. Host 2 runs one replica. The result of this experiment is illustrated with the *dark/red bar*.

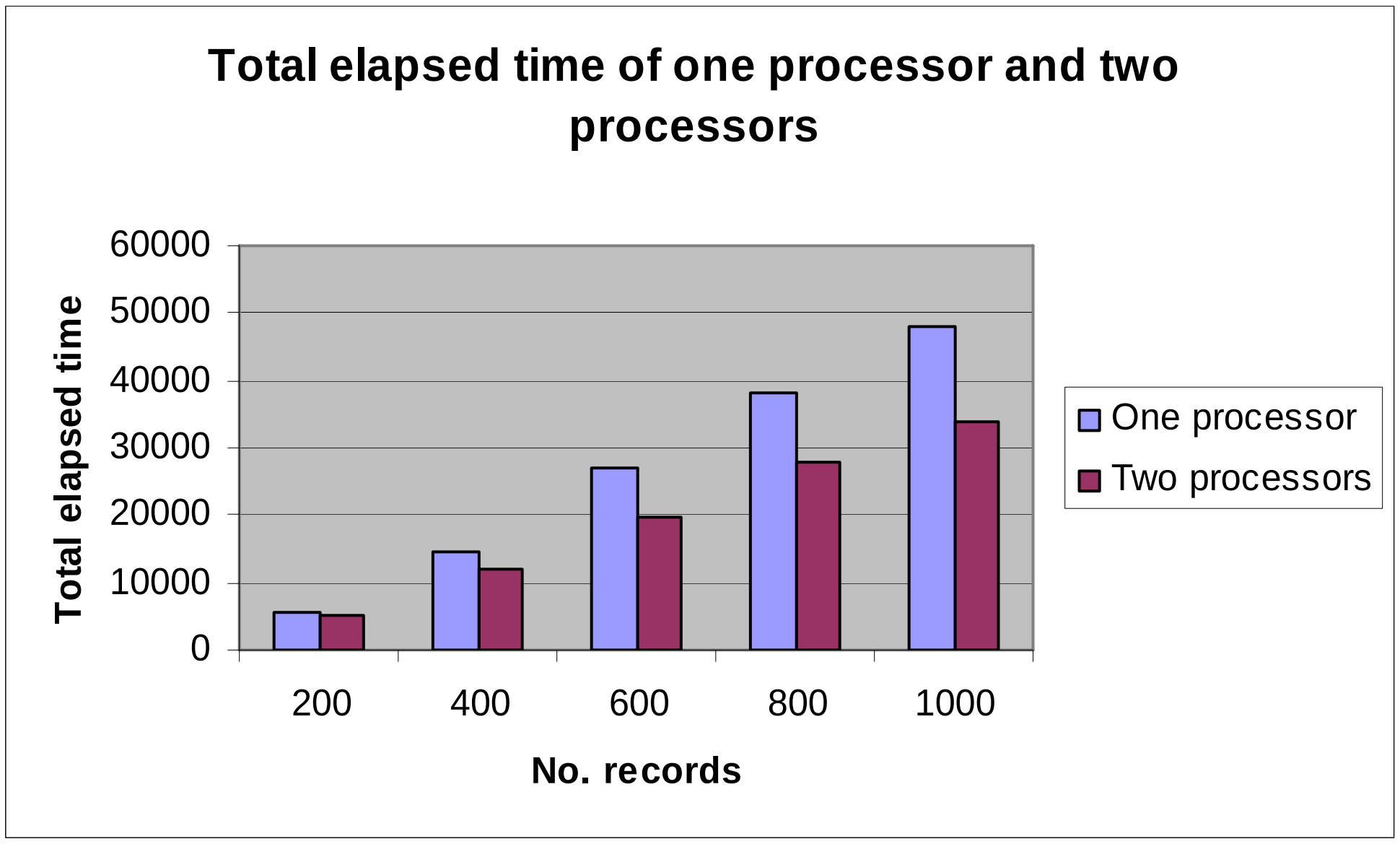

**Figure 33: Case study 1 - Total elapsed time (TET) of one processor and two processors**

Figure 33 shows the comparison between Case 1 and Case 2 when running for 200, 400, 600, 800 and 1000 records. It illustrates clearly that *Case 2* always has *less* processing time than Case 1. Specifically, note that the larger the number of records to be processed, *bigger* the *gap* between Case 1 and Case 2.

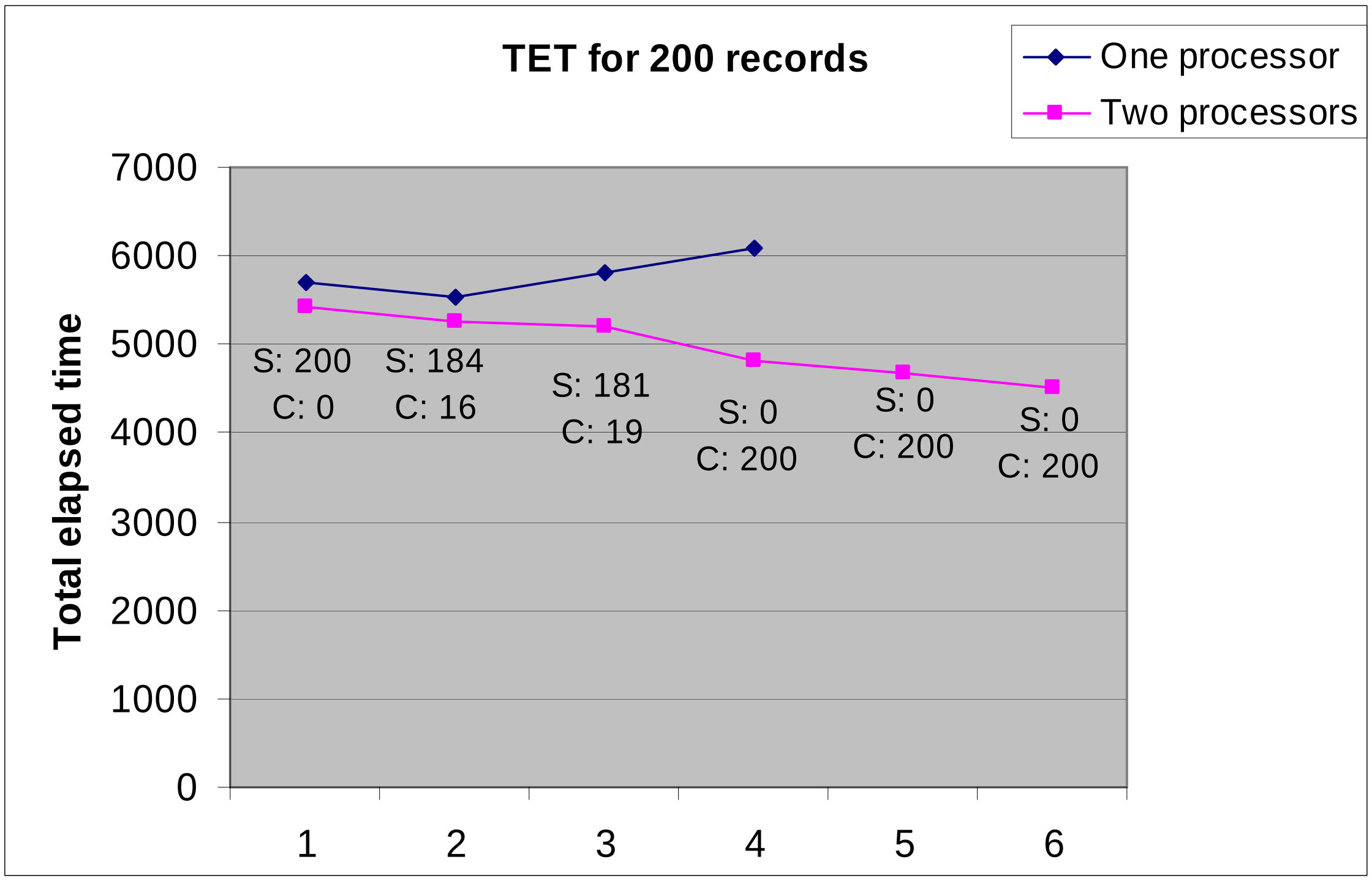

**Figure 34: Case study 1 - Total elapsed time for 200 records of one processor and two processors**

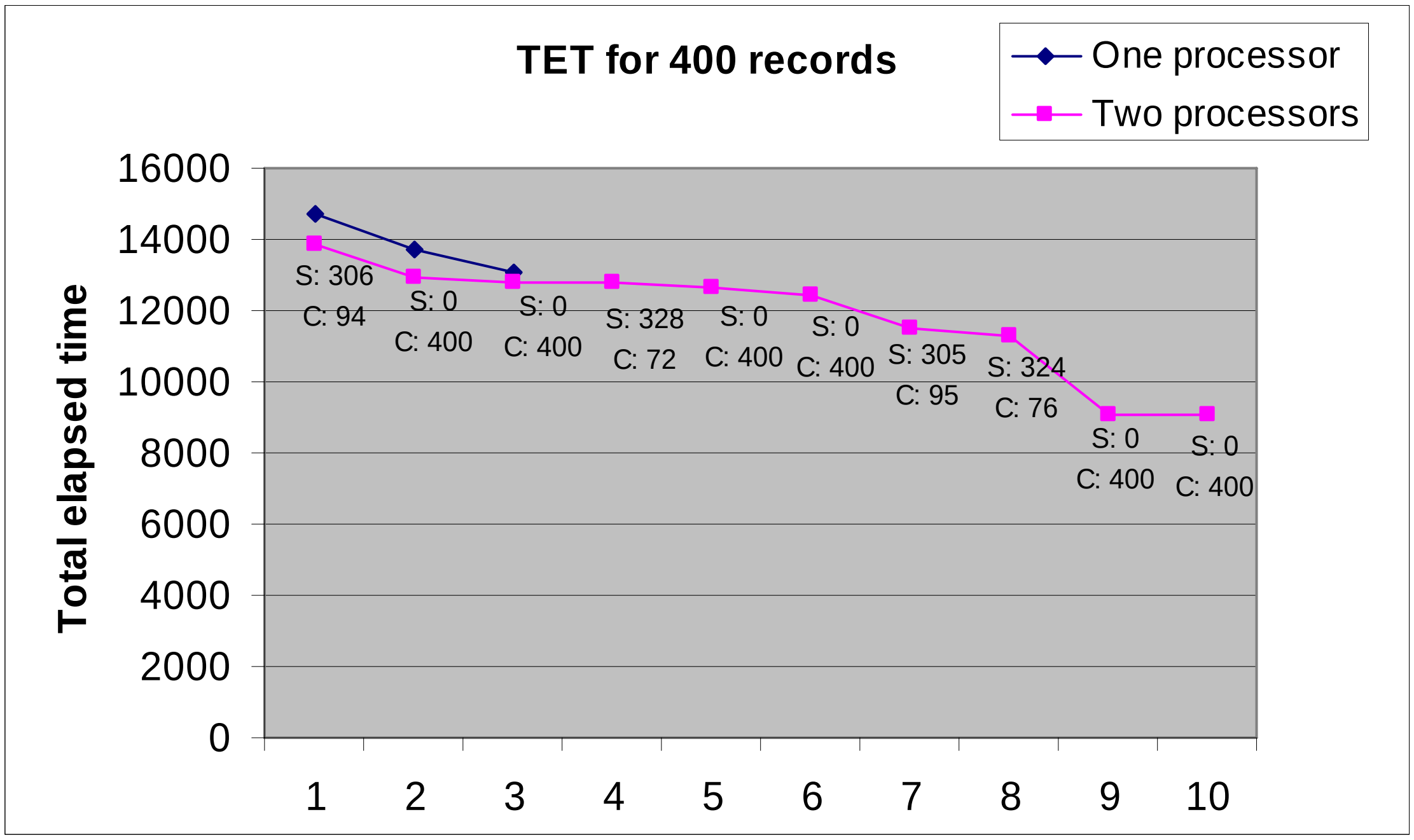

**Figure 35: Case study 1 - Total elapsed time for 400 records of one processor and two processors**

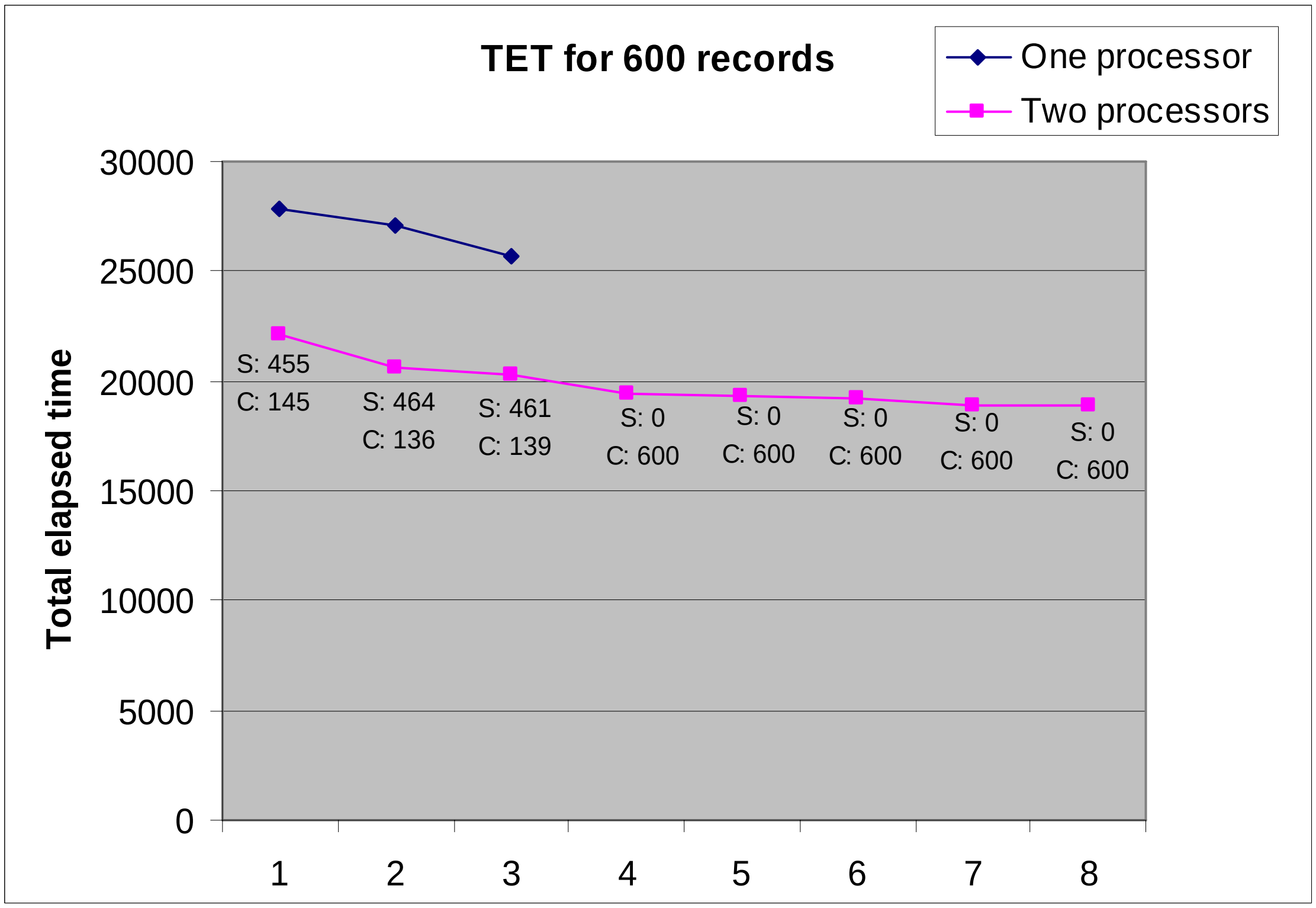

**Figure 36: Case study 1 - Total elapsed time for 600 records of one processor and two processors**

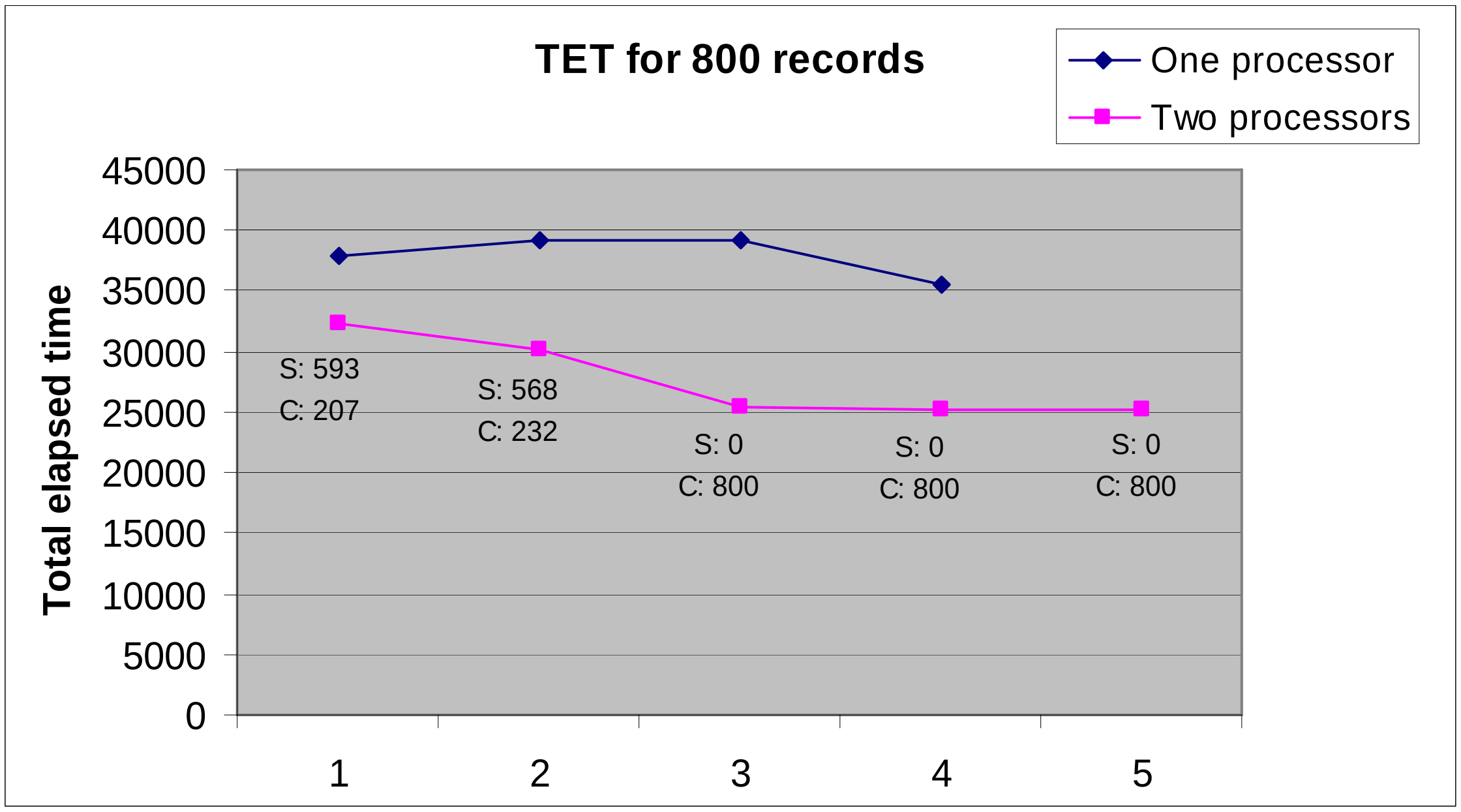

**Figure 37: Case study 1 - Total elapsed time for 800 records of one processor and two processors**

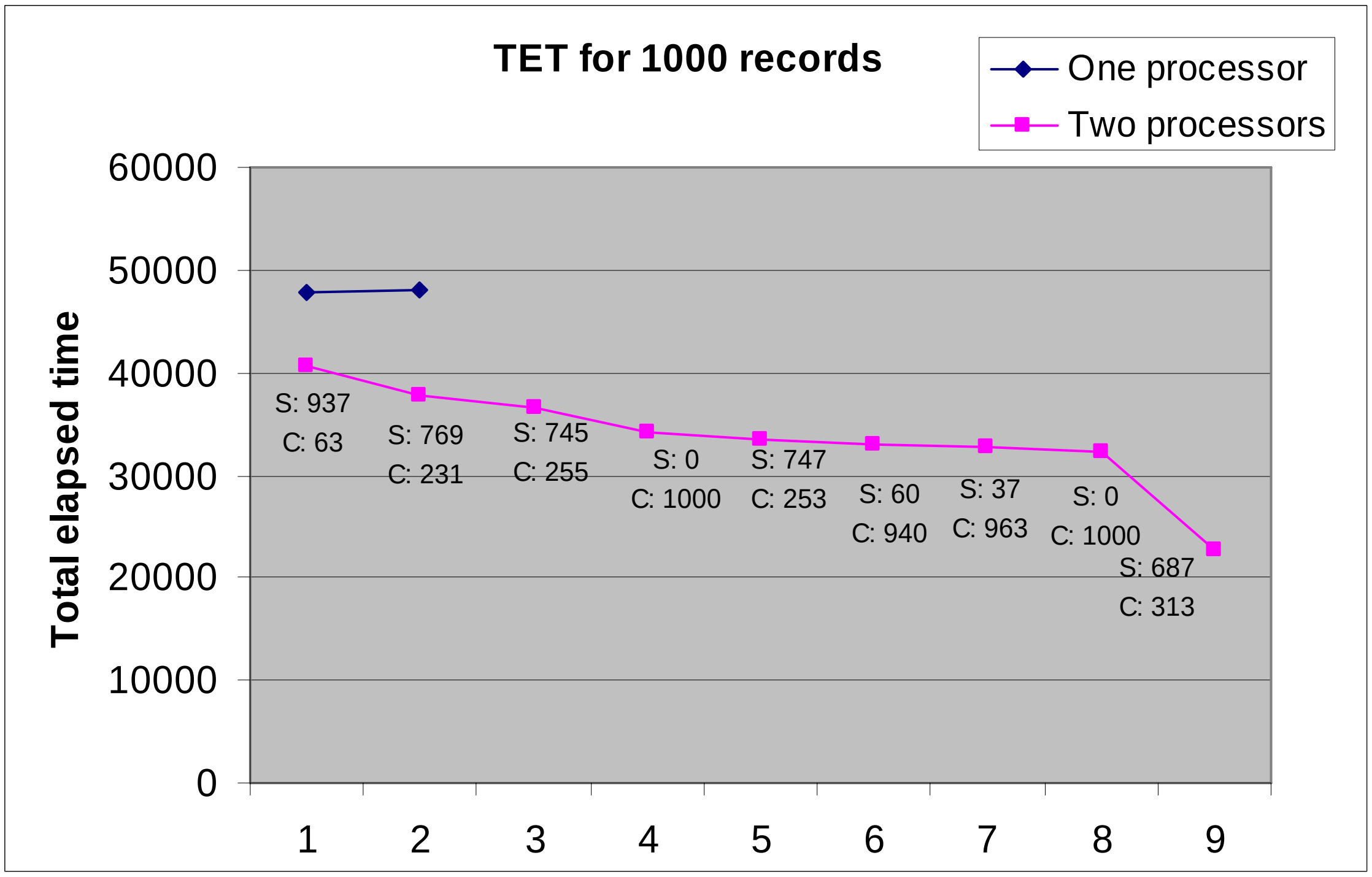

**Figure 38: Case study 1 - Total elapsed time for 1000 records of one processor and two processors**

Figures 34–38 above show the results for 200, 400, 600, 800, 1000 records separately. It can be noted that the time to process for one processor (the dark/blue line on diamond points) is always greater than the time to process for two processors (the light/pink line

on square points). Depending on their availability when they join the middleware, the clients can share different percentage of jobs with the server. From these figures, it is clear that the processing time reduction is higher when a client shares higher percentage of work with the server.

## 5.2 Case study 2: Iozone, a filesystem benchmark tool

### 5.2.1 Application overview

Iozone [12] is a filesystem benchmark tool which generates and measures a variety of file operations. Iozone has been ported to many platforms and runs under many operating systems. Iozone is useful for performing a broad filesystem analysis of a vendor's computer platform. The benchmark tests file I/O performance for the following operations: read, write, re-read, re-write, read backwards, read strided, fread, fwrite, random read, pread, mmap, aio_read, aio_write.

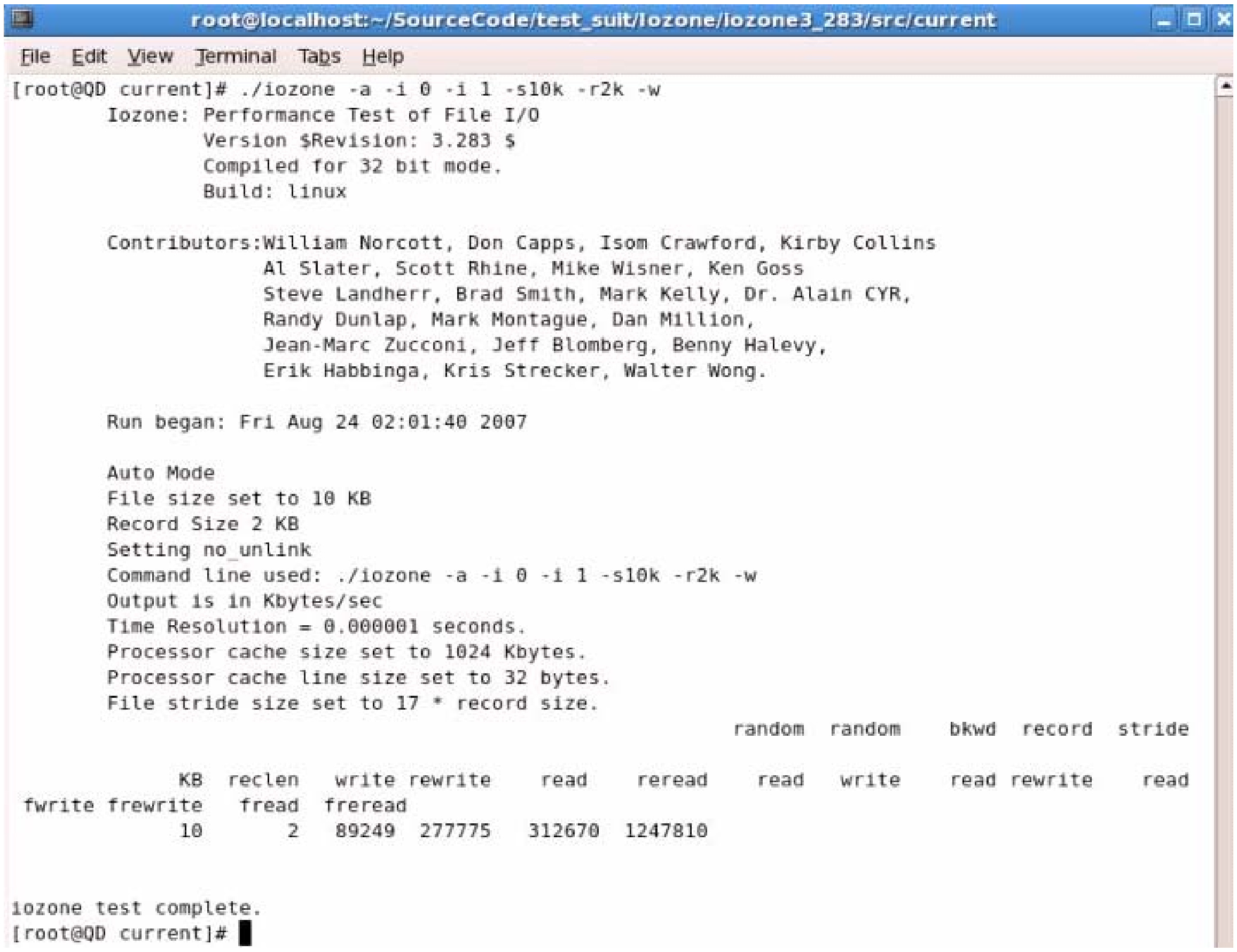

**Figure 39: Case study 2 - Iozone shows the result after it finished processing.**

This benchmark is experimented with multiple writing and then reading to a file of size 10KB with each record size of 2KB. The experiment was done with the following command line:

```
./iozone –a –i 0 –I 1 –s10K –r2k -w
```

The idea is that the benchmark can be applied as an application with a lot of data to be written to the data storage and then delivered to any consumer.

### 5.2.2 Application operations

Application opens iozone.tmp file to write in and then read data from this file. The middleware can recognize and observe the application's operations with data access write/read system calls.

```
open("iozone.tmp", O_RDWR|O_LARGEFILE)  = 3
fsync(3)                                = 0
gettimeofday({1183588317, 276622}, NULL) = 0
write(3, "::::::::\0\0\0\0\0\0\0\0\0\0\0\0\0\0\0\0\0\0\0\0\0\0\0\0"..., 2048) = 2048
write(3, "::::::::\0\0\0\0\0\0\0\0\0\0\0\0\0\0\0\0\0\0\0\0\0\0\0\0"..., 2048) = 2048
write(3, "::::::::\0\0\0\0\0\0\0\0\0\0\0\0\0\0\0\0\0\0\0\0\0\0\0\0"..., 2048) = 2048
write(3, "::::::::\0\0\0\0\0\0\0\0\0\0\0\0\0\0\0\0\0\0\0\0\0\0\0\0"..., 2048) = 2048
write(3, "::::::::\0\0\0\0\0\0\0\0\0\0\0\0\0\0\0\0\0\0\0\0\0\0\0\0"..., 2048) = 2048
gettimeofday({1183588317, 277017}, NULL) = 0
fsync(3)                                = 0
close(3)                                = 0
write(1, "   25250", 8   25250)                   = 8
write(1, "   28494", 8   28494)                   = 8
open("iozone.tmp", O_RDONLY|O_LARGEFILE) = 3
fsync(3)                                = 0
read(3, "::::::::\0\0\0\0\0\0\0\0\0\0\0\0\0\0\0\0\0\0\0\0\0\0\0\0"..., 4096) = 4096
_llseek(3, 0, [0], SEEK_SET)            = 0
gettimeofday({1183588317, 278221}, NULL) = 0
read(3, "::::::::\0\0\0\0\0\0\0\0\0\0\0\0\0\0\0\0\0\0\0\0\0\0\0\0"..., 2048) = 2048
read(3, "::::::::\0\0\0\0\0\0\0\0\0\0\0\0\0\0\0\0\0\0\0\0\0\0\0\0"..., 2048) = 2048
read(3, "::::::::\0\0\0\0\0\0\0\0\0\0\0\0\0\0\0\0\0\0\0\0\0\0\0\0"..., 2048) = 2048
read(3, "::::::::\0\0\0\0\0\0\0\0\0\0\0\0\0\0\0\0\0\0\0\0\0\0\0\0"..., 2048) = 2048
read(3, "::::::::\0\0\0\0\0\0\0\0\0\0\0\0\0\0\0\0\0\0\0\0\0\0\0\0"..., 2048) = 2048
gettimeofday({1183588317, 278606}, NULL) = 0
fsync(3)                                = 0
close(3)                                = 0
```

**Figure 40: Case study 2 - Iozone is observed at low level of the wrapper**

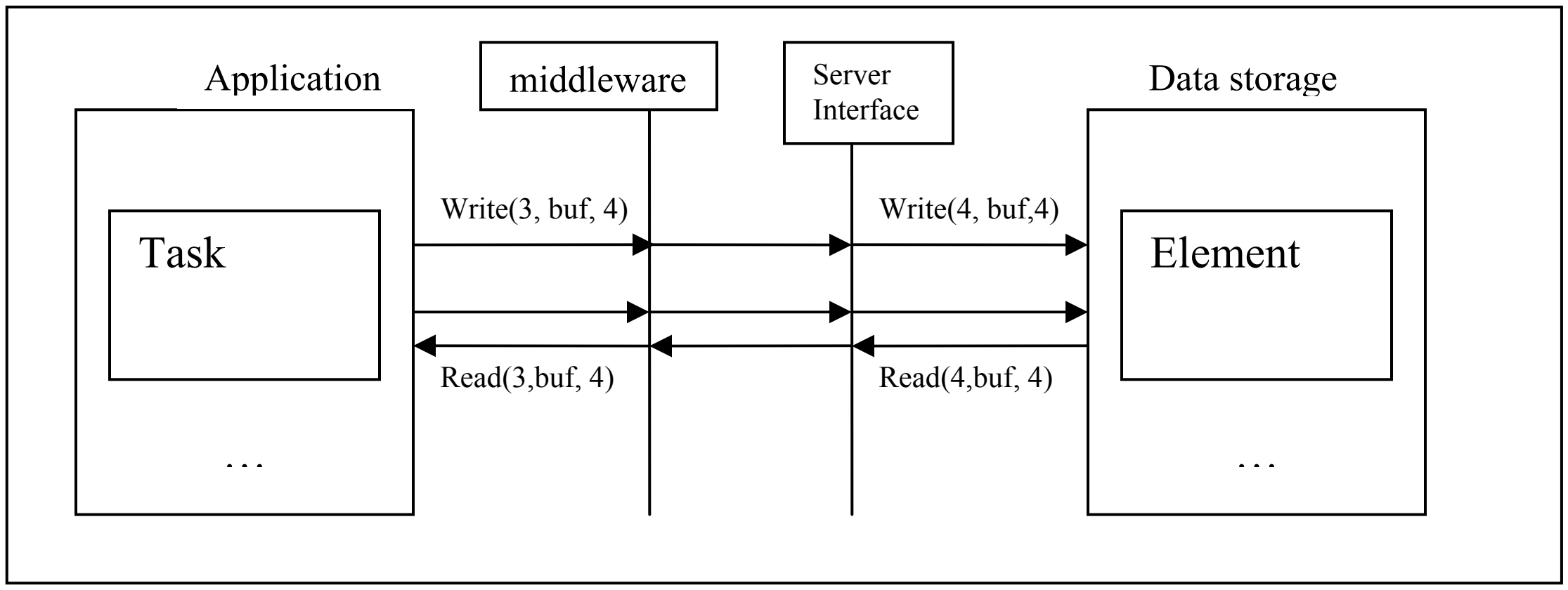

**Figure 41: Case study 2 - Iozone is processed for data request**

This example was intended as a proof of concept for the middleware. It showed that the read and write system calls can be captured by our wrapper, directed to the server side and the data request was processed correctly as per the design.

# 5.3 Case study 3: File backup

## 5.3.1 Application overview

This application updates records from a file. This application does pretty much the same activities like the application in Case study 1 that reads a record, does a computation and updates that record. The result will be written back into another file for back-up purpose. This case study illustrates some points mentioned in the partitioning discussion. This application is written in Java in which a *read* system call reads a big chunk of data and fetch it into memory to be shared with other requests (unlike the C program in which each *read* system call represents one data request and the returned result is equal to one data record). The case study is written in two versions: the first one is a multi-threaded application and the second one is a single-threaded application.

### Observation

```
[pid 31524] open("README_InputFile.txt", O_RDONLY|O_LARGEFILE) = 3
[pid 31524] fstat64(3, {st_mode=S_IFREG|0644, st_size=20, ...}) = 0
[pid 31524] gettimeofday({1191534486, 327268}, NULL) = 0
[pid 31524] gettimeofday({1191534486, 327323}, NULL) = 0
[pid 31524] gettimeofday({1191534486, 327372}, NULL) = 0
[pid 31524] gettimeofday({1191534486, 327420}, NULL) = 0
[pid 31524] gettimeofday({1191534486, 327475}, NULL) = 0
[pid 31524] gettimeofday({1191534486, 327524}, NULL) = 0
[pid 31524] gettimeofday({1191534486, 327571}, NULL) = 0
[pid 31524] gettimeofday({1191534486, 327640}, NULL) = 0
[pid 31524] open("ReadWriteConcurrent.out", O_WRONLY|O_CREAT|O_TRUNC|O_LARGEFILE, 0666) = 4
[pid 31524] fstat64(4, {st_mode=S_IFREG|0644, st_size=0, ...}) = 0
[pid 31524] gettimeofday({1191534486, 328733}, NULL) = 0
[pid 31524] gettimeofday({1191534486, 328796}, NULL) = 0
[pid 31524] stat64("/root/JavaProgram/accessfile/RWConcurrency/ReadWriteConcurrent
$ChildConcurrent.class", {st_mode=S_IFREG|0644, st_size=1902, ...}) = 0
[pid 31524] open("/root/JavaProgram/accessfile/RWConcurrency/ReadWriteConcurrent
$ChildConcurrent.class", O_RDONLY|O_LARGEFILE) = 5
[pid 31524] fstat64(5, {st_mode=S_IFREG|0644, st_size=1902, ...}) = 0
[pid 31524] stat64("/root/JavaProgram/accessfile/RWConcurrency/ReadWriteConcurrent
$ChildConcurrent.class", {st_mode=S_IFREG|0644, st_size=1902, ...}) = 0
[pid 31524] read(5, "\312\376\272\276\0\0\0002\0q\t\0\35\0002\n\0\36\0003\t"..., 1902) = 1902
[pid 31524] close(5)                    = 0
[pid 31524] gettimeofday({1191534486, 330263}, NULL) = 0
[pid 31524] gettimeofday({1191534486, 330718}, NULL) = 0
[pid 31524] gettimeofday({1191534486, 330858}, NULL) = 0
[pid 31524] gettimeofday({1191534486, 330908}, NULL) = 0
[pid 31524] gettimeofday({1191534486, 331224}, NULL) = 0
[pid 31524] gettimeofday({1191534486, 331315}, NULL) = 0
[pid 31524] gettimeofday({1191534486, 331365}, NULL) = 0
[pid 31524] gettimeofday({1191534486, 331413}, NULL) = 0
[pid 31524] mmap2(NULL, 331776, PROT_READ|PROT_WRITE|PROT_EXEC, MAP_PRIVATE|MAP_ANONYMOUS, -1, 0)
= 0xb585f000

[pid 31532] read(3,  <unfinished ...>
[pid 31534] <... sched_getaffinity resumed>  { 1 }) = 4
[pid 31533] <... sched_getaffinity resumed>  { 1 }) = 4
[pid 31532] <... read resumed> "101\n102\n103\n104\n105\n", 8192) = 20
```

File is opened and starts to observe the shared file

map file to memory to share

read data > one record_data size

**Figure 42: Case study 3 - The Java multi-threaded application is observed at the wrapper level**

With a multi-threaded application, the read-in data fetched into memory will be shared with other processes. The middleware detects the *mmap* system call. This happens before the first *read* system call since the time the shared data storage is opened for concurrent data requests. Note that this mapped memory will be shared among processes in this replica, meaning local shared memory. The case can also be realized when the size of the first read system call is greater than the size of one data record size.

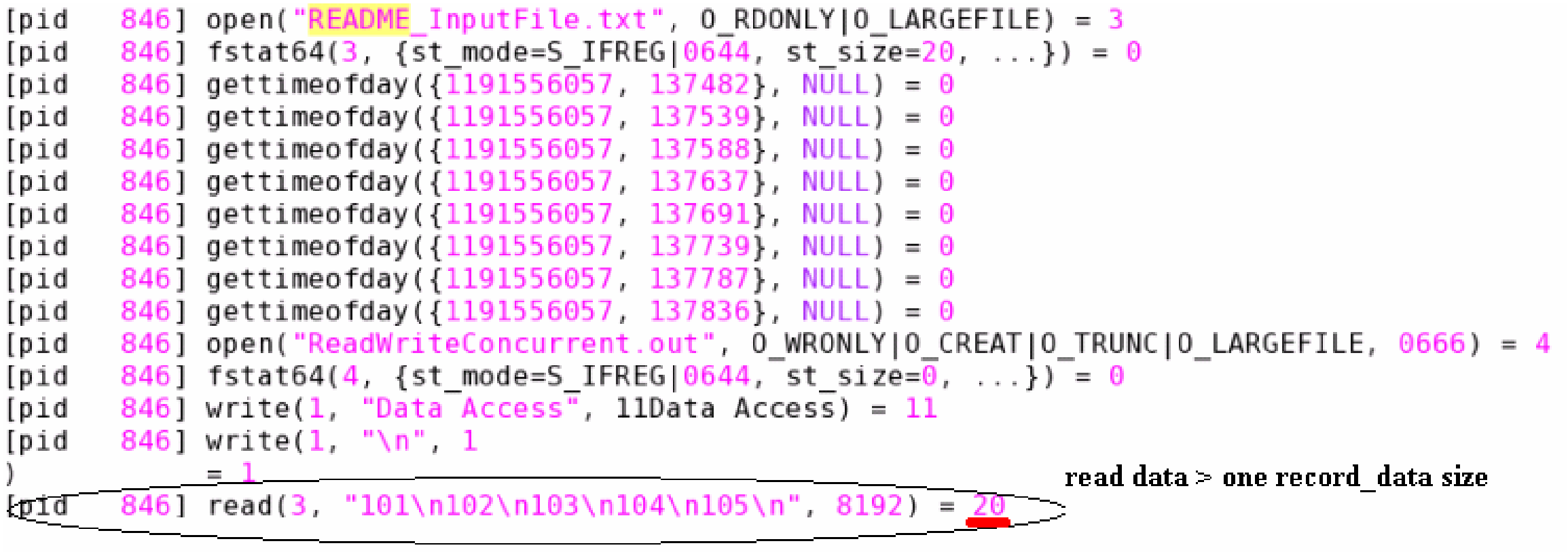

```
[pid   846] open("README_InputFile.txt", O_RDONLY|O_LARGEFILE) = 3
[pid   846] fstat64(3, {st_mode=S_IFREG|0644, st_size=20, ...}) = 0
[pid   846] gettimeofday({1191556057, 137482}, NULL) = 0
[pid   846] gettimeofday({1191556057, 137539}, NULL) = 0
[pid   846] gettimeofday({1191556057, 137588}, NULL) = 0
[pid   846] gettimeofday({1191556057, 137637}, NULL) = 0
[pid   846] gettimeofday({1191556057, 137691}, NULL) = 0
[pid   846] gettimeofday({1191556057, 137739}, NULL) = 0
[pid   846] gettimeofday({1191556057, 137787}, NULL) = 0
[pid   846] gettimeofday({1191556057, 137836}, NULL) = 0
[pid   846] open("ReadWriteConcurrent.out", O_WRONLY|O_CREAT|O_TRUNC|O_LARGEFILE, 0666) = 4
[pid   846] fstat64(4, {st_mode=S_IFREG|0644, st_size=0, ...}) = 0
[pid   846] write(1, "Data Access", 11Data Access) = 11
[pid   846] write(1, "\n", 1
)           = 1
[pid   846] read(3, "101\n102\n103\n104\n105\n", 8192) = 20
```

**Figure 43: Cases study 3 - The Java single-threaded application is observed at the wrapper level**

With a single-threaded application, the middleware does not detect *mmap*; but still the observation showed that the first *read* system call reads in a big chunk of data into memory to share with subsequent data requests. This shared memory is also local space being used only by consecutive data requests of this replica, not shared with other replicas.

For either case, the middleware realizes that there is shared data among data requests of a replica based on the first *read* system call and the size of the read-in data chunk. And the returned data from the read system call will be fetched into memory to be shared among data requests of that replica. Thus, each replica still works independent of the other and data partitioning is handled appropriately.

# CHAPTER 6: Conclusion and Future work

## 6.1 Summary

This thesis presented an approach to automatically execute a legacy application (for which the source code is not available) on multiple hosts within an enterprise LAN distributively. This approach is implemented in a distributed middleware which runs the target legacy application in a dynamic network in which the number of hosts available for the computation and their CPU availability change over time. However, for the legacy application to run correctly, the server data storage of the application has to be split into disjoint partitions and each replica of the target application can operate on one partition without any conflict. We evaluated the implemented middleware on three typical test cases and found it to work as expected.

## 6.2 Contribution

We developed two approaches, namely the internal analysis and the external analysis, which analyzed an application at runtime to extract the parallel tasks in that application. Existing techniques use either one of these approaches with some specific application, but we applied the combination of the two on general applications and with fewer assumptions.

We developed a dynamic scheduling algorithm to exploit the dynamic CPU scavenging in a LAN. Existing dynamic scheduling algorithms assume that each host in the LAN has the same processing capacity, but we permit the hosts to have different processing capacities in an enterprise environment. Furthermore, we also allow the status of the hosts dynamically change from time to time. Our dynamic scheduling adapts to these situations.

We built a distributed middleware which can automatically wrap up an existing legacy application and execute it using multiple hosts in a LAN. This middleware also partitions the data storage in a transparent way to support such parallel execution maintaining the semantics of the application. We evaluated the middleware on a couple of applications to demonstrate the feasibility of the design. Also, the test results showed significant improvement in utilization/performance as expected.

## 6.3 Future work

Since the required information about the application is very limited, the middleware can be used on very broad types of applications. If more information about the application is available, the design can be further improved. For applications with internal repeated computation, our design cannot observe internal shared variables among those computations. Alternatively, if the shared variables are mapped into memory, a shared memory wrapper abstraction needs to be integrated on the server side. For applications with independent components, we already examined component dependency after replicating.

Another interesting point is that in Windows, there is a fairly general way to interpose an object providing identical interfaces. But within the operating Windows, there are a few mechanisms, including file systems, to interpose the virtual memory subsystem, which allow the developers to reuse functionality of the OS. This can bring out a couple of ways to handle the shared memory which are not discussed in this thesis.

On a slightly different note, this thesis provides a wrapper to overcome limitations of the observed application and build up the server side with some distributed features. This approach can also be applied to achieve other features, such as reliability, availability, etc.